\documentclass[aps,prb,amssymb,reprint,twocolumn,superscriptaddressgroupaddress]{revtex4-1}
\usepackage{amsmath}
\usepackage{hyperref}
\usepackage[english]{babel}
\usepackage[utf8]{inputenc}
\usepackage{float}
\usepackage{graphicx}
\usepackage{braket}
\usepackage{xcolor}
\allowdisplaybreaks\begin{document}
	\title{Collective ground states in small lattices of coupled quantum dots}
	\author{DinhDuy Vu}
	\author{S. Das Sarma}
	\affiliation{Condensed Matter Theory Center and Joint Quantum Institute, Department of Physics, University of Maryland, College Park, Maryland 20742, USA}
	\begin{abstract}
		Motivated by recent developments on the fabrication and control of semiconductor-based quantum dots, we theoretically study a finite system of tunnel-coupled quantum dots with the electrons interacting through the long-range Coulomb interaction.  When the inter-electron separation is large and the quantum dot confinement potential is weak, the system behaves as an effective Wigner crystal with a period determined by the electron average density with considerable electron hopping throughout the system.  For stronger periodic confinement potentials, however, the system makes a crossover to a Mott-type ground state where the electrons are completely localized at the individual dots with little inter-dot tunneling. In between these two phases, the system is essentially a strongly correlated electron liquid with inter-site electron hopping constrained by strong Coulomb interaction.  We characterize this Wigner-Mott-liquid quantum crossover with detailed numerical finite-size diagonalization calculations of the coupled interacting quantum dot system, showing that these phases can be smoothly connected by tuning the system parameters. Experimental feasibility of observing such a hopping-tuned Wigner-Mott-liquid crossover in currently available semiconductor quantum dots is discussed. In particular, we connect our theoretical results to recent quantum-dot-based quantum emulation experiments where collective Coulomb blockade was demonstrated. We discuss realistic disorder effects on our theoretical findings. One conclusion of our work is that experiments must explore lower density quantum dot arrays in order to clearly observe the Wigner phase although the Mott-liquid crossover phenomenon should already manifest itself in the currently available quantum dot arrays. We also suggest a direct experimental electron density probe, such as atomic force microscopy or scanning tunneling microscopy, for a clear observation of the effective Wigner crystal phase.
	\end{abstract}
	\maketitle
	\section{Introduction}
	Wigner pointed out in 1934 \cite{Wigner} that free electrons interacting via the long-range Coulomb interaction (and in the presence of a compensating positive charge background to keep the system stable) may condense into a quantum crystal solid phase provided the electron density is low enough so that the quantum kinetic energy is overwhelmed by the Coulomb potential energy, leading to a periodic spatial density modulation instead of a uniform density distribution preferred by the noninteracting or the weakly interacting usual electron liquid system. The Wigner crystal is simply a crystal of electrons just as ordinary ions form a crystalline solid driven by their potential energy at not too high temperatures. The tricky issue for a quantum Wigner crystal is that the electron effective mass being very low compared with ionic masses, the quantum Wigner crystallization necessitates very low carrier densities (as well as very low temperatures) in order to overcome quantum fluctuations which prefer the electron liquid (or gas) phase so as to minimize the kinetic energy. In one dimensional systems where electrons interact via a long-range potential, a 1D quantum Wigner crystal manifesting true long-range order is not allowed but finite systems still have signatures of short range order associated with Wigner crystallization \cite{Wigner2018} and references therein. Indeed, finite size 1D Wigner crystals have been measured, most recently in Ref. \cite{exp0} in a carbon nanotube. In addition, a classical 2D Wigner crystal has been observed in low density electrons confined to the surface of He-4 \cite{wigner2d}.	
	 
	Fifteen years after Wigner’s prediction for quantum crystallization tuned by Coulomb interaction, Mott in 1949 conjectured a new type of ‘electron solid’, the Mott insulator, where band electrons could localize at the lattice sites when the hopping or inter-site-tunneling amplitude is suppressed strongly by increasing the lattice period \cite{mott}.  The underlying mechanism is that for weak enough electron hopping (i.e. for large enough lattice period) compared to the interaction strength, the metallic band electrons would simply find it energetically unfavorable to hop between lattice sites and become localized at individual sites to minimize their potential energy.  Although both Wigner and Mott solids are driven by electron-electron interactions with the itinerant metallic electron liquid phase going over to the localized electron solid phase \cite{Kohn, metal_insulator}, there are significant conceptual differences between the two localized phases, and the possible relationships between the two are virtually unexplored. The existence of the Wigner crystallization depends crucially on the long-range nature of Coulomb interaction in a free electron type model whereas all modern discussions of the Mott insulator are typically based on the Hubbard model, which is a tight binding description of a short-range on-site interaction between electrons of opposite spins.  The original 1949 idea of the Mott insulator did, however, invoke the Coulomb interaction between the electrons as the driving force for electron localization at the lattice sites. Wigner crystallization spontaneously breaks the translational invariance whereas the Mott metal-insulator transition obeys the periodic symmetry of the underlying lattice. Another conceptual difference between the two is that the Wigner solid can, in principle, conduct since the whole solid can move in the presence of an external electric field (unless pinned by impurities or defects) whereas the Mott solid is an insulator by definition (``Mott insulator") since the external lattice obviously pins the system allowing no electric charge conduction. In real life, Wigner crystals are almost always pinned since there is always some external pinning potential, thus typical Wigner crystals are also insulators although conceptually they do not have to be. In the current work, the electrons are always subjected to a background potential (albeit very weak in the Wigner case), and therefore, the Wigner and Mott solids in the context of our work are both insulators. In fact, even the finite-size electron liquid state in our work is strictly speaking an insulator with a very small energy gap because of the physics of Coulomb blockade associated with Coulomb interaction in any finite systems.
	
	In the current work, we theoretically study the crossover between Mott and Wigner insulators, specifically in the context of small arrays of coupled quantum dots in semiconductor structures.  Such small analog solid state quantum emulator systems consisting of a few (2-9) quantum dots have recently been developed in Delft \cite{Delft2017,Volk2019} and Princeton \cite{Princeton2019}. In fact, recent experiments and theory have emphasized the possibility of studying quantum ferromagnetism using quantum dot arrays \cite{nagaoka2019,ferromagnetism,nagaokanumerical}. In particular, exquisite quantum control and precise fabrication enable experimentalists to control the electrostatic environments of these coupled quantum dot arrays to such a degree that both the inter-dot tunneling and the number of electrons per dot can be tuned at will.  We emphasize that the fabrication and control of these coupled quantum dot arrays serve as the primary motivation of our work, and we firmly believe that it should be possible to eventually observe the evolution from a Wigner solid phase to a Mott insulator phase by appropriately tuning the electrostatic environment in such quantum dot arrays although it appears that an observation of the Wigner-like phase would be a challenge in the currently available quantum dot arrays where only the Mott and the liquid phase are accessible right now.  Our hope is to motivate such measurements in the future.
	
	It may be useful for us to recapitulate the elements of Wigner and Mott physics to motivate our work.  Conceptually, it is easier to start by discussing the quantum Wigner solid although it is much more difficult to achieve the crystal phase experimentally.  We imagine a collection of $N$ electrons interacting via the Coulomb interaction in a $d$-dimensional space of linear size $L$.  There must be some external potential confining the electrons so that they do not fly apart, and we assume that such a confining potential is present defining the total system size $L$.  Now, the problem is defined at $T=0$ simply by the dimensionless length $r_s =1/(na_B)$, where $n=N/L^d$ is the effective electron density and ‘$a_B$’ is a unit of length, conventionally taken to be the effective Bohr radius of the system.  Here $r_s$, sometimes called the Wigner-Seitz radius, is the average dimensionless separation between two electrons. Note that we are skipping over factors of various powers $\pi$ in defining $r_s$ for the sake of simplicity in the discussion. The Coulomb interaction between electrons goes as $1/r_s$  and the quantum kinetic energy goes as $n^2$ or $1/r_s^2$ based on the uncertainty principle.  This means that for low density or large $r_s$, the system would crystallize keeping the inter-electron separation maximal so as to minimize the potential energy cost whereas for small $r_s$, the system is an electron liquid to minimize the quantum fluctuations associated with the kinetic energy.  There have been many detailed numerical calculations evaluating the critical value $r_c$ of the $r_s$ parameter separating the Wigner solid phase from the electron liquid phase, and $r_c\sim 30$ in $d=2$ and $r_c\sim 100$ in $d=3$ \cite{Wignercrystal1,Wignercrystal2,Wignercrystal3}, making Wigner crystallization an extreme low-density phenomenon not of any relevance to ordinary metals which have $r_s\sim 5$.  In addition to having a low critical density, Wigner crystallization also necessitates a very low temperature ($\ll T_F$ where $T_F$ is corresponding Fermi temperature).  Note that Wigner crystallization automatically comes with a length scale of the Wigner crystal period, which is commensurate with $k_F$ since the crystal must have a period connected to the corresponding electron density. 
	
	While the Wigner physics arises from the competition between the free electron kinetic energy and the Coulomb potential energy (in the absence of any background lattice potential), the Mott insulator arises from the physics of narrow bands in a background periodic lattice where the quantum kinetic energy is typically the inter-site electron hopping energy $t$.  When this hopping energy is much smaller than the typical Coulomb energy between the electrons on neighboring sites, the system, according to Mott, will minimize the Coulomb energy by becoming completely localized on the lattice sites.  So the condition for Mott transition from a band metal to a Mott insulator on a tight binding lattice is that $t \ll E_c$.  Given that $t$ typically decays exponentially with the lattice spacing $a$ while $E_c \sim 1/a$, dimensional arguments suggest such a Mott transition at large values of $a$ although such an argument does not prove that this transition must happen and does not provide the critical lattice separation defining the Mott transition. We also note that strictly speaking a 1D system cannot have a true long-range ordered Wigner crystal although the situation would resemble a crystal on finite scale \cite{Schulz}. This distinction would not be important for our work since we manifestly consider small systems of current experimental interest in the semiconductor quantum dot qubit community.
	
	Since our motivation is to study coupled quantum dots, we specifically consider a one-dimensional (1D) array of $N_d$ dots, which we model by a periodic background potential. (Our system thus has two independent length scales defined by the lattice period and the average electron density.) The 1D nature of the system is both a great simplification and a substantial complication.  Many things are known exactly in 1D (unlike in higher dimensions) about interacting electron systems, both on a tight-binding lattice and in the continuum (i.e. free electron-like).  In particular, it is known that there is no Wigner crystallization phase transition in a Coulomb interacting 1D electron system in the thermodynamic limit (i.e. there is no unique $r_c$ for 1D Wigner crystallization), but the system develops strong short-range periodic order at the length scale of average electron separation which falls off very slowly (slower than any power law) \cite{Schulz,Coulomb_Luttinger1,Coulomb_Luttinger2}, and as such, it appears to be a crystal on finite length scale (although there is no long range order). By contrast,  a 1D half-filled periodic system with short-range interactions is known to be insulating \cite{Lieb}, which for our purpose can be construed to be a Mott insulator. Also, the `metallic' electron liquid phase is a Luttinger liquid in one dimension in contrast to a normal Fermi liquid as in higher dimensions although the Luttinger liquid aspects of the underlying physics do not enter  explicitly into our finite size theory where we use exact diagonalization to obtain our results.
	
	In our work, we vary the strength of the periodic potential in a 1D array in the presence of Coulomb interaction to smoothly interpolate between the strong tight binding limit, where the periodic potential is very strong, and the free electron limit, where the periodic potential is vanishingly small.  We calculate the electron density in the collective ground state as a function of system parameters to discern the Wigner and Mott limit in order to understand the Mott to Wigner crossover in small 1D coupled quantum dot arrays. The electron liquid phase (with a small but finite Coulomb gap) arises as the generic crossover phase in our results with Wigner and Mott phases showing up in the limit of very weak and very strong lattice potentials, respectively.
	
	The rest of the work is organized as follows.  In sec. II we provide our Hamiltonian, explain the various parameters controlling the crossover physics and present our exact diagonalization results.  In sec. III we describe the Mott-Wigner-liquid crossover in the system and provide an effective phase diagram. We also consider the classical situation briefly in sec. III. In sec. IV, we connect our results with a recent experiment reporting the observation of collective Coulomb blockade in quantum dot arrays. We conclude in sec. V discussing future experimental implications of our results and summarizing our main findings. We provide additional detailed results in the Appendix to complement the results in the main text.
	
	\section{Ground state of a coupled quantum dot array}	

  \begin{figure*}
  	\centering
  	\begin{minipage}{0.04\textwidth}
  		\rotatebox{90}{$\rho(x)a$}
  	\end{minipage}	
  	\begin{minipage}{0.95\textwidth}
  		\centering
  		\includegraphics[width=0.32\textwidth]{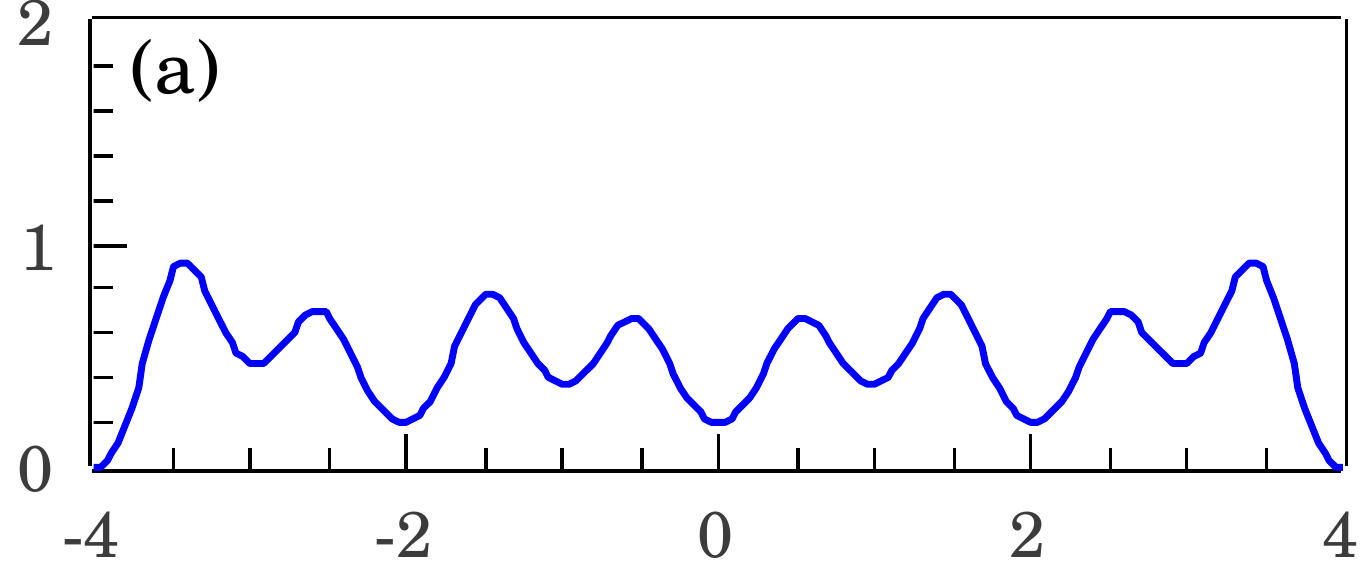}  	
  		\includegraphics[width=0.32\textwidth]{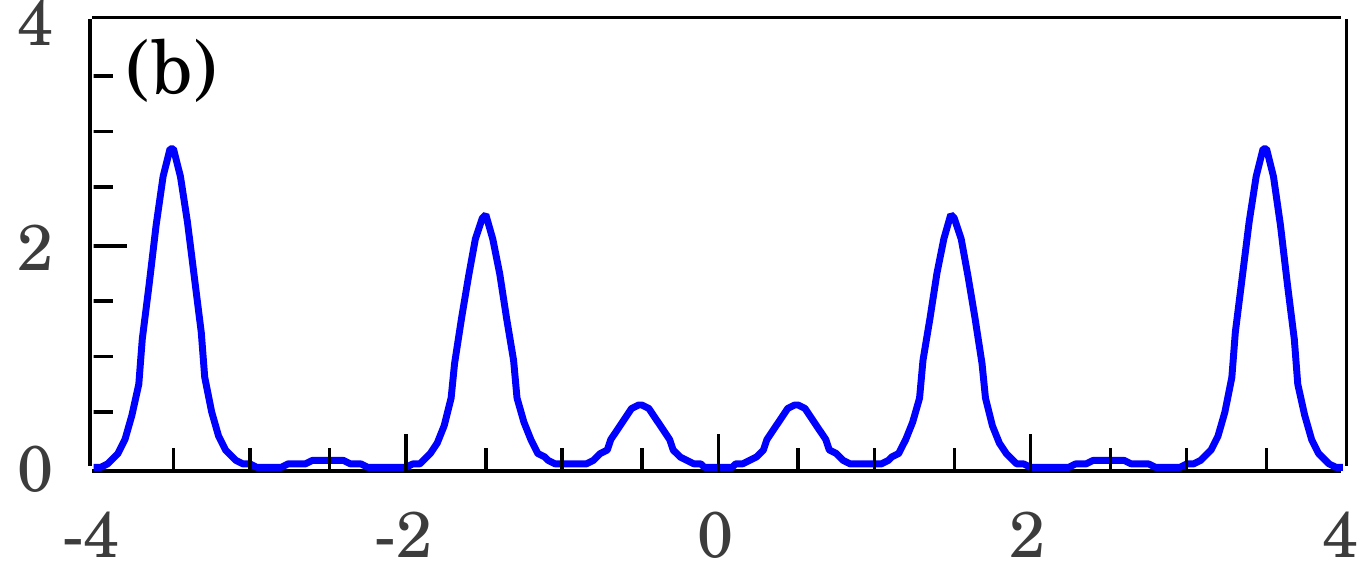}
  		\includegraphics[width=0.32\textwidth]{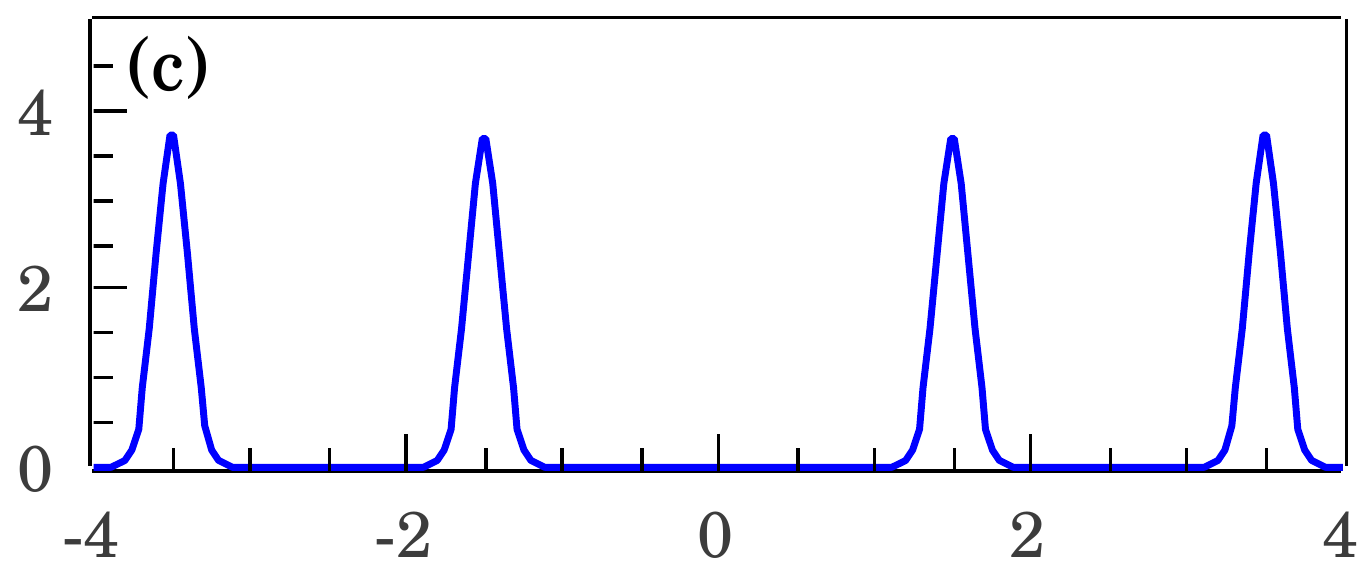} 
  		
  		\includegraphics[width=0.32\textwidth]{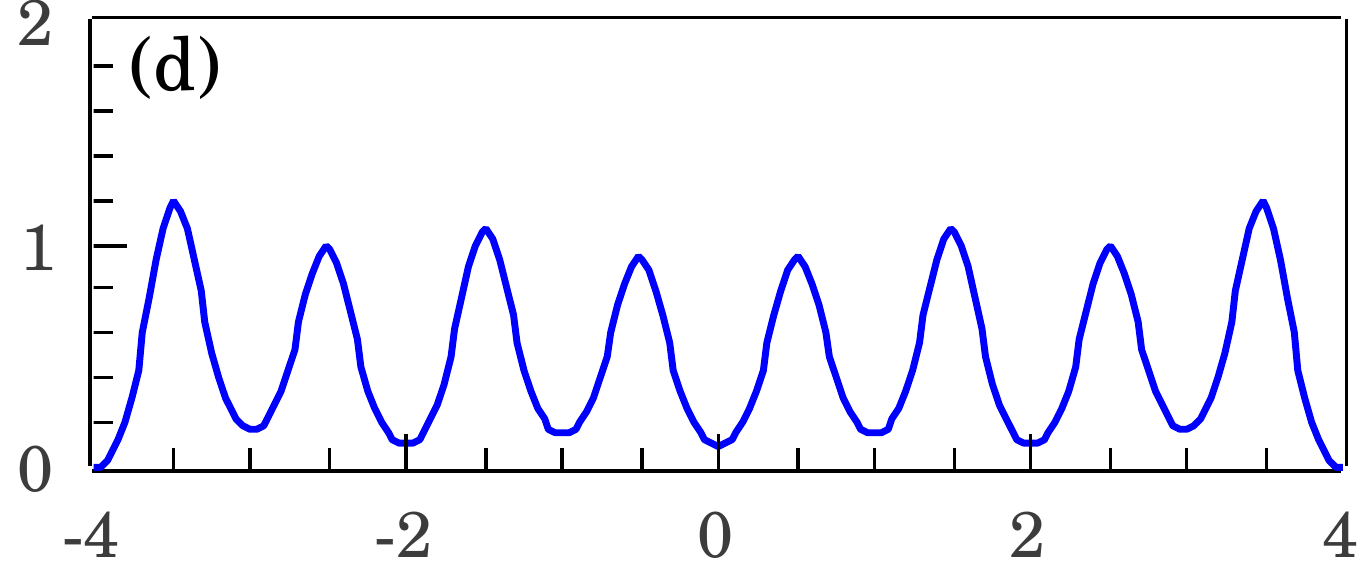} 
  		\includegraphics[width=0.32\textwidth]{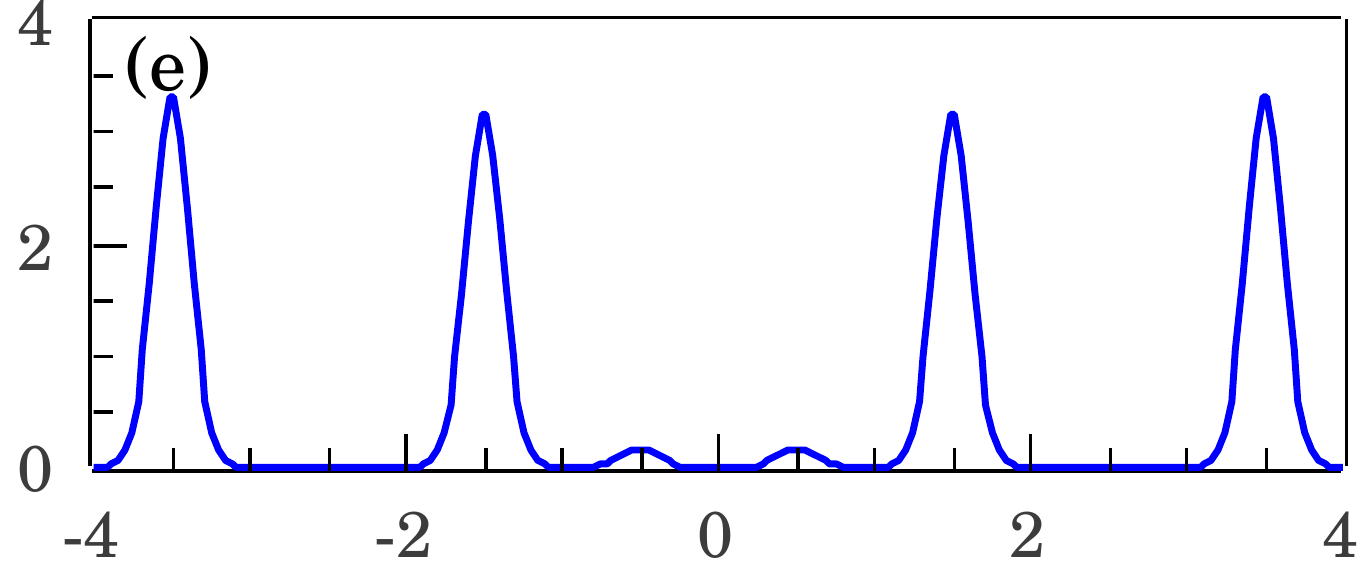} 
  		\includegraphics[width=0.32\textwidth]{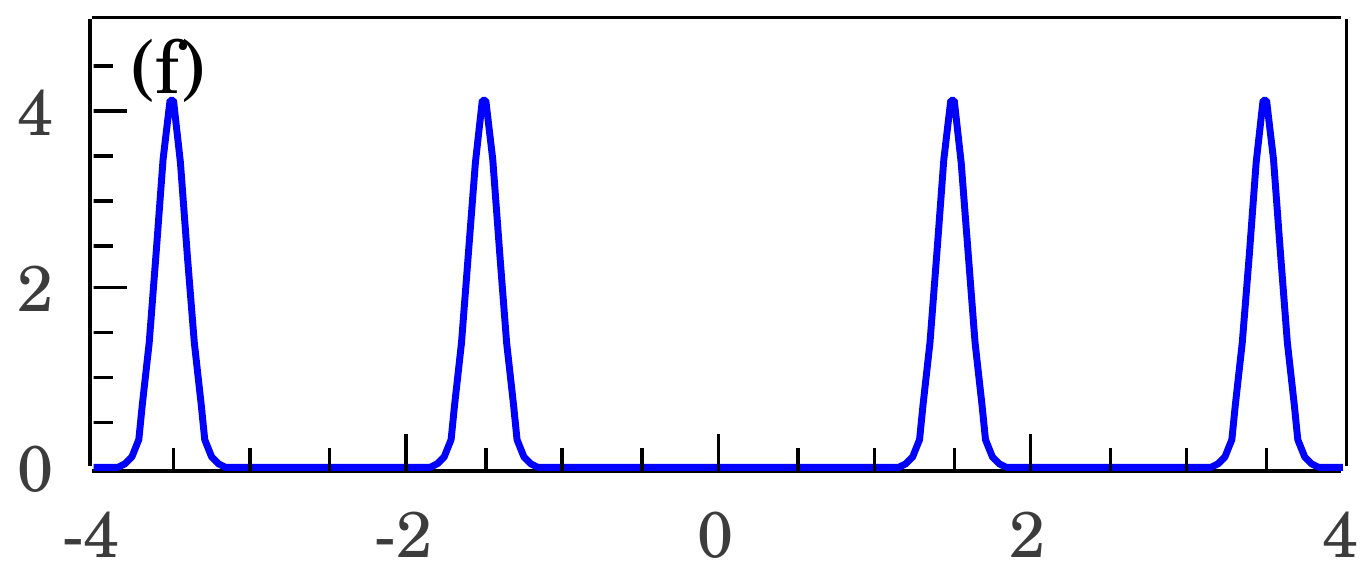} 
  		
  		\includegraphics[width=0.32\textwidth]{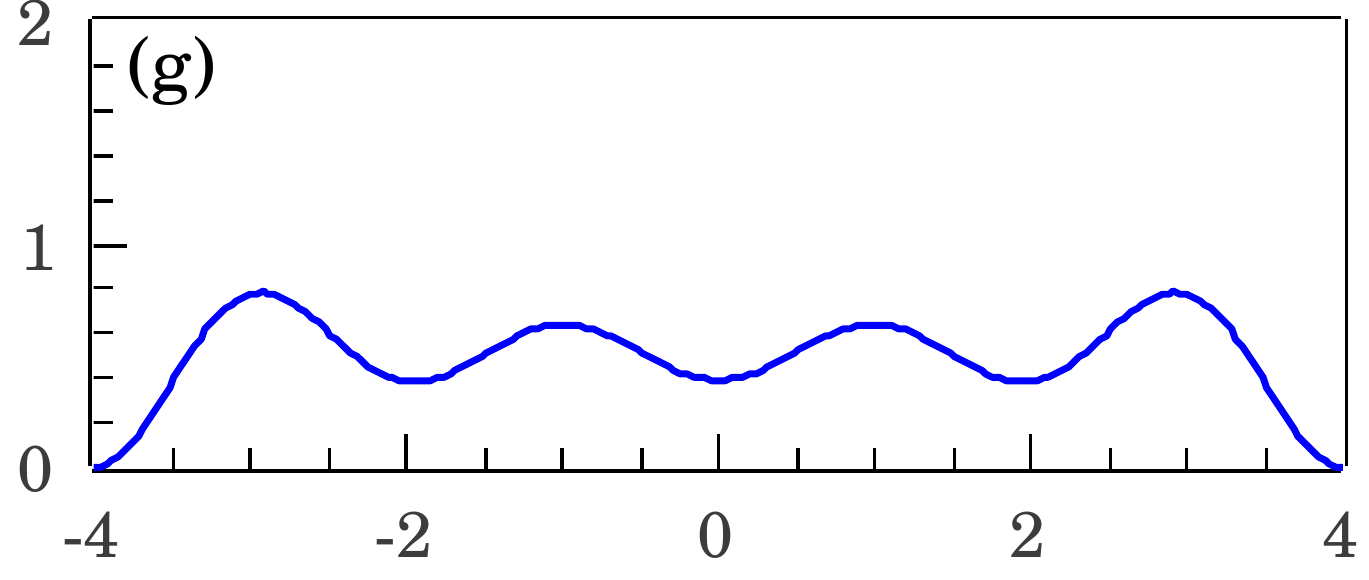} 
  		\includegraphics[width=0.32\textwidth]{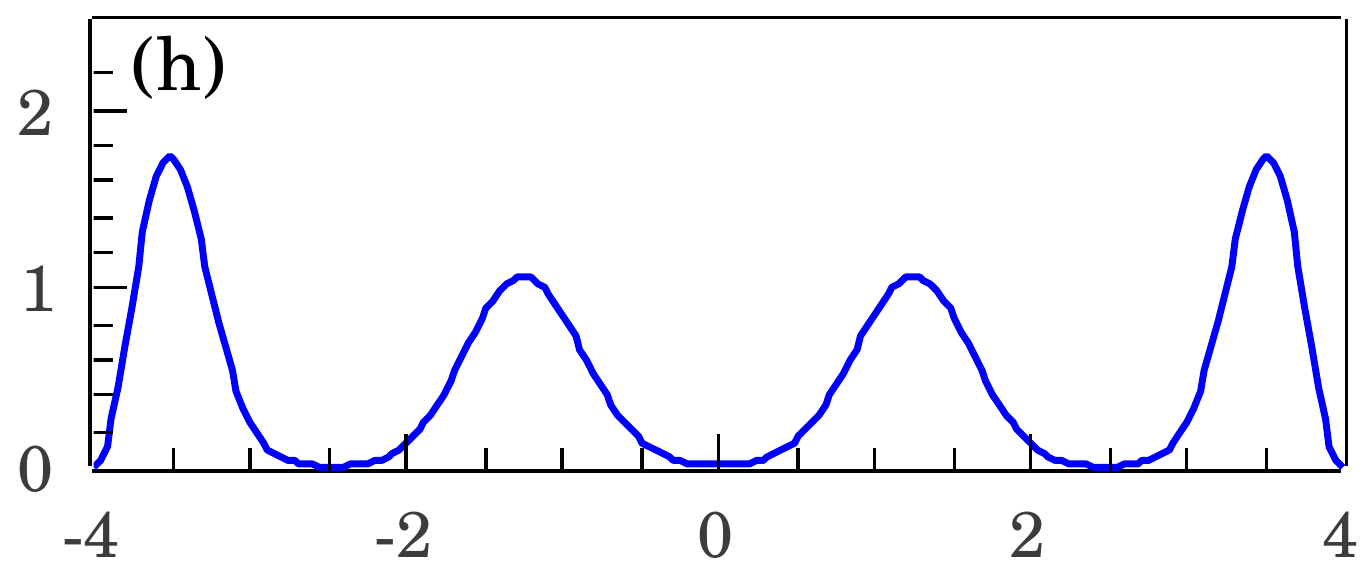} 
  		\includegraphics[width=0.32\textwidth]{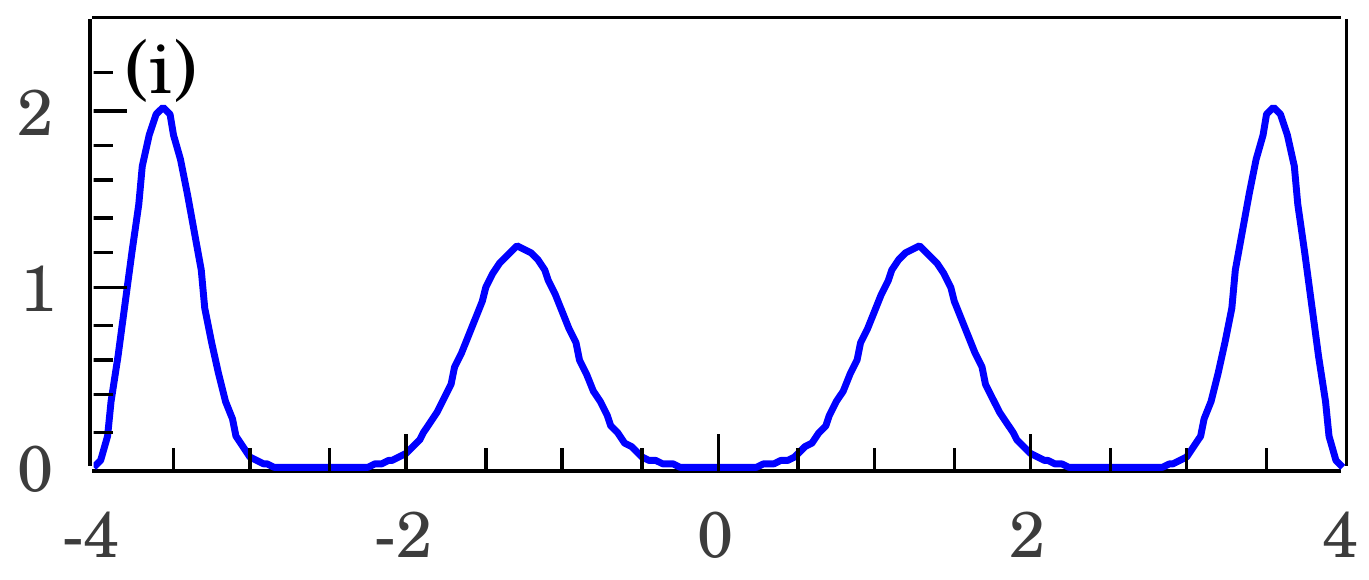}   		
  	\end{minipage}
  	\vspace{0.1in}
  	$x/a$
  	\caption{Calculated quantum spatial density profile of 4 spinless electrons in an eight-dot array at $T=0$.  The upper row: fixed $a=2~$a\textsubscript{B} and increasing background potential (a) $V_0=2$~Ry, (b) $V_0=12~$Ry, (c) $V_0=30$~Ry. The middle row: fixed $V_0=20$~Ry and increasing inter-dot spacing (d) $a=1~$a\textsubscript{B}, (e) $a=2~$a\textsubscript{B}, (f) $a=3~$a\textsubscript{B}. Both rows equivalently describe the tuning to Mott phase. The lower row: Wigner crystallization with no background potential $V_0=0$ and increasing electron spacing (g) $a=1~$a\textsubscript{B}, (h) $a=40~$a\textsubscript{B}, (i) $a=80~$a\textsubscript{B}. The spatial density profile of the Winger crystal (i) is qualitatively similar to the Mott phase of (c) and (f).}\label{fig1}
  \end{figure*}	 
	
	We model the quantum dot array by a 1D cosine potential whose period is the spacing between neighboring dots. We emphasize that our notations of insulator (conductor) refer to the localized (extended) state of the electron spatial density profile. The Hamiltonian reads
	\begin{equation}\label{eq1}
	\begin{split}
	H&=\sum_{i=1}^{N}\frac{-\hbar^2}{2m}\frac{\partial^2}{\partial{x_i}^2} \pm V_0\cos\left( \frac{2\pi x_i}{a} \right)\\
	&\quad +  \frac{\hbar^2}{ma_B}\sum_{i<j}\frac{1}{\sqrt{(x_i-x_j)^2+d^2}},
	\end{split}
	\end{equation}
	with $a$ and $V_0$ being the period and strength of the background potential, and we use open boundary conditions at $x=\pm N_da/2$ where $N_d>N$ is the number of dots. The plus (minus) sign is for even (odd) $N_d$. For GaAs electrons, the effective electron mass is $m\sim 0.06 m_e$ and the dielectric constant is $\varepsilon\sim 10$; making the Bohr radius $a_B\sim 10$~nm, which is 200 times as large as the vacuum value, and the Ryberg energy $\text{Ry}\sim 5$~meV, which is 2000 times less than the hydrogen atom ionization energy. The cut-off $d$ in the Coulomb term regulates the short-distance behavior of the interaction potential (and could represent the lateral confinement size of the system in the direction transverse to the 1D array). However, as we are interested in states where the electrons are localized apart from each other, $d$ is not an important parameter as long as $d\ll a$. Throughout this paper, we keep $d=0.05 a$ only for the sake of numerical computation - varying $d$ does not change the result as long as $d\ll a$. The system is then controlled by two parameters: the inter-dot spacing $a$ and the potential height $V_0$. For convenience, we express $a$ and $V_0$ in terms of a\textsubscript{B} and Ry respectively. 
	
	To simulate the quantum ground state for each pair of ($a$, $V_0$), we use the configuration interaction method to diagonalize the Hamiltonian \eqref{eq1}. The algorithm consists of two steps. First, we solve the non-interacting single-particle Hamiltonian (Eq.~\ref{eq1} with $N=1$) to find the single-particle wavefunctions and the corresponding eigenvalues. This Hamiltonian is diagonalized in the basis of sine functions satisfying the open-boundary condition. We keep up to 2000 sine functions in this step, testing and ensuring convergence. In the second step, we construct the many-body basis from Slater determinants of single-particle wavefunctions obtained in the first step. We use up to 25 single-particle wavefunctions and 20000 Slater determinants to find the many-body interacting ground state. When the Hamiltonian has inversion symmetry (in the absence of disorder), the Hamiltonian can be diagonalized in blocks, which saves significant computational resources. The numerical work is carried out in a large high performance computing cluster. Most of the simulations are performed with spinless electrons (unless explicitly stated otherwise, e.g. sec IV) as the exchange energy is exponentially small for highly localized states, and including spin in the calculations does not change anything except for a corresponding change in the band filling factor because of the spin degeneracy. We refer to the minima of our cosine potentials as `quantum dots' as they define the individual lattice sites in the 1D system. Thus, our `dots' are essentially lattice sites. Nominally, without any interaction, the system is a band metal when the lowest band is partially-filled (i.e. number of electrons is less than number of dots or sites) whereas it is an ordinary band insulator when the number of dots equals the number of electrons (since we consider spinless electrons).
	 \subsection{Spatial density profile}
	 
   \begin{figure*}
   	\centering
   	\begin{minipage}{0.04\textwidth}
   		\rotatebox{90}{$\rho(x)a$}
   	\end{minipage}	
   	\begin{minipage}{0.95\textwidth}
   		\centering
   		\includegraphics[width=0.32\textwidth]{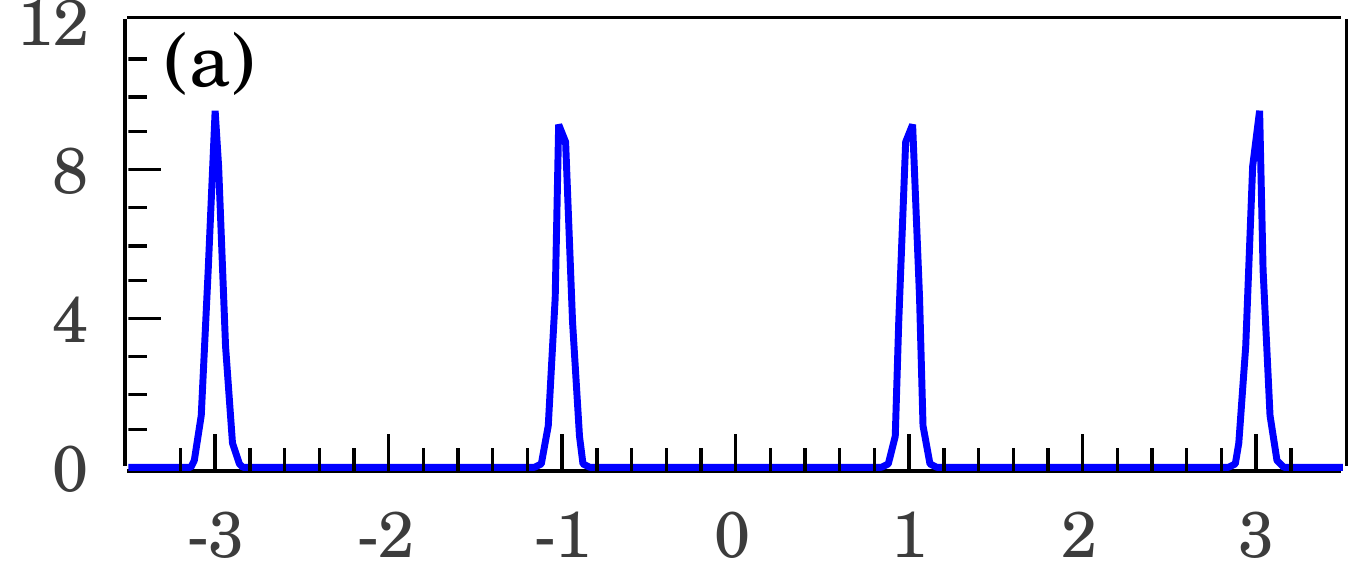}  	
   		\includegraphics[width=0.32\textwidth]{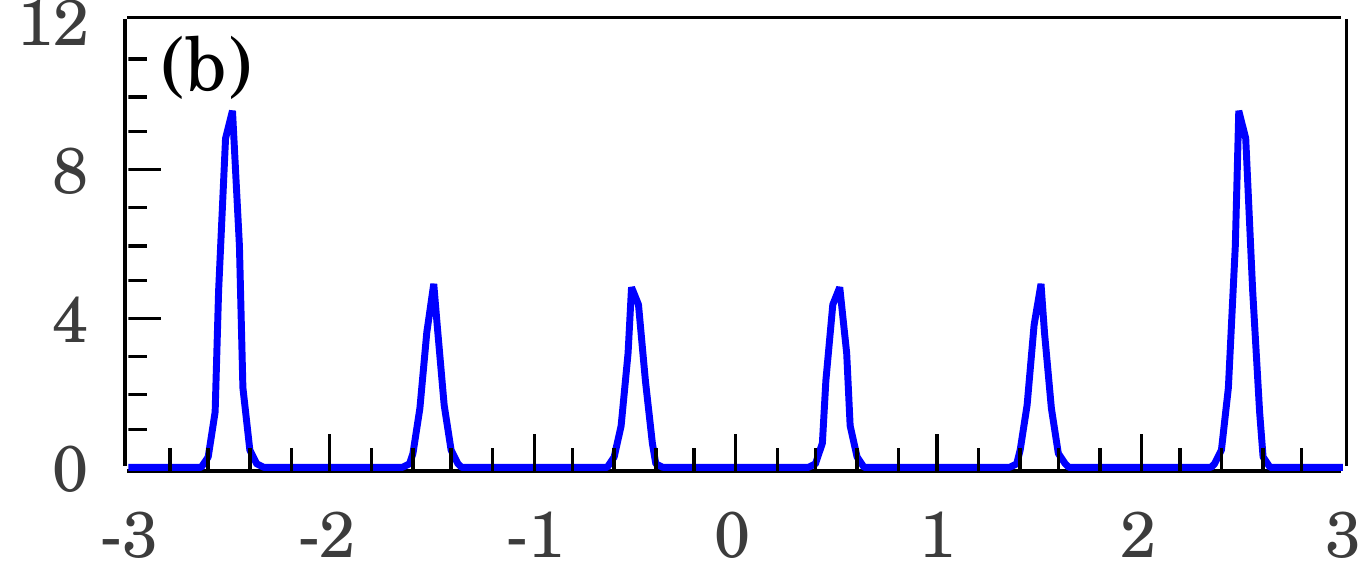}
   		\includegraphics[width=0.32\textwidth]{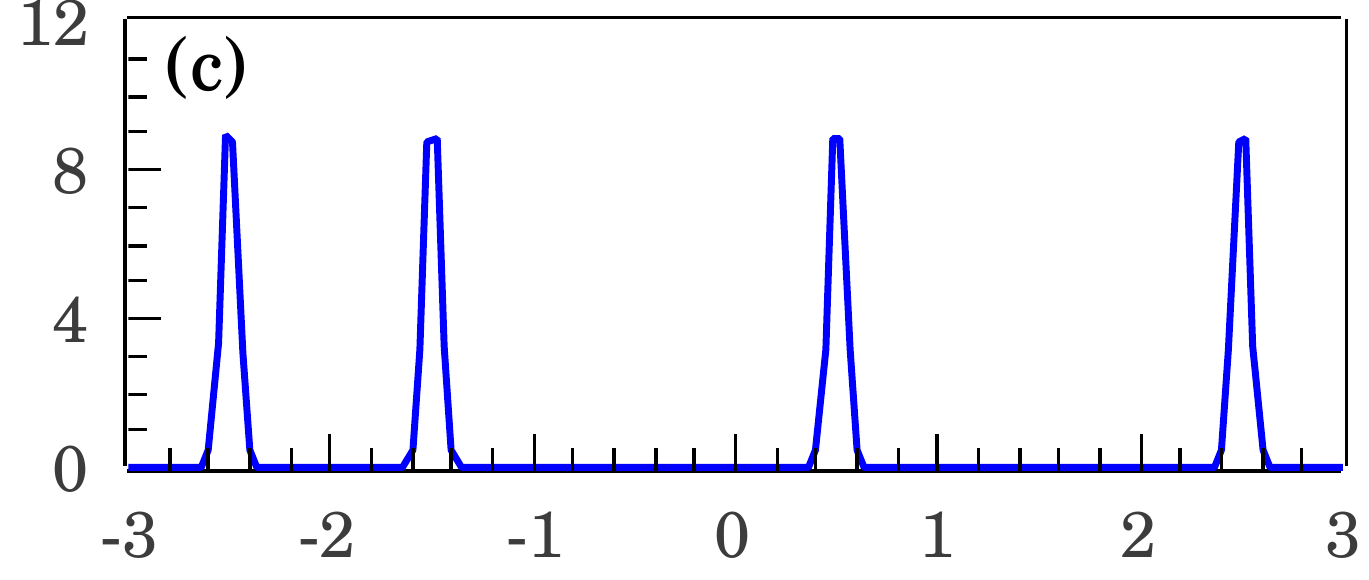} 
   		
   		\includegraphics[width=0.32\textwidth]{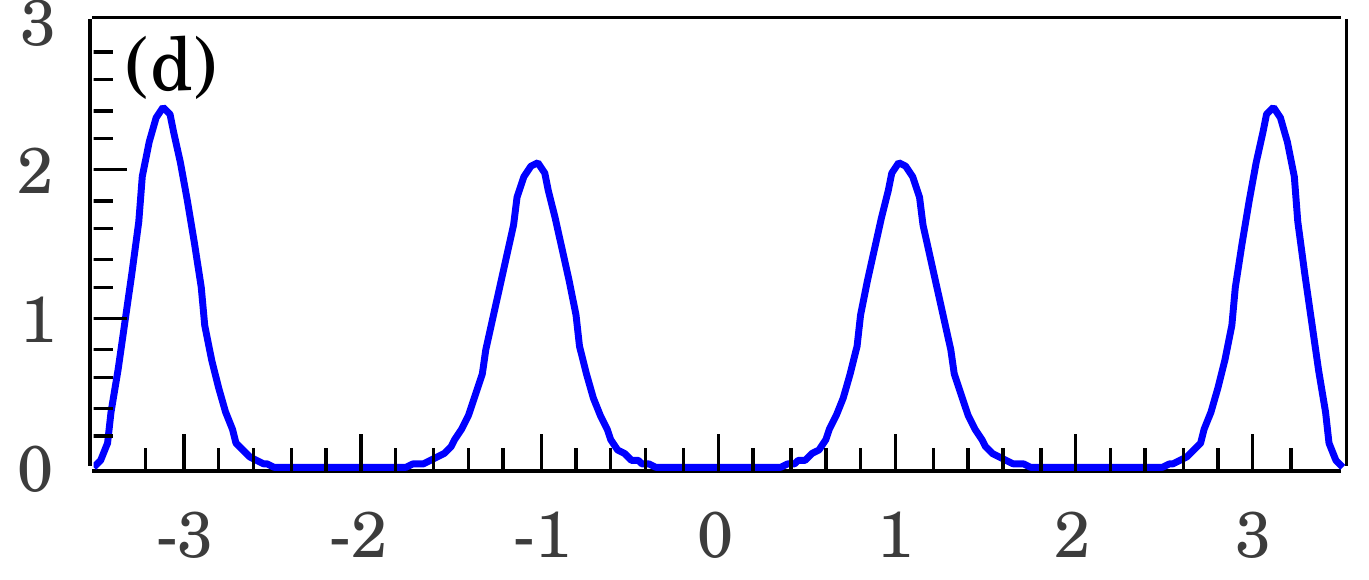} 
   		\includegraphics[width=0.32\textwidth]{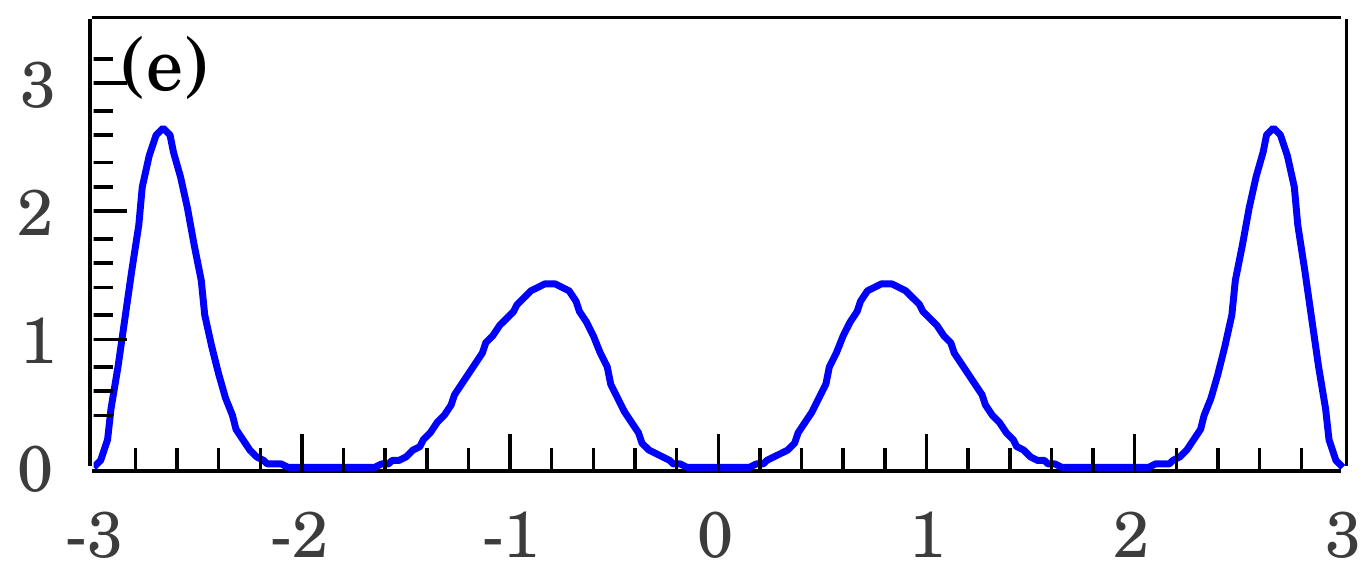} 
   		\includegraphics[width=0.32\textwidth]{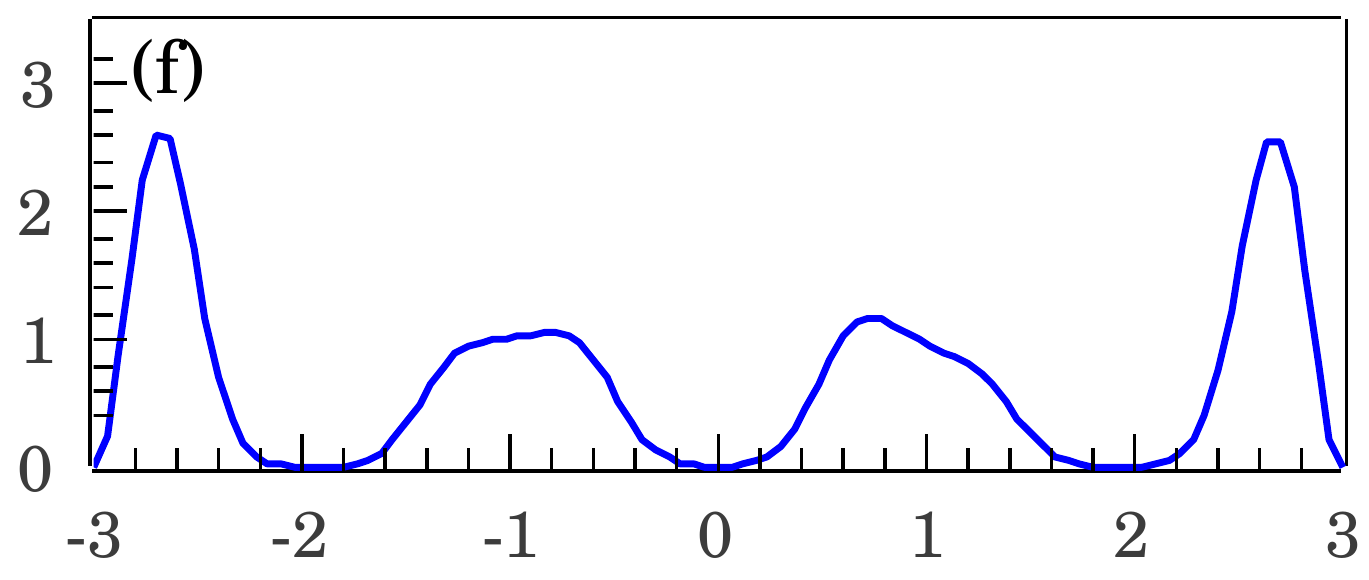} 
   		
   	\end{minipage}
   	\vspace{0.1in}
   	$x/a$
   	\caption{Ground state spatial density profile of 4 spinless electrons. The upper row: Mott  insulator phase at $V_0=1~\text{Ry}, a=70$~a\textsubscript{B}. The lower row: Wigner crystal phase at $V_0=0.001~\text{Ry}, a=100$~a\textsubscript{B}. For Figs.(a) and (d), $N_d=7$; for Figs.(b) and (e), $N_d=6$; for Figs.(c) and (f), $N_d=6$ with the potential on the fourth dot enhanced by 10\%.}\label{fig2}
   \end{figure*}	 	 
	 
	There are now three energy scales in the system: the kinetic energy that goes as $1/a^2$, the periodic background potential $V_0$, and the Coulomb potential that goes as $1/a$. Our continuum Hamiltonian with periodic trapping potential provides a minimal model for controlling the interplay among these three energies, thus inducing electron liquid-Wigner crystal or electron liquid-Mott insulator crossovers. We consider 4 electrons in 8 dots in Fig.~\ref{fig1}, which is a half-filled system since each dot can be occupied by one spinless electron.  Such a half-filled system is by definition a metal in the noninteracting limit since all 8 dots will have equal amplitudes for electron occupation if there is no interaction.
	
	In Figs.~\ref{fig1}(a)-(c), with increasing $V_0$, the electrons start to localize on the dots but are still able to hop between them, thus leaving an effective non-zero charge on all the dots for $V_0$ not too large. At higher $V_0$, the hopping strength is suppressed below the Coulomb repulsion, leading to electrons localizing on a subset of the dots so that they can stay as far away from each other as possible for the given filling, thus minimizing the potential energy. Such a localization is the effective Mott insulator state in contrast to the electron liquid (or the metallic) state where the electron hopping is equally probable on all the dots. Figures~\ref{fig1}(d)-(f) describe the process of increasing the dot spacing at fixed $V_0$. The interaction energy decreases as $1/a$ but the hopping strength is suppressed exponentially with $a$; therefore the hopping among the dots is essentially blocked at sufficiently large $a$, leaving only 4 dots occupied similar to the previous case of increasing $V_0$. Thus, our model can be tuned to an effective Mott insulator either by increasing the background potential $V_0$ or by increasing the inter-dot separation $a$. 
	
	In addition, by tuning to $V_0=0$, we can simulate the Wigner crystal formation (low average density) from an electron liquid (high average density) in Figs.~\ref{fig1}(g)-(i). The result of increasing the potential energy (either by increasing $a$ or by increasing $V_0$) is a qualitatively similar state with 4 distinct localized density peaks, representing an insulator. Conventionally, the state corresponding to Figs.~\ref{fig1}(c) and (f) are called Mott insulator while the state in Fig.~\ref{fig1}(i) is the Wigner crystal. We note that there is no phase transition between these two localized phases, only a smooth crossover as a function of $V_0$. We also emphasize that for the noninteracting situation, where the Coulomb interaction is weak (or absent), the system in Fig.~\ref{fig1} is always a half-filled metal with all 8 dots equally likely to be occupied by electrons - the electron localization on 4 dots (out of 8 in the system) leading to the insulating phase is a direct result of Coulomb interaction among the electrons. The metallic liquid state is clearly visible with all 8 dots partially occupied by electrons in Figs.~\ref{fig1}(a), (d), (g) where interaction effects are weak compared with the hopping kinetic energy. By contrast, Figs.~\ref{fig1}(c), (f), (i) represent strongly interacting insulating states in spite of the system being half-filled. 

    \subsection{Mott-Wigner incommensurability}	   
    
    In the previous example of 4 electrons in an eight-dot array (Fig.~\ref{fig1}), the Mott insulator and Wigner crystal phases are qualitatively similar with the only difference being that the Mott state is typically more strongly localized (although the same would be true to the Wigner phase for very large dot separations). This is not always the case and there can in fact be a qualitative difference between these two phases even within the crossover physics being studied here. For simplicity, we first study the extreme Mott insulator, i.e. the hopping strength $t\to 0$. Then, the ground state energy is mostly the Coulomb potential energy
    \begin{equation}\label{eq2}
    E=\sum_{i=1}^{N-1}\sum_{j=i+1}^{N} W(x_i,x_j),
    \end{equation}
    where $W$ is the interaction potential and $\{x\} \subset \{a, 2a, ..., N_d a\}$. If the lowest $E$ is non-degenerate, i.e. there is only one set $\{x\}$ that gives the minimum energy, the Mott insulator corresponds to a single occupancy configuration and the number of density peaks is clearly the number of electrons, similar to a Wigner crystal. Otherwise, the Mott state is a superposition of multiple occupancy configurations, resulting in a higher number of density peaks. We call these cases `incommensurate' - here the incommensuration is with respect to the ratio of the number of spatial density peaks in the ground state to the number of electrons (e.g. 4 peaks for 4 electrons as in Fig.~\ref{fig1} is commensurate). By numerical trials on Eq.~\eqref{eq2} with Coulomb potential $W(x_i,x_j)\propto 1/|x_i-x_j|$, we find that $N=4$ is incommensurate with $N_d=3m$ and $N=3$ is incommensurate with $N_d=2m$ with $m$ an integer. For larger $N$, the rule of incommensurability is more complex. In Figs.~\ref{fig2}(a)-(b) and (d)-(e), we show the spatial density profiles of 4 spinless electrons in $N_d$-dot arrays in the Mott insulator (high $V_0$) and the Wigner crystal (low $V_0$) phases by varying $N_d$ ($> 4$, with the number of electrons fixed at 4). While the Wigner crystal phase always has 4 density peaks regardless of the number of dots (corresponding to 4 electrons in the system - the Wigner phase must always be commensurate with the electron number), the Mott phase has 4 peaks for $N_d=7$ and 6 peaks for $N_d=6$ due to the degeneracy of the Coulomb potential when $N_d$ is a multiple of 3 (other cases of $N_d$ are shown in the Appendix). We note that this incommensuration arising from degeneracy  is only possible when all the dots are identical. As a result, the incommensurate state is fragile and may not be observed experimentally unless disorder is minimal in the system. Specifically, in Figs.~\ref{fig2}(c) and (f), the fourth dot from the left has a 10\% enhanced local trapping potential, thus slightly altering the periodic cosine potential of the model, leading to the density profile of the incommensurate Mott phase at $N_d=6$ to shift dramatically from 6 to 4 density peaks. Additional results emphasizing the Mott incommensurability and comparing Mott/Wigner phases are provided in the Appendix. 
    
\begin{figure}
	\centering
	\begin{minipage}{0.01\textwidth}
		\rotatebox{90}{$\rho(x)a$}
	\end{minipage}
	\begin{minipage}{0.46\textwidth}   
		\centering
		\includegraphics[width=0.49\textwidth]{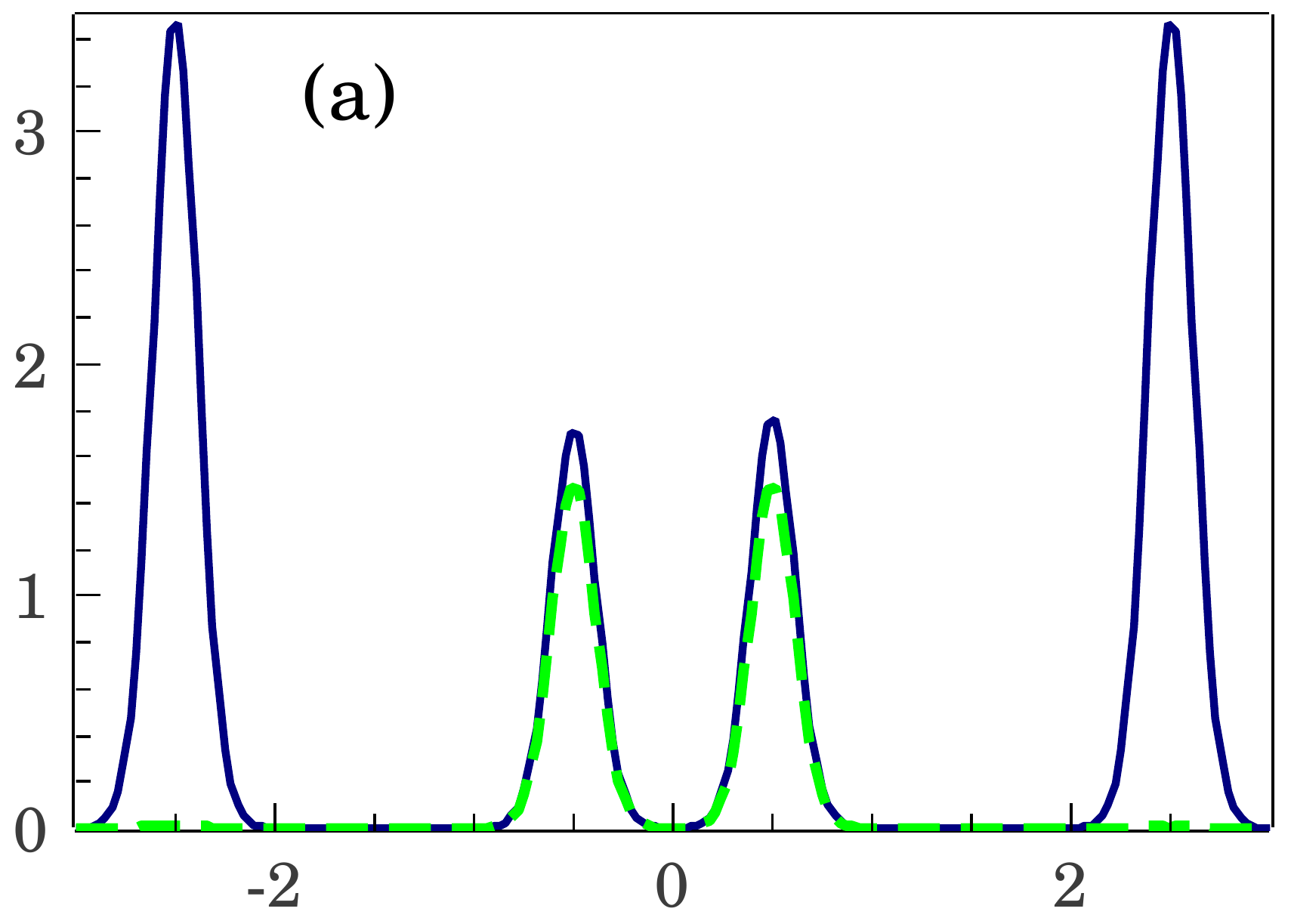}
		\includegraphics[width=0.49\textwidth]{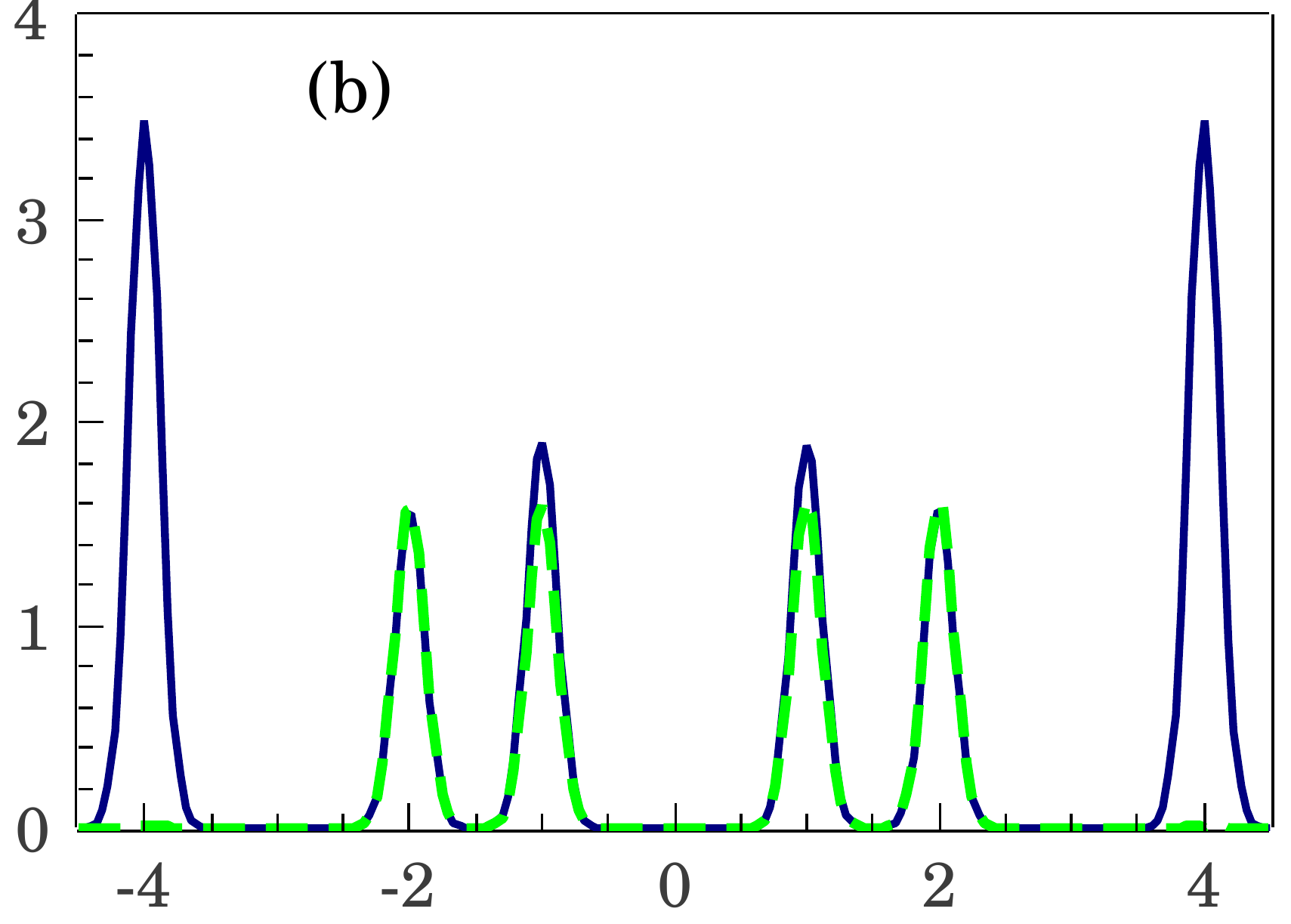}
		\includegraphics[width=0.49\textwidth]{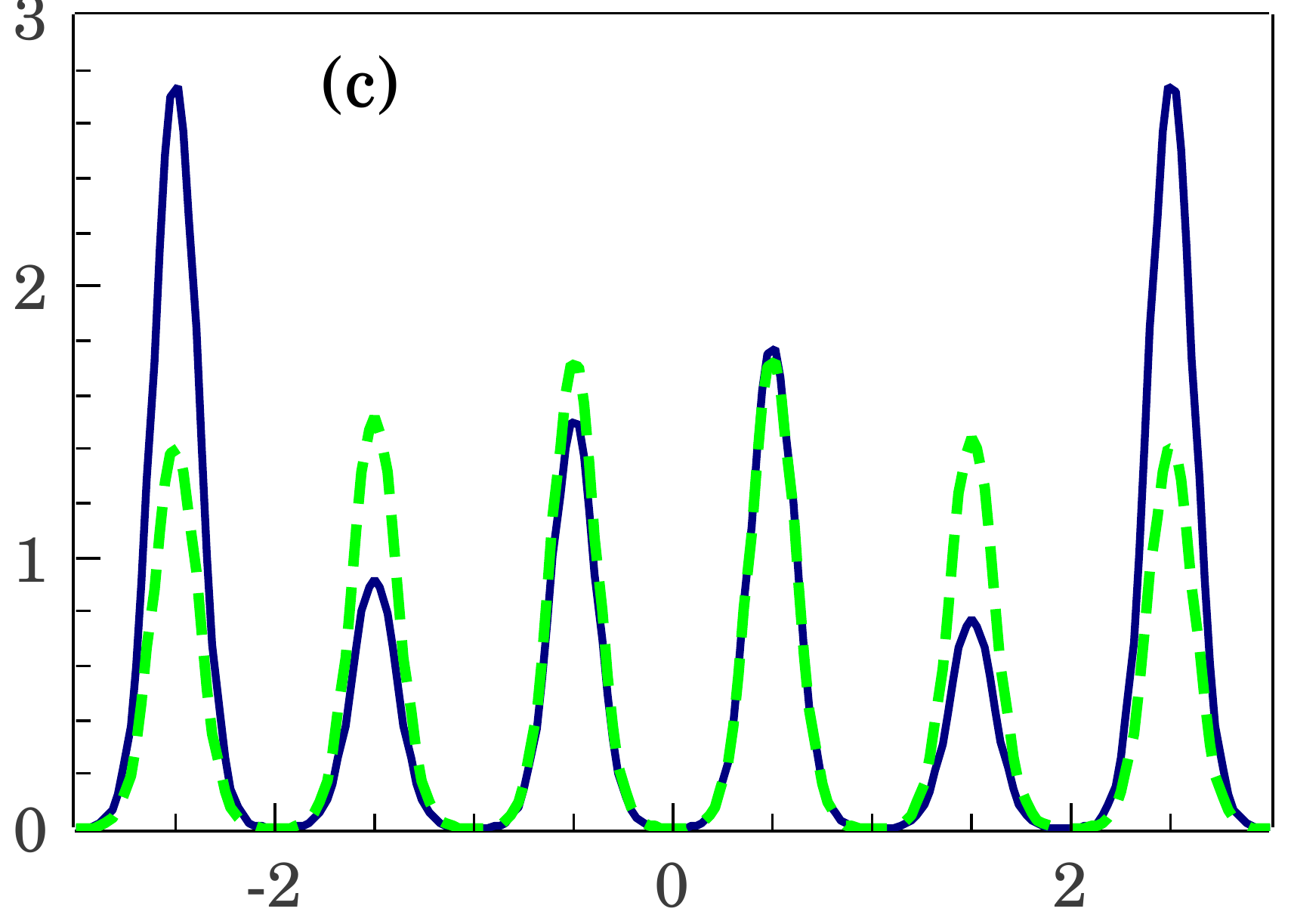}
		\includegraphics[width=0.49\textwidth]{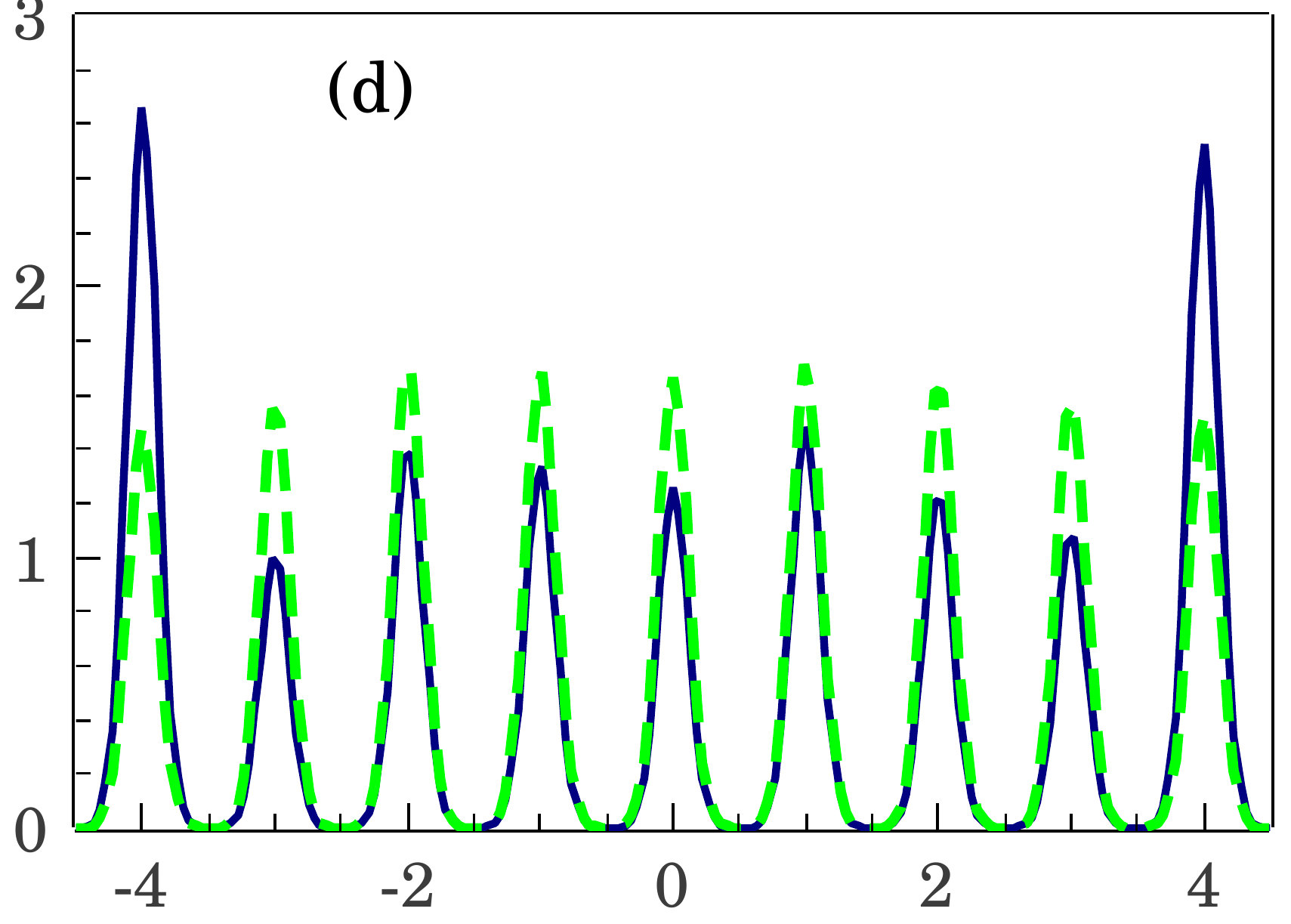}

		$x/a$
	\end{minipage}
	\caption{Average density profiles (blue (dark) lines) and the corresponding standard deviations (green (gray) dashed lines) of incommensurate Mott phases having $a=3$~a\textsubscript{B} ($E_c\approx 0.67$~Ry) and $V_0=15~$Ry. For Figs.(a) and (c), $N=3$ and $N_d=6$; for Figs.(b) and (d), $N=4$ and $N_d=9$. With $V_\text{rms}=0.01V$ for Figs.(a) and (b), the incommensurate Mott phase is recovered as $V_\text{rms}<E_c$. With  $V_\text{rms}=0.1V_0$ for Figs.(c) and (d), the number of peaks equals the number of dots due to the disorder-induced localization as $V_\text{rms}>E_c$.}\label{disorder1}
\end{figure}

    Although it seems that a slight disorder can wash out the Mott incommensurate phase, this phase can be recovered by averaging density profiles from different randomized disorder configurations. In Fig.~\ref{disorder1}, we show the calculated effect on the Mott phase ($a=3$~a\textsubscript{B}, $V_0=15~$Ry) of adding a randomized disorder potential to the background lattice confining periodic potential. The background potential in the presence of the random potential $F(x)$ is
    \begin{equation}
    V(x) = V_0\cos(2\pi x/a) + F(x) ,
    \end{equation}	
    where the disorder $F(x)$ is chosen to have a Gaussian random distribution with zero mean and a variance given by $V_\text{rms}$. The displayed density profiles are averaged over 100 disorder configurations. For the given parameters, the typical Coulomb interaction energy is $E_c=e^2/a\approx 0.67$~Ry. The quantum dot array parameters for Fig.\ref{disorder1} (i.e. values of $a$, $V_0$, etc.) have been chosen to correspond to a realistic GaAs-based experimental quantum-dot system being studied at Delft University \cite{communication}. When $V_\text{rms}<E_c$ (see Figs.~\ref{disorder1}(a-b)), the ground state degeneracy is lifted and the system retreats to one particular occupancy configuration, thus destroying the incommensurate state. However, by averaging over different disorder configurations, this state can be observed. On the other hand, when $V_\text{rms}>E_c$ (see Figs.~\ref{disorder1}(c-d)), the localization is driven by the disorder instead of the interaction. The result is the averaged density profile with all the dots occupied. Moreover, the density standard deviation in each dot is almost identical, as opposed to the disordered Mott phase. This should be construed more as Anderson-Mott localized phase in the presence of both disorder and interaction rather than the Mott-Hubbard localized phase driven solely by interaction.

\begin{figure}
	\centering
	\begin{minipage}{0.01\textwidth}
		\rotatebox{90}{\hspace{0.15in} $\rho(x)a$}
	\end{minipage}
	\begin{minipage}{0.46\textwidth}   
		\centering
		\includegraphics[width=0.49\textwidth]{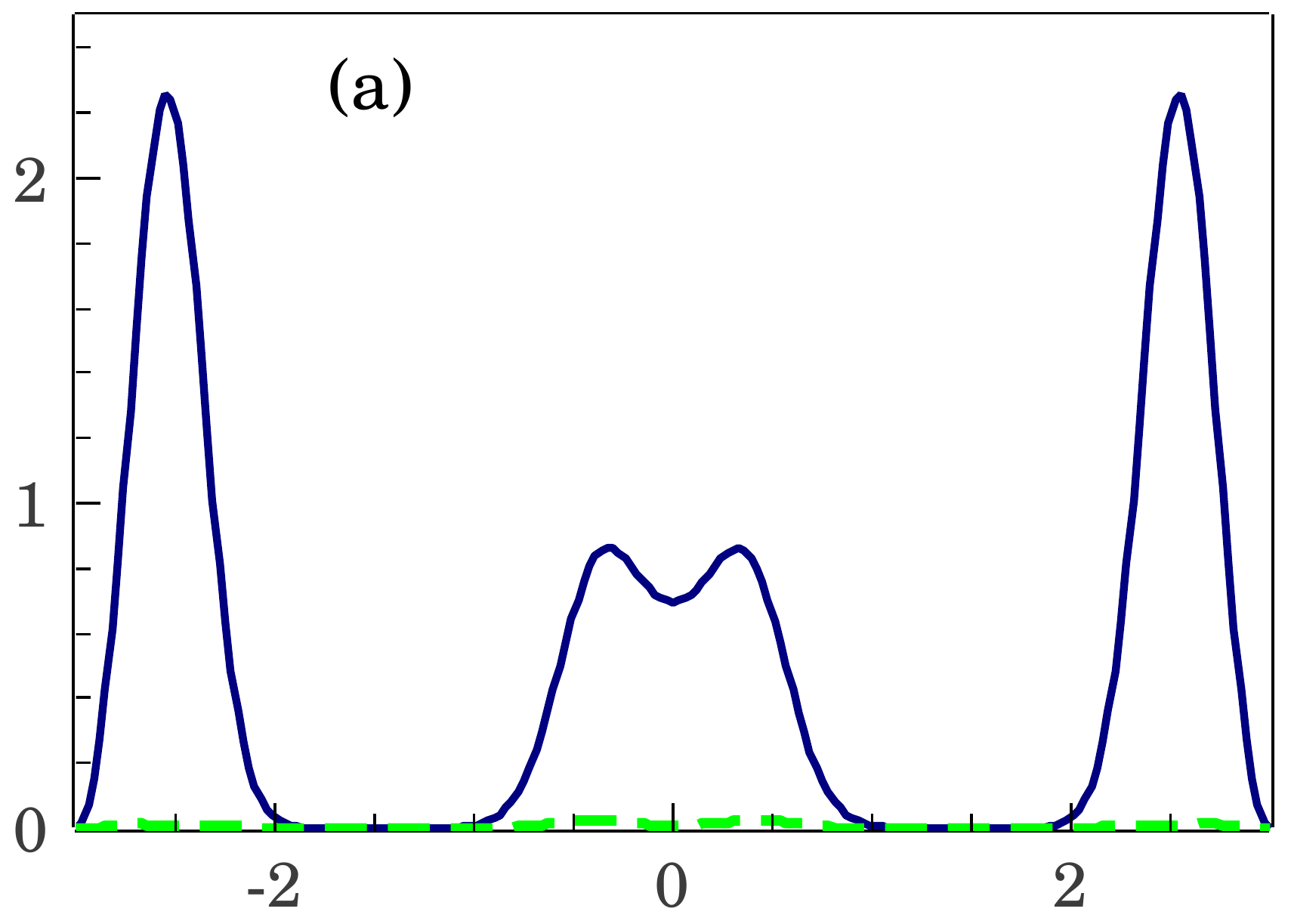}
		\includegraphics[width=0.49\textwidth]{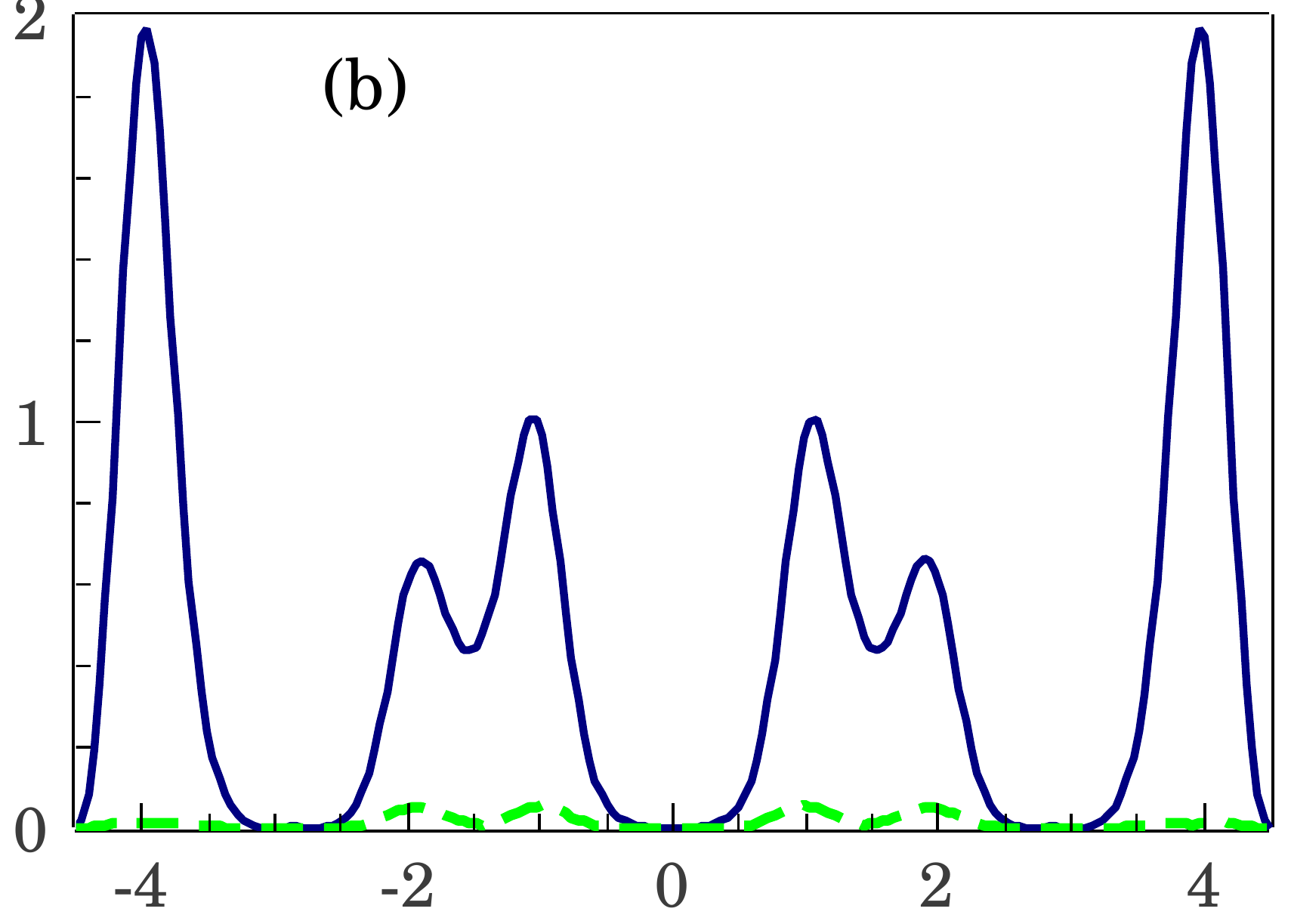}
		
		$x/a$
	\end{minipage}
	\caption{Similar to Fig.~\ref{disorder1} but for Wigner phases having $a=100$~a\textsubscript{B} and $V_0=10^{-3}~$Ry. (a) $N=3$, $N_d=6$ and $V_\text{rms}=0.01V_0$. (b) $N=4$, $N_d=9$ and $V_\text{rms}=0.1V_0$. The number of peaks always equals the number of electrons and the density standard deviation is much smaller than the average value, showing that disorder, in general, has little influence on the Wigner phase.}\label{disorder2}
  \end{figure}

    Proceeding similarly for the Wigner phase, we show in Fig.~\ref{disorder2} that disorder hardly affects the Wigner density profile through the fact that the standard deviation is much smaller than the average density, allowing a clear identification of the density peaks associated with the effective Wigner crystal. In conclusion, the possibility of a qualitative distinction between the Mott phase and the Wigner phase in a slightly disordered background potential based on commensurability is one important new finding of our work. We do believe that the incommensuration physics shown in Fig.~\ref{fig2} is the most decisive way to distinguish between the Mott and the Wigner phase. As emphasized above, the incommensuration Mott physics would manifest itself directly in clean samples with very little disorder or in disordered samples through averaging whereas the Wigner phase remains relatively immune to disorder and incommensuration.

	\subsection{Charge gap}
	
    \begin{figure}
    	\centering
    	\begin{minipage}{0.02\textwidth}
    		\rotatebox{90}{\hspace{0.4 in} $d\ln(\Delta E)/d\ln(n)$}
    	\end{minipage}
    	\begin{minipage}{0.4\textwidth}   
    		\centering
    		\includegraphics[width=\textwidth]{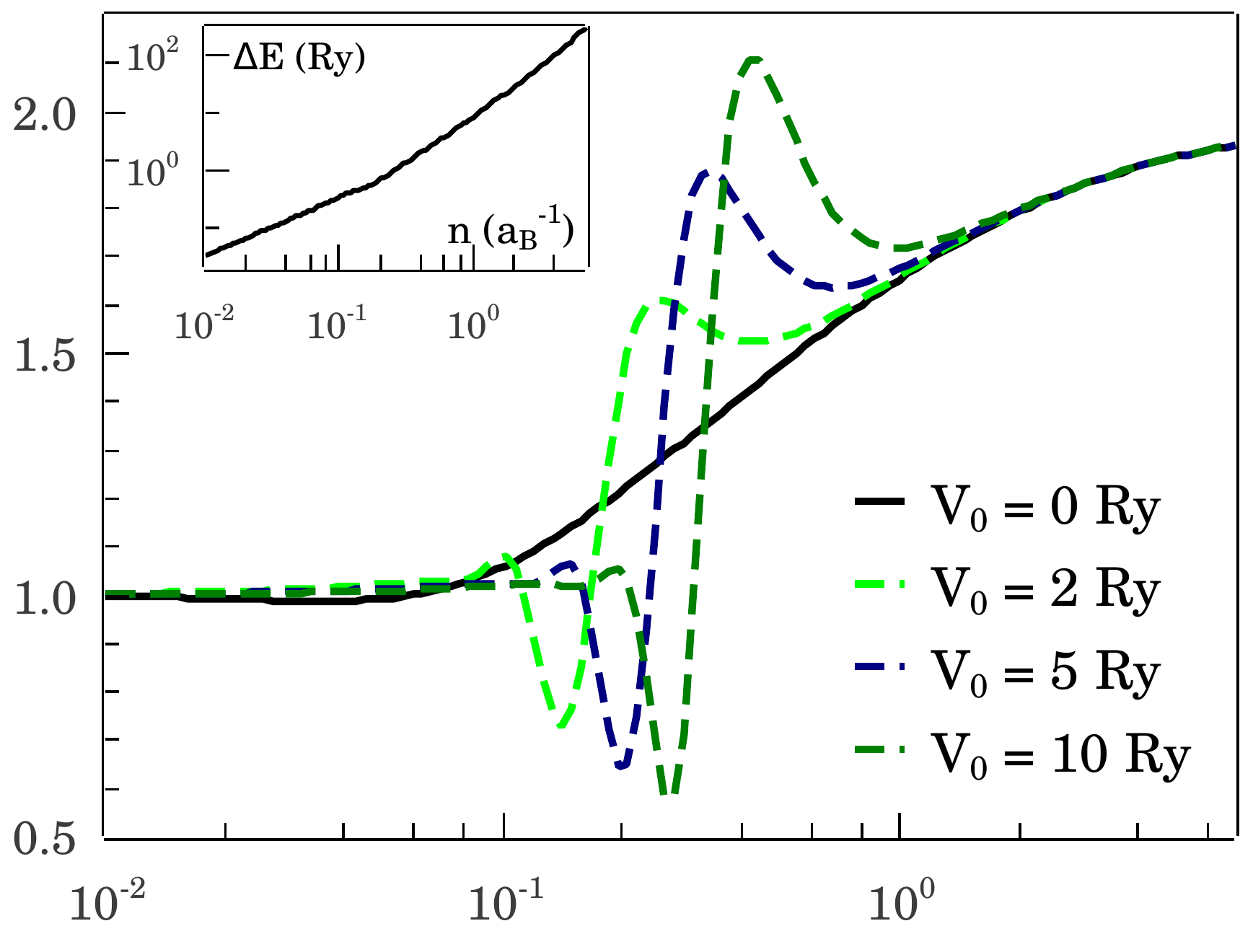}
    		
    		$n$ (a\textsubscript{B}\textsuperscript{-1})
    	\end{minipage}
    	\caption{The effective exponent of the charge gap $\Delta E$ of a four-electron-eight-dot array with respect to the average density $n=2/a$. This exponent approaches 1 at low density and 2 at high density. The inset shows the charge gap with varying average density $n$ at fixed $V_0=2$~Ry in logarithmic scales.}\label{fig3}
    \end{figure}
    
	\begin{figure}
		\centering
		\begin{minipage}{0.02\textwidth}
			\rotatebox{90}{$\Delta E$ (Ry)}
		\end{minipage}
		\begin{minipage}{0.4\textwidth}   
			\centering
			\includegraphics[width=\textwidth]{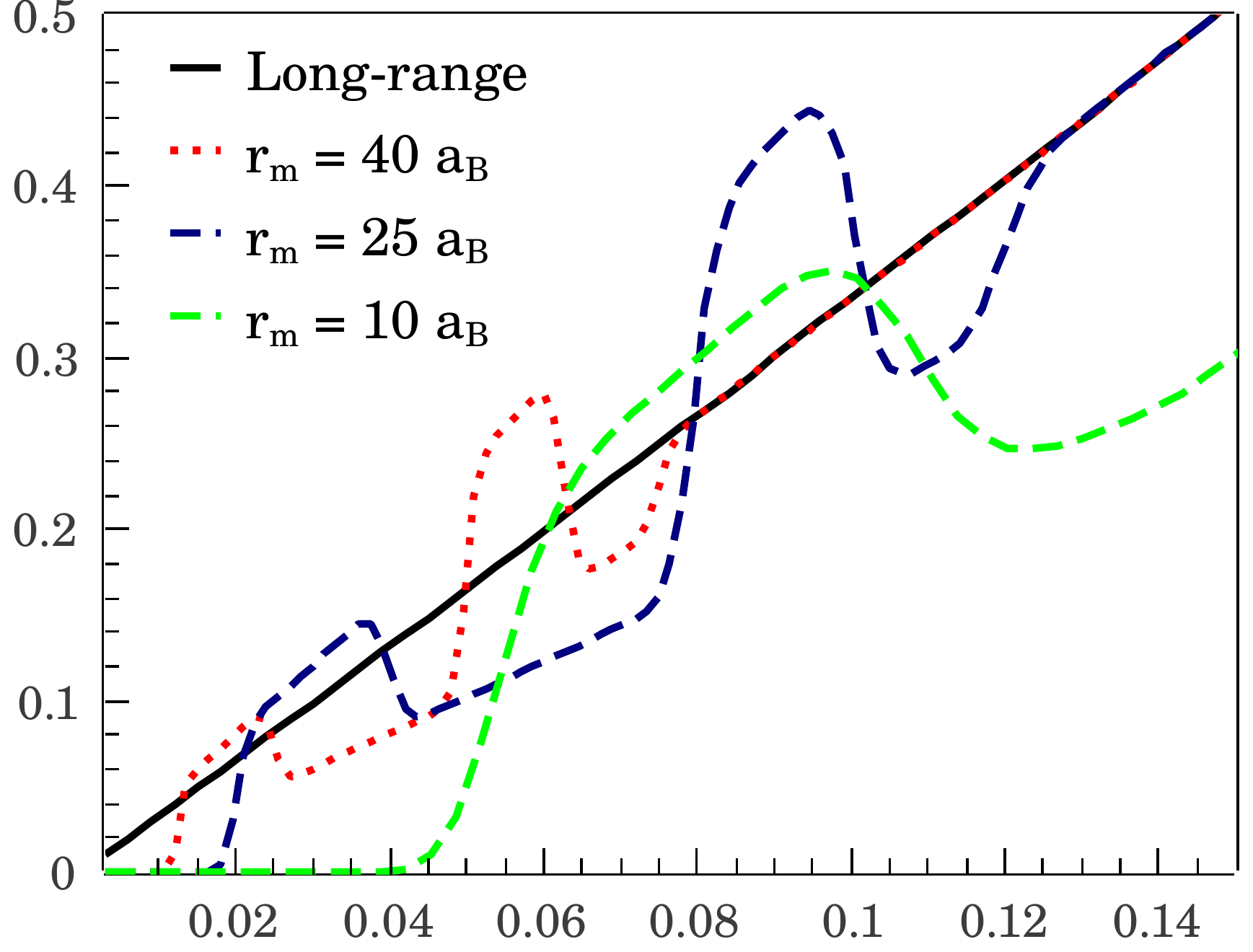}
			
			$n$ (a\textsubscript{B}\textsuperscript{-1})
		\end{minipage}
		\caption{Charge gap of 4 electrons in an eight-dot array with long-range and modified short-range interactions (with a range $r_m$) at $V_0 = 5~$Ry. The gaps of short-range cases first peak at $n\approx 0.025, 0.04$ and 0.1~a\textsubscript{B}\textsuperscript{-1} for $r_m=40$, 25 and 10~a\textsubscript{B} respectively.}\label{fig4}
	\end{figure}    	
	
	In addition to the nature of the spatial density profiles (localized or extended) discussed above (Figs.~\ref{fig1} and \ref{fig2}), insulators can also be differentiated from metals by the existence of a non-zero charge gap - the energy needed to add a single electron to the system, analogous to the Coulomb blockade effect \cite{Cblockage, Delft2017}. The charge gap of a system that already has $N$ electrons with the $(N+1)$th electron added is defined as
    \begin{equation}
    \Delta E(N) = E(N+1)+E(N-1)-2E(N),
    \end{equation}
    where $E(N)$ is the ground state energy of the $N$-electron array. In the inset of Fig.~\ref{fig3}, we show the charge gap of an eight-dot array with 4 electrons at fixed $V_0 = 2$~Ry as a function of the average density $n=2/a$. At low average density, the gap grows linearly with $n$ up to some value $n_c$; then the slope increases signaling a change in the exponent of $n$, corresponding to a switch in the dominant energy scale. In the main Fig.~\ref{fig3}, we compute the effective exponent $d\ln(\Delta E)/d\ln(n)$ for different $V_0$. There is a universal behavior: at low $n$ (i.e. large $a$), the exponent approaches 1 reflecting the dominance of the Coulomb interaction in the strongly localized insulating regime; at high $n$ (i.e. small $a$), the exponent approaches 2 as the kinetic energy takes over the system in the strong metallic extended regime. However, the crossover density $n_c$ increases with $V_0$. This is because $V_0$ helps effective localization compared with the free electron situation by constraining the kinetic energy of electron motion.
	
	The charge gap increases with the density and grows linearly at low average density. Then it can be inferred that the charge gap is always non-zero for any pairs of $(a,V_0)$ defining our model. This is the direct result of the long-range nature of the interaction. To show the relevance of the interaction range in this context, we repeat the calculation for a model short-range interaction with a range $r_m$ such that $W(x_1,x_2)\propto 1/\sqrt{(x_1-x_2)^2+d^2}$ for $|x_1-x_2|\le r_m$ and $W=0$ otherwise, focusing on the low-$n$ regime where the charge gap is mostly due to the interaction. In Fig.~\ref{fig4}, contrary to the ever increasing charge gap in the long-range case, short-range interacting systems have a finite region of zero charge gap. Specifically, the charge gap first peaks at $n=1/r_m$, corresponding to one electron per interaction range. Other peaks occur at higher electron fillings over the interaction range, i.e. at the average densities of $2/r_m$ and $3/r_m$. This result is qualitatively consistent with exact results for the short-range 1D Hubbard model. In the limit $V_0\to \infty$ and $r_m\to 0$, i.e. the zero-range spinful interacting Hubbard model, the exact solution \cite{Lieb} shows that the charge gap starts to appear when every site is occupied (half filling) so that the added electron must occupy a filled site and interact with the electron already located there. For the finite-range spinless model used for our results in Fig.~\ref{fig4}, we may therefore expect that, the charge gap should emerge when each interaction range $r_m$ is filled by more than one electron. As a result, in a system interacting via the infinite-range Coulomb potential, the charge gap is always non-zero. In the thermodynamic limit, i.e. $N,N_d\to \infty$ but $N/N_d$ and the hopping strength $t$ staying finite, within the long-range Hubbard model, the insulator-metal crossover has been hypothesized to happen when the interaction range increases which is consistent with our finding for the model with $r_m$ \cite{Hubbard1,Hubbard2}.
	
	We mention that our finding of a small charge gap in the liquid phase and a large charge gap in the Mott phase is analogous to what was termed `collective Coulomb blockade' and `Mott gap' (or just ordinary Coulomb blockade') in the context of Hubbard model-based studies of coupled quantum dots \cite{Cblockage, Delft2017}.  In particular, strong inter-dot tunneling in the liquid phase leads to the delocalized  behavior that all the dots are undergoing Coulomb blockade together whereas in the strongly localized  Mott phase, each electron is strictly localized in individual dots leading to a large charge gap associated with Coulomb blockade in a single small dot.  Thus, Mott to liquid crossover is also a crossover between individual Coulomb blockade to collective Coulomb blockade.  Interestingly, the Wigner phase also manifests the collective Coulomb blockade.
	
	\section{Mott-Wigner-liquid crossover}
	\subsection{Localized phase}
	
	\begin{figure}
		\centering
		\begin{minipage}{0.02\textwidth}
			\rotatebox{90}{Occupancy}
		\end{minipage}
		\begin{minipage}{0.4\textwidth}   
			\centering
			\includegraphics[width=\textwidth]{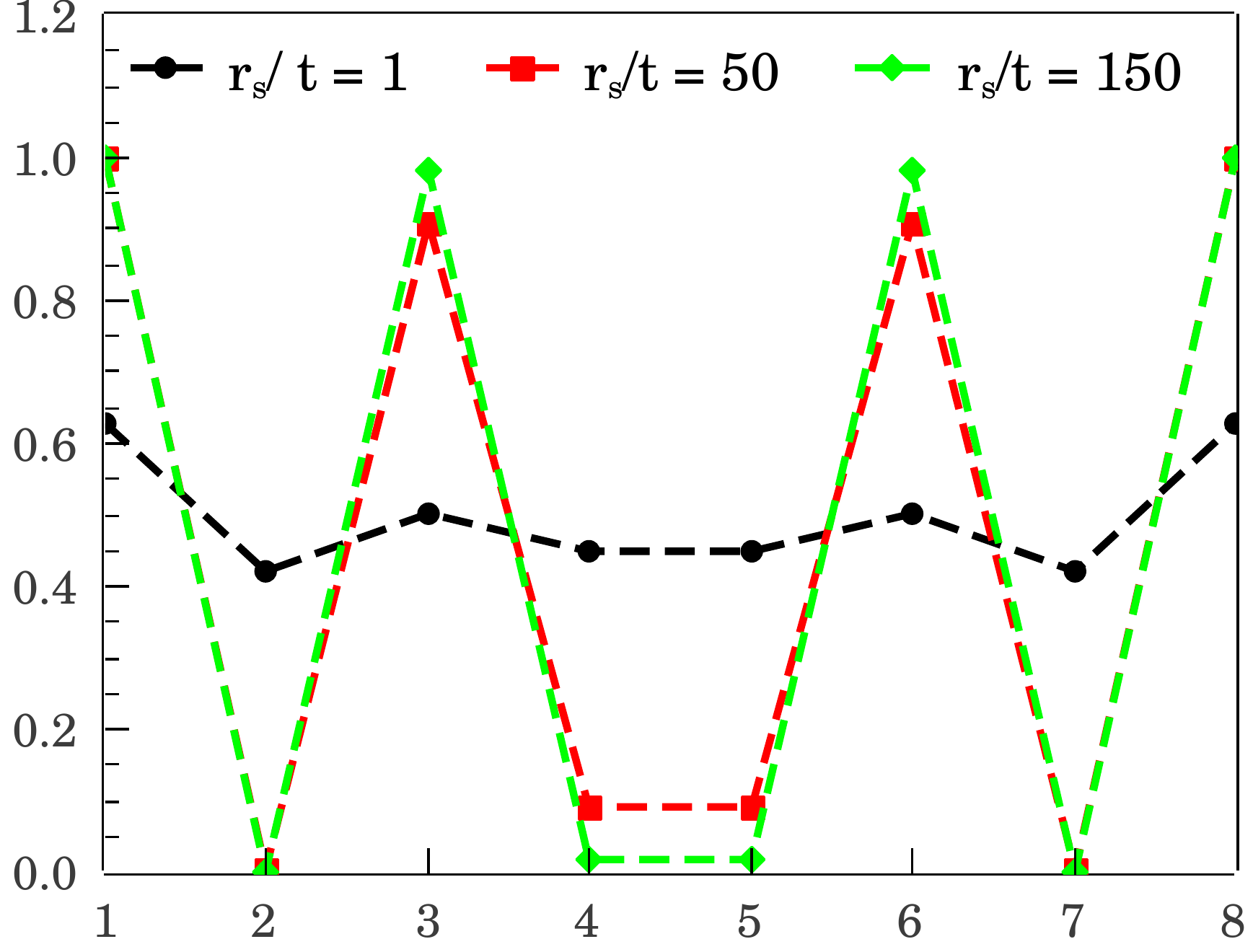}
			
			site
		\end{minipage}
		\caption{Occupancy of a spinless single-band tight-binding model with 4 electrons in 8 sites interacting through Coulomb force.}\label{fig5}
	\end{figure}
	
	\begin{figure}
		\centering
		\begin{minipage}{0.02\textwidth}
			\rotatebox{90}{$\ln(r_s)$}
		\end{minipage}
		\begin{minipage}{0.40\textwidth}   
			\centering
			\includegraphics[width=\textwidth]{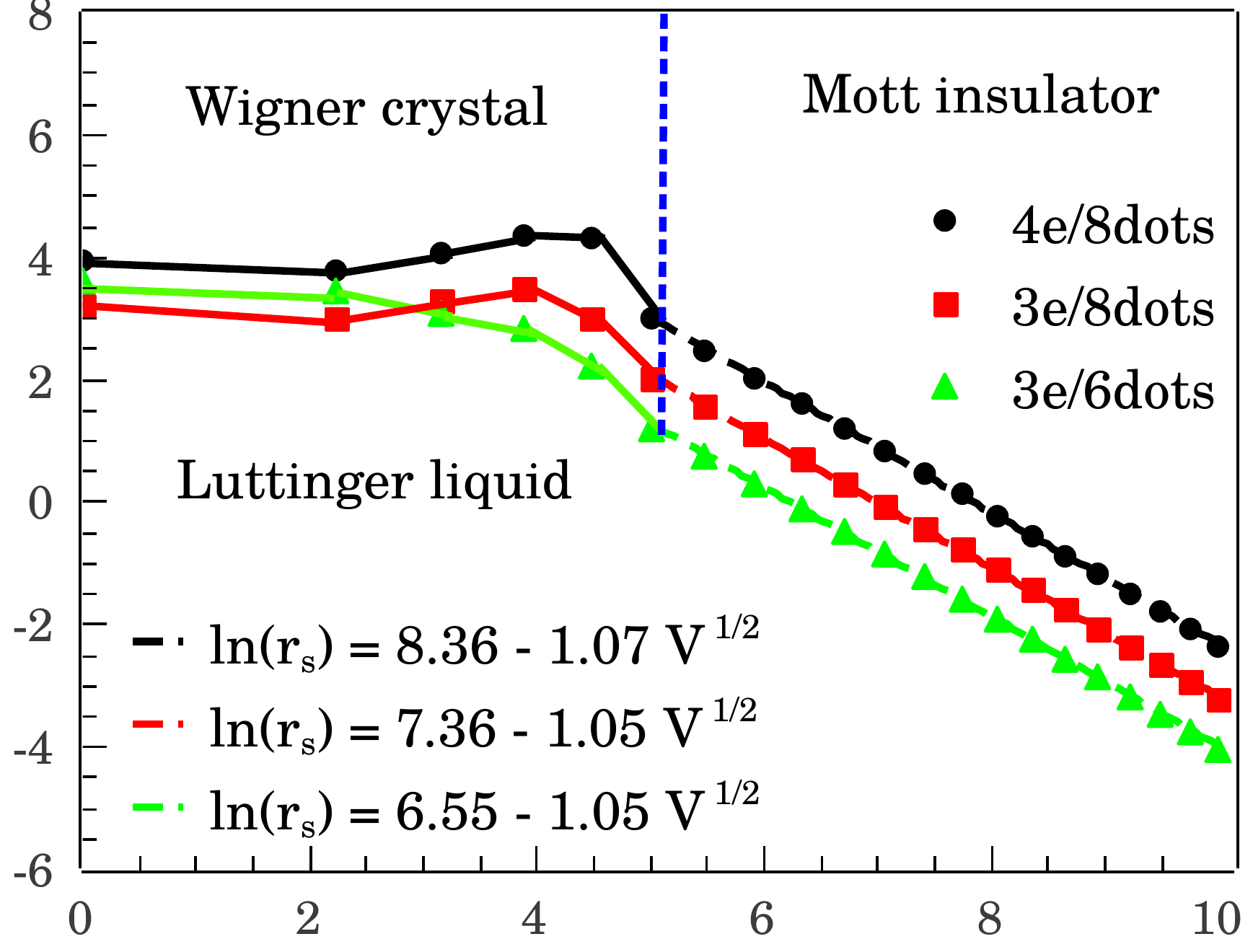}
			
			$V^{1/2}$
		\end{minipage}
		\caption{Critical $r_s$ and $V$ at which the strong localization starts to happen. The dashed lines are fitted against the numerical data for $V>25$. The solid lines for $V<25$ are to aid the vision. The strong localization regime is divided into two parts: Mott insulator for $V>25$ and Wigner crystal for $V<25$.}\label{fig6}
	\end{figure}	
	We now discuss the crossover between the Mott phase, with a strong lattice potential, and the Wigner phase where the lattice potential is absent (or very weak).  We emphasize that there is no quantum phase transition here, and the difference between the two phases is purely qualitative with the Mott phase being commensurate (incommensurate) with the lattice (electron density) and the Wigner phase being the opposite in the thermodynamic limit.  In fact, even the electron liquid to the Mott (or Wigner) phase in our finite 1D system is a crossover as a function of the lattice potential (and/or the inter-particle separation) with no true phase transition although qualitatively the spatial electron density is localized (extended) in the Mott/Wigner (liquid) phase as shown in Figs.~\ref{fig1} and \ref{fig2}.

	In the limit of strong lattice trapping potential, we expect the crossover to depend strongly on $V_0$. In this regime, the system can be mapped into a tight-binding model \cite{Hubbard1,Hubbard2,Hubbard3}. We note that this approximation already excludes the Wigner crystal phase, which exists only for very low $V_0$ where the system is essentially free-electron-like. Within the tight binding model, by definition, there can be no Wigner crystal phase, only Mott and liquid phases. We first discuss the crossover between the liquid and the Mott phase (e.g. Fig.~\ref{fig1}(a) to \ref{fig1}(c) or \ref{fig1}(d) to \ref{fig1}(f)) - here the interacting electron liquid is a Luttinger liquid because the system is one dimensional, and therefore, the crossover we are discussing is a Luttinger-Mott crossover (although this is entirely academic and of no particular significance to our consideration). Within the on-site interacting Hubbard model and the thermodynamic limit $N,N_d\to \infty$, the conditions for a Luttinger-Mott crossover are as follows: (i) the number of particles is commensurate with lattice sites $\bar{n}=N_d/N$ is an integer, (ii) the Luttinger interaction constant $K<1/\bar{n}^2$ \cite{Mott_Luttinger,Mott_Luttinger2}. We have already shown that the commensurability for finite sizes and long-range interaction is defined by more complex conditions, and therefore we expect the finite-size Luttinger-Mott criteria to be modified from these infinite system conditions.
	
	 The tight binding model uses site indices $i=1,..., N_d$ instead of the continuum position variable, hence we first need to rescale our Hamiltonian \eqref{eq1} with respect to the inter-dot spacing in order to obtain the tight binding limit
	\begin{equation}\label{eq4}
	\begin{split}
	H&=\frac{\hbar^2}{ma^2}\left[\sum_{i=1}^{N}\frac{\partial^2}{2\partial{X_i}^2}\pm V\cos(2\pi X_i) \right.\\
		&\quad \left.+ \sum_{i<j}\frac{r_s}{\sqrt{(X_i-X_j)^2+\eta^2}}\right];
	\end{split}
	\end{equation}
	where $X=x/a$ and the two controlling dimensionless parameters are
	\begin{equation}
	V=ma^2V_0/\hbar^2,\quad r_s=a/a_B.
	\end{equation}
	If $V_0$ is in Ryberg energy and $a$ is in Bohr radius, the relation reduces to $V=V_0a^2/2$ and $r_s=a$. The soft Coulomb constant $\eta=d/a$ is chosen to be 0.05 as in the previous section. The corresponding spinless single-band tight-binding model is then
	\begin{equation}\label{eq6}
	H_\text{tb}=\left(\sum_{i} tc_i^\dagger c_{i+1} +h.c. \right)+ \sum_{i<j}\frac{r_s}{|i-j|} n_in_j .
	\end{equation} 
    As the number of electrons is chosen less than the number of dots and electrons repel each other, double occupancy is unlikely (we can always impose it as a constraint). Therefore, we ignore both the spin degree and the on-site interaction, thus considerably simplifying the problem. However, electrons at different sites can interact through the long-range Coulomb interaction described by the second term in Eq.~\eqref{eq6}. As we already point out in subsection IIC, this term qualitatively resembles the on-site interaction strength $U$ in the Hubbard model. The hopping strength $t$ can be estimated by the WKB approximation. For large $V$, the kinetic energy of the particle inside the dot is $k^2\sim \sqrt{ V} \ll V$, so we can estimate $k^2\approx 0$ and classical turning points as $X_1\approx-0.5$ and $ X_2\approx 0.5$. The tunneling amplitude is approximately
	\begin{equation}\label{eq7}
	\begin{split}
	t&\propto \exp\left[\int_{X_1}^{X_2} -\sqrt{2(V(X)-k^2)}dX\right]\\
	&\approx \exp[-1.27\sqrt{V}].
	\end{split}
	\end{equation}
	
	To validate Eq.\eqref{eq7}, we consider a translationally invariant single-particle 1D Hamiltonian with a cosine potential $H=-\partial_X^2/2 + V\cos(2\pi X)$. This Hamiltonian has exact solutions - the Mathieu function, allowing us to calculate the bandwidth $\delta = E(k=\pi) - E(k=0)$. We obtain the derivatives of $d\ln \delta/d\sqrt{V}$ at $V=25, 100, 400, 600$ as $0.92, 1.12, 1.20, 1.22$ respectively. Therefore, even though the correct asymptote is $t\propto e^{-1.27\sqrt{V}}$, for the range $V<100$ used in our simulation, it is reasonable to approximate $t\propto e^{-1.0\sqrt{V}}$. Then, by fitting the ground state energy of the 8 dots - 4 electrons continuum model given in Eq.\eqref{eq4} to the corresponding tight-binding model for $r_s=0$, we have the approximate relation
	 $t=24\exp(-1.0\sqrt{V})$ connecting the continuum and the tight-binding model in the large $V$ limit of interest. Note that the approximate nature of this free electron to tight binding mapping is not of much significance since our interest is a general understanding of the Mott-liquid crossover.
	
	In Fig.~\ref{fig5}, we present the simulation results for the tight-binding model of 4 electrons in 8 sites with different values of the ratio $r_s/t$ (recall that $r_s=a/a_B$). For a low value of $r_s/t$, all the sites are almost equally occupied, and the system is an effectively `metallic' electron liquid. As $r_s/t$ increases, the occupancies on sites 1-3-6-8 are enhanced while other sites are less likely to have electrons due to the suppressed hopping. Beyond a certain large critical value of $r_s/t$, we achieve complete localization, i.e. 4 sites with occupancy close to unity ($>0.99$) and the rest are empty (with occupancy $<0.01$). As a result, we have the following empirical rule for the conductor-insulator or liquid-Mott transition
	\begin{equation}\label{eq8}
	\frac{t}{r_s} =\text{const}=24e^{-C} \Rightarrow \ln(r_s)= C-1.0\sqrt{V}.
	\end{equation}
	
	By numerically searching for the lowest $r_s/t$ where any occupied site (or the multiple sites in the incommensurate cases) has occupancy $> 0.99$ , we obtain the constant $C$ of Eq.~\eqref{eq8} as $C=8.5, 7.5 \text{ and } 6.7$ for different configurations: 4e/8 dots, 3e/8 dots and 3e/6 dots, respectively. To test this finding, we carry out a simulation for $V$ ranging from 0 to 100 in the continuum model described by Eq.\eqref{eq4}. At each value of $V$, we obtain the minimum $r_s$ such that the density profile shows 4 distinct peaks with occupancy larger than 0.99 each for the four-electron case; for the three-electron cases, the middle peak is allowed to split into two sub-peaks because of the incommensurability. The results are shown in Fig.~\ref{fig6}. For $V > 25$, all the boundaries show similar linear relation between $\ln(r_s)$ and $\sqrt{V}$ and the fitted coefficients are consistent quantitatively with the prediction from the tight binding model. For $V < 25$, the localization is weakly dependent on $V$, which suggests that the process is driven mostly by the interaction rather than the underlying lattice potential. We emphasize that this interaction-driven `localization-delocalization' transition is the crux of Mott physics, which should be observable in all arrays of coupled quantum dots according to our simulations. 
	
	We conclude that for low $V$ ($V<25$), the delocalized-to-localized crossover is more related to the Wigner crystallization at $V=0$. For high $V$ ($V>25$), the crossover is more like Mott transition (which only needs short-range interaction) and the long-range nature of the interaction is less important since the required value of $r_s$ decays exponentially with $\sqrt{V}$. Based on these purely qualitative arguments, we divide the localization regime into (somewhat arbitrarily) two parts: Mott insulator for $V>25$ where the lattice plays a dominant role and Wigner crystal for $V<25$ where the lattice does not play a dominant role. Phenomenologically, both Wigner crystal and Mott insulator look similar with sharply distinguishable spatial density peaks, and we could very easily define all non-zero $V$ insulating situation as the Mott phase calling only the $V=0$ localized phase insulator the Wigner phase. However, as we will show in the next subsection, there are some key differences between Mott and Wigner phases in terms of density correlations. We emphasize that a smooth Mott-Wigner crossover can be caused by continuously varying the coupling strength $r_s$ relative to $V$ with the large $V$ ($r_s$) phase being comparatively more Mott (Wigner)-like. The specific `critical' parameter separating these phases is arbitrary and ill-defined since the phenomenon is purely crossover physics. 	
	
	\subsection{Correlated phase}
	\begin{figure}
		\centering
		\begin{minipage}{0.02\textwidth}
			\rotatebox{90}{\hspace{0.3in} $R(\Delta x)a$ $(10^{-2})$}
		\end{minipage}
		\begin{minipage}{0.45\textwidth}   
			\centering
			\includegraphics[width=\textwidth]{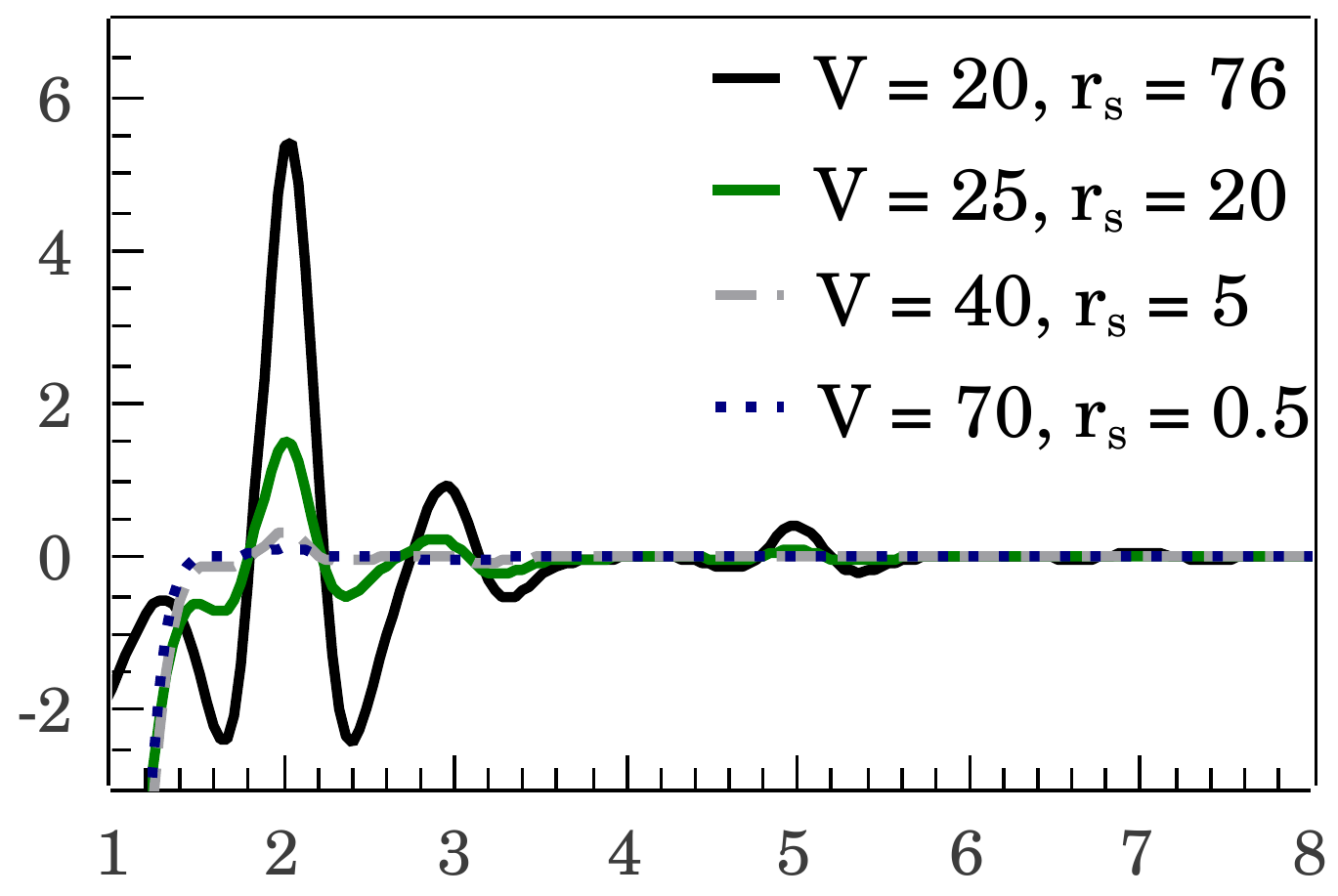}
			
			$\Delta x/a$
		\end{minipage}		
		\caption{Average density correlation compared to the average density $(1/a)$ with respect to distance $\Delta x$ at points along the conductor-insulator phase boundary in Fig~\ref{fig6} for the four-electron-eight-dot array. The correlation essentially vanishes for $V>25$ justifying that being the effective Mott-Wigner boundary.}\label{fig7}
	\end{figure}
	
\begin{figure}
		\begin{minipage}{0.02\textwidth}
			\rotatebox{90}{\hspace{0.2in} $\ln(r_s)$}
		\end{minipage}
		\begin{minipage}{0.42\textwidth}	
			\includegraphics[width=\textwidth]{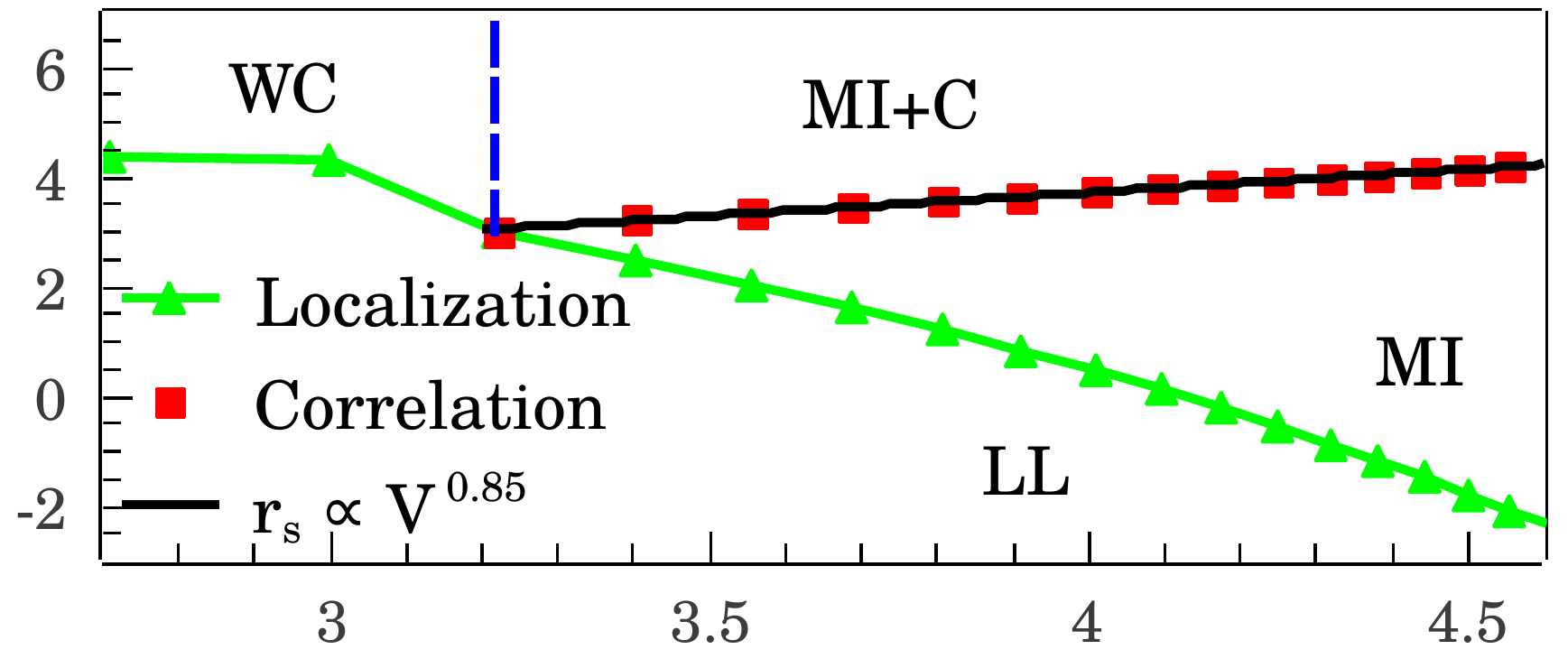}
			
			$\ln(V)$
		\end{minipage}
		\caption{Phase diagram of 4 electrons in an eight-dot array showing the correlated Wigner crystal (WC) phase, uncorrelated Mott insulator (MI), correlated Mott insulator (MI+C) and Luttinger liquid (LL) phases.}\label{fig8}
	\end{figure}

	As mentioned above, Wigner crystals and Mott insulators are similar in terms of their density profiles, both manifesting strongly localized density peaks.. However, Mott insulators only need weak interaction when $V$ is sufficiently large (as can be seen in the exponentially decaying $r_s$) while Wigner crystal formation requires large $r_s$ long-range interaction. Therefore, the density correlation function can provide a signature to distinguish the Wigner phase \cite{collectivemode, wignercorrelation, noorder}. The correlation function is defined as
	\begin{equation}
	R(\Delta x) = \int dx \braket{\rho(x)\rho(x+\Delta x)} - \braket{\rho(x)}\braket{\rho(x+\Delta x)}.
	\end{equation}
	In Fig.~\ref{fig7}, we show the calculated density correlation scaled to average density ($\sim 1/a$) for various $(V,r_s)$ values along the localization phase boundary shown in Fig.~\ref{fig6} for the 4 electrons/8 dots case. For $V>25$ where the localization is mostly aided by the interactions in the presence of the periodic potential, the correlation is significantly suppressed, while for $V<25$, the Wigner crystal region, the density correlation is noticeable. In the Mott insulator phase, the electrons can be considered as individual oscillators trapped at the sites; whereas in the Wigner crystal phase the excitation is always collective. This also reinforces the collective (individual) Coulomb blockade property of the Wigner (Mott) phase. An equivalent qualitative description is that the Wigner crystal phase is essentially the `correlated' Mott phase where lowering the lattice potential and decreasing electron density enhances the density correlations, inducing a crossover in the system from individually site-localized Mott phase to a `correlated' Mott phase, and eventually, to the Wigner crystal phase with the correlations being maximum in the crystalline phase. This individual/collective behavior can be estimated using a classical model of a system of coupled oscillators \cite{collectivemode}. The potential energy (including the periodic background and the Coulomb potential) tensor is given by expanding around the equilibrium position $X_i = n+1/2$ with $n$ being an integer
	\begin{equation}\label{eq10}
	\begin{split}
	A_{i,i} &= \frac{\partial^2}{\partial X_i^2} \left(-V\cos(2\pi X_i) + \sum_{j \ne i} \frac{r_s}{|X_i-X_j|} \right)\\ &= 4\pi^2 V +  2r_s\sum_{j\ne i}\frac{1}{|X_i-X_j|^3};\\
	A_{i,j} &=\frac{\partial^2}{\partial X_i\partial X_j}  \frac{r_s}{|X_i-X_j|} = \frac{-2r_s}{|X_i-X_j|^3}.
	\end{split}
	\end{equation}
	The maximum correlation is obtained when the off-diagonal elements are much larger than the diagonal ones. As a result, the correlation properties of the system should depend on the ratio $r_s/V$. Hence, we are able to tune the system from uncorrelated to correlated Mott insulator phase by increasing $r_s$ at a fixed $V$ or equivalently decreasing $V$ at fixed $r_s$. To study the crossover to the correlated Mott phase, we increase $r_s$ at each value of $V>25$ until the maximum of the correlation function is $0.015/a$ - the maximum value of the correlation function at $V=25$. The boundary of the completely-correlated Wigner crystal (WC), uncorrelated Mott insulator (MI), and correlated Mott insulator (MI+C) as well as the extended metallic Luttinger liquid (LL) phase are shown together in Fig.~\ref{fig8}. From the numerical simulation on the continuum model with tunable $V$ and $r_s$ , the correlation boundary is best described by $r_s\propto V^{0.85}$ which is close to the crude estimation using coupled oscillators model.
	
	We note that our qualitative phase digram depicted in Figs.~\ref{fig8} and \ref{fig9} (to be described and discussed below) is qualitative since all the phases here are simply crossover phenomena.  In a finite system, we do not expect strict quantum phases, nevertheless we believe that these three phases (Wigner, Luttinger, Mott) are meaningful to explore experimentally.  After all it is well-known that 1D Coulomb systems do not have any long range crystalline order \cite{Schulz}, but the recent experimental observation of the effective 1D Wigner crystal in a finite 1D system is still a useful advance. \cite{exp0}
	
	\subsection{Correlation physics in the tight binding model}
	
	\begin{figure}
		\begin{minipage}{0.02\textwidth}
			\rotatebox{90}{\hspace{0.2in} $a$ (a\textsubscript{B})}
		\end{minipage}
		\begin{minipage}{0.42\textwidth}	
			\includegraphics[width=\textwidth]{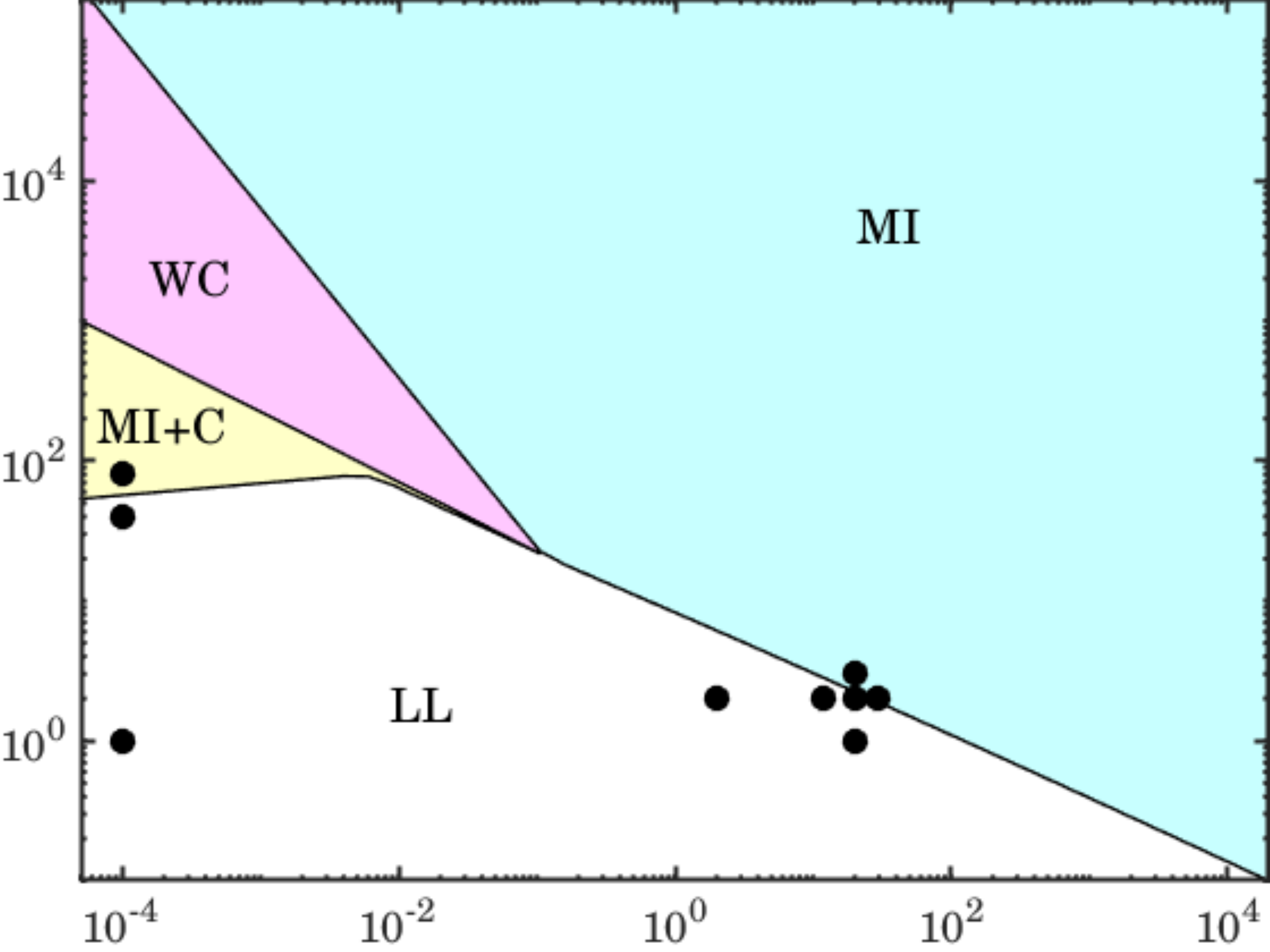}
			
			$V_0$ (Ry)
		\end{minipage}
		\caption{Phase diagram same as Fig.~\ref{fig8} but in physical parameters: the inter-dot spacing $a$ and the potential barrier height $V_0$. The black dots represent states in Fig.~\ref{fig1}. The value $V_0=0$ in Figs.~\ref{fig1}(g)-(i) is approximated as $10^{-4}$ to display on the logarithmic scale.}\label{fig9}
	\end{figure}
	To derive the condition for the delocalized/localized crossover, we  make the connection between the physical system and the tight-binding model, implying the possibility of studying the effect in a semiconductor quantum emulator in the laboratory \cite{Delft2017}. The same question arises with the correlation property: whether it can be studied by a quantum emulator. Our first task is to understand how the continuum model in MI+C phase is mapped into the tight-binding model. The fitted equation of the correlated phase boundary suggests that the interaction strength $r_s$ is comparable to the band gap (controlled by $V$). Thus, one has to consider multiple bands in the tight-binding model. In that case, each dot possesses not only charge but also dipole and multipoles. As we are interested only in the region of strong localization, the wavefunction overlap between different dots is negligible and the Hamiltonian up to the second order contains only the following interactions: charge-charge, charge-dipole, dipole-dipole and charge-quadrupole. Moreover, in the large $V$ regime, we can approximate the electrons as oscillating inside a harmonic potential with frequency $\omega \propto V^{1/2}$. As a result, the band gap, dipole and quadrupole pole go as $\Delta E \propto V^{1/2}, \braket{x} \propto V^{-1/4} \text{ and } \braket{x^2} \propto V^{-1/2}$. The detailed interactions are
	\begin{equation*}
	\begin{split}
	&\text{charge-charge: }V_0(i-j) =r_s n_in_j/|i-j|;\\
	&\text{charge-dipole: }V_1(i-j)=r_s\gamma_{\alpha,\beta}\frac{ n_ic_{j,\beta}^+c_{j,\alpha}\text{sgn}(i-j)}{V^{1/4}|i-j|^2};\\
	&\text{dipole-dipole: }V_2(i-j)=r_s\gamma_{\alpha,\beta}\gamma_{\delta, \sigma} \frac{ c_{j,\delta}^+c_{j,\sigma}c_{j,\beta}^+c_{j,\alpha}}{V^{1/2}|i-j|^3};\\
	&\text{charge-quadrupole pole: }V_2'(i-j)=r_s\gamma'_{\alpha,\beta}\frac{ n_ic_{j,\beta}^+c_{j,\alpha}}{V^{1/2}|i-j|^3};
	\end{split}
	\end{equation*}  
	where the Greek indices indicate dot energy levels and coefficients $\gamma$, $\gamma'$ depend only on the level indices. We emphasize that all the terms in the Hamiltonian commute with the dot occupancy operator $n_i = \sum_{\alpha} c_{i,\alpha}^\dagger c_{i,\alpha}$. Therefore, individual dot occupancies are good quantum numbers and will not exhibit the correlation. On the other hand, the Hamiltonian does not commute with level occupancy operators. Thus, if one can resolve the occupancy of each dot excitation level (e.g. measure the dot dipole), the correlation physics can be directly seen in a quantum dot emulator. The ability of inducing inter-dot correlation depends on the ratio of non-commuting interactions to the band gap, i.e. $V_1/\Delta E \propto r_s/V^{3/4}$ and $V_2,V_2'/\Delta E \propto r_s/V$. It should be noted that in a symmetric system, charge-dipole interactions between dots cancel each other (in fact in our approximation using coupled oscillators, we assume that the equilibrium positions of the Coulomb and trapping potentials coincide, which is a consequence of symmetry and therefore the total charge-dipole interaction disappears automatically). For a system with only a few dots and electrons, the symmetry is broken and charge-dipole interaction might be non-zero. This may explain why the extracted numerical exponent in Fig.~\ref{fig8} is between 3/4 and 1. 
	
	To give more insight into reproducing different collective ground states in a quantum dot emulator, we convert  the phase diagram expressed (Fig.~\ref{fig8})  in dimensionless parameters $r_s$ and $V$ back to the physical parameters $a$ and $V_0$ (we remind that $V$ depends on both $a$ and $V_0$) in Fig.~\ref{fig9}. Figure~\ref{fig9} using experimental parameters (average inter-electron distance and the effective lattice potential) shows that the most accessible phases in the coupled dot system are Mott insulator and Luttinger liquid. The Wigner crystal phase requires very large $a$ (low average density) and very small $V_0$, while the correlated Mott insulator exists at low $V_0$ and even larger $a$.
	
	The currently available quantum dot arrays are incapable of manifesting the Wigner phase as shown in our Fig.~\ref{fig9}, but improvement in fabrication and control may led to the observation of the Wigner phase in quantum dot arrays.
	
	\subsection{Crossover to classical phases}
\begin{figure*}
	  	\centering
	  	\begin{minipage}{0.04\textwidth}
	  		\rotatebox{90}{$\rho(x)a$}
	  	\end{minipage}	
	  	\begin{minipage}{0.95\textwidth}
	  		\centering
	  		\includegraphics[width=0.32\textwidth]{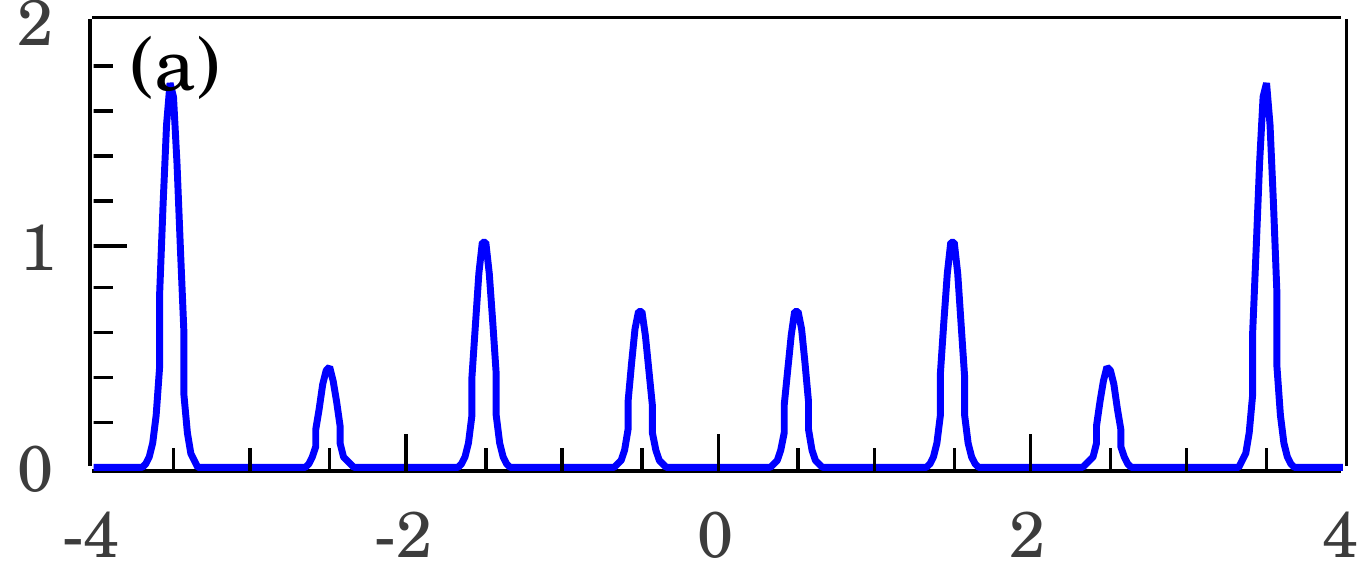}  	
	  		\includegraphics[width=0.32\textwidth]{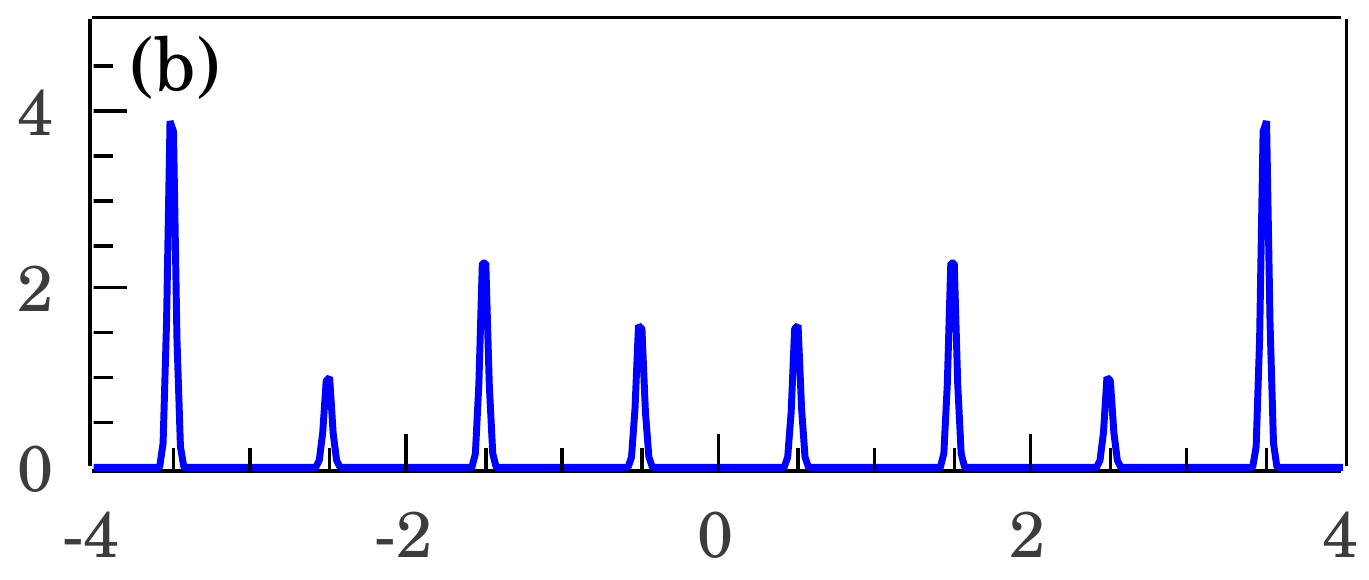}
	  		\includegraphics[width=0.32\textwidth]{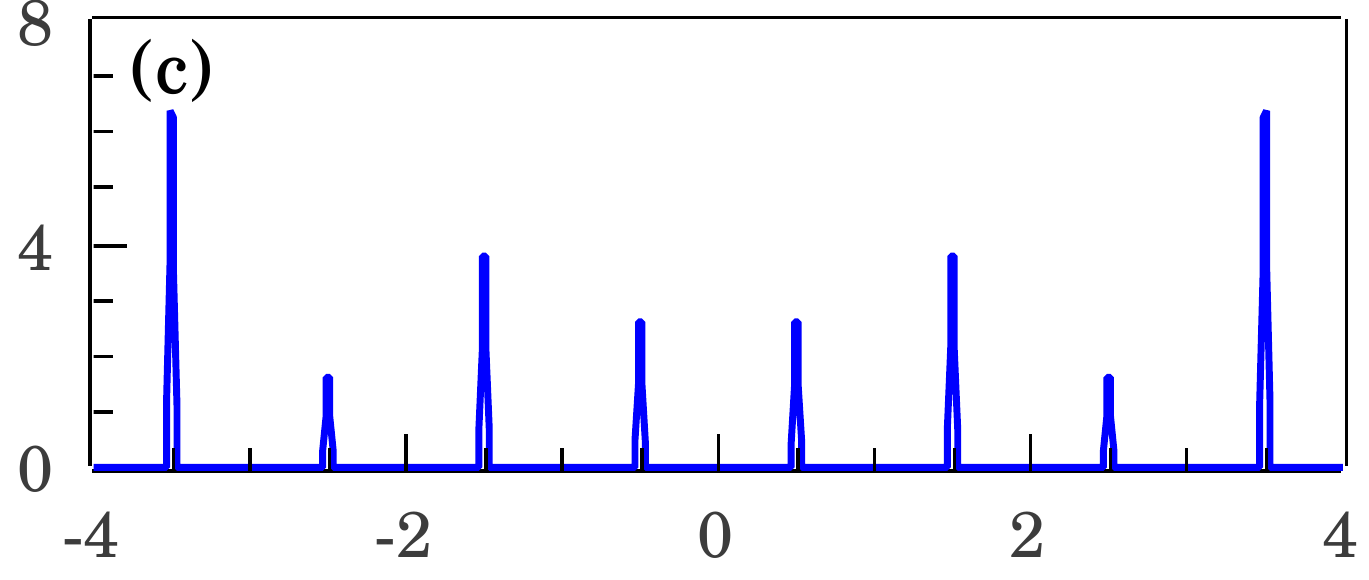} 
	  		
	  		\includegraphics[width=0.32\textwidth]{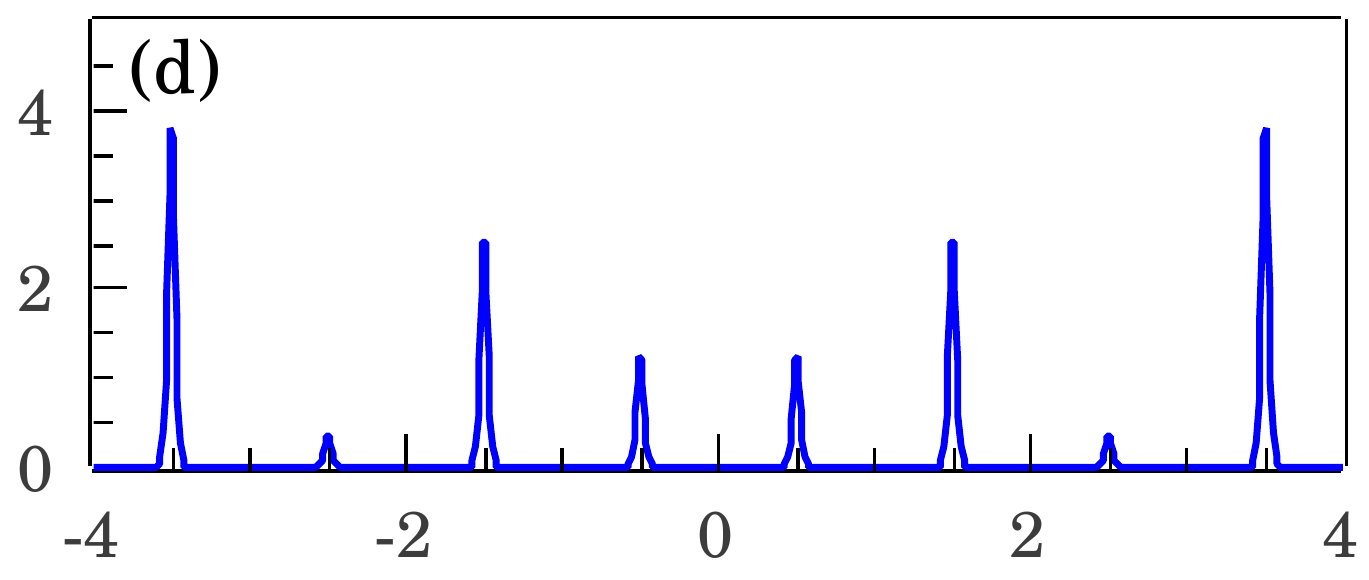}  	
	  		\includegraphics[width=0.32\textwidth]{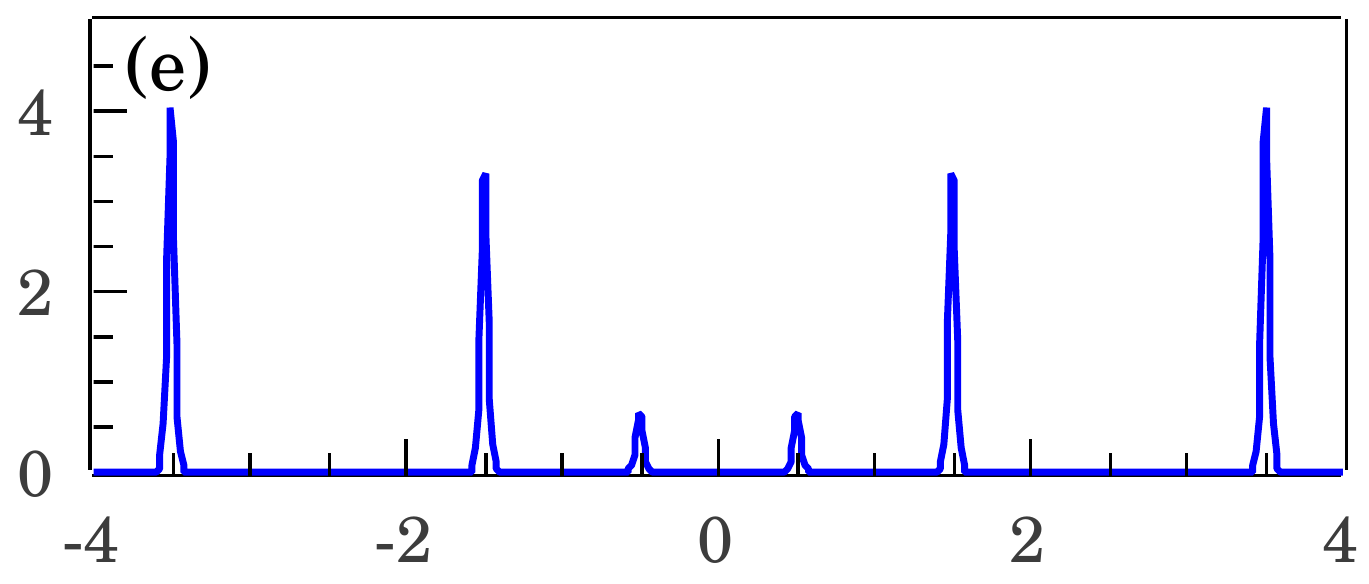}
	  		\includegraphics[width=0.32\textwidth]{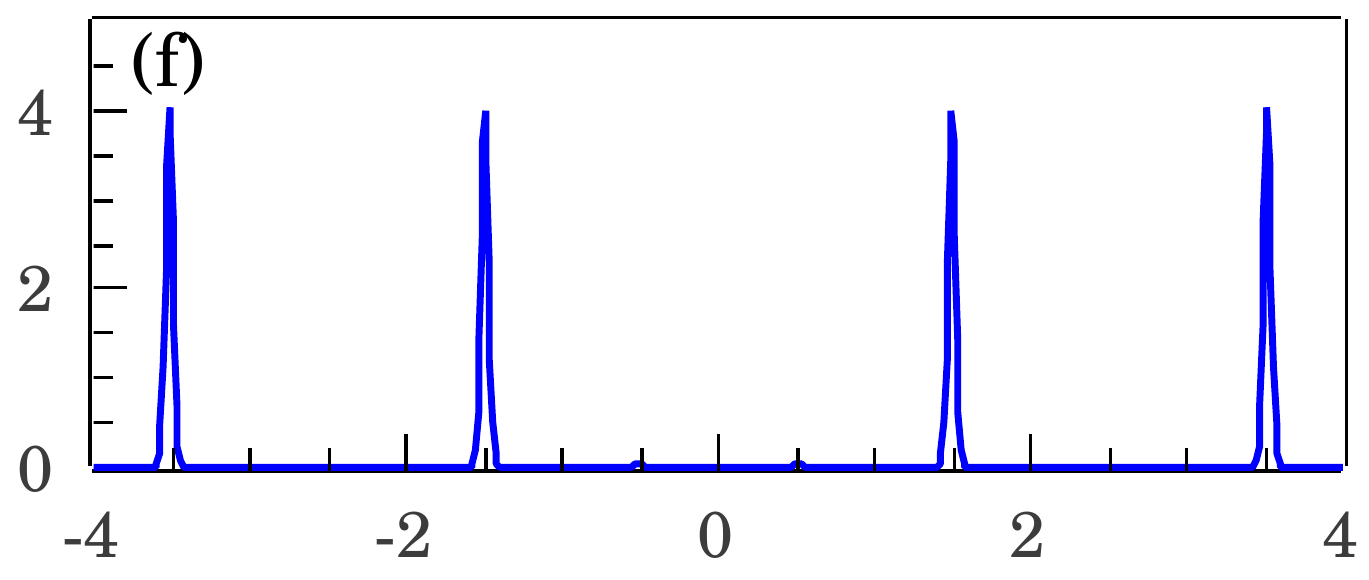}  		
	  	\end{minipage}
	  	
	  	$x/a$
	  	\caption{The classical limit of the spatial density profile of 4 electrons in an eight-dot array. The upper row: $k_BT=0.2$~Ry, $k_BT_F = 0.08$~Ry, $a=2~$a\textsubscript{B} and (a) $V_0=2$~Ry, (b) $V_0=12~$Ry, (c) $V_0=30~$Ry. The occupancy pattern stays unchanged despite the increasing $V_0$ at fixed $a$. The lower row: $k_BT=1.2\times 10^{-3}$~Ry, $V_0=0.5$~Ry and (d) $k_BT_F = 8.0\times10^{-6}$~Ry, $a=200$~a\textsubscript{B}, (e) $k_BT_F=8.9\times10^{-5}$~Ry, $a=60$~a\textsubscript{B}, (f) $k_BT_F=8.0\times 10^{-4}$~Ry, $a=20$~a\textsubscript{B}. The occupancy pattern shifts from 8 peaks to 4 peaks when decreasing $a$ at fixed $V_0$ but not when increasing $V_0$ at fixed $a$.}\label{fig10}
	  \end{figure*}

	Strictly speaking, our previous results on the collective quantum ground states are only valid at zero temperature. However, these results are still applicable when the temperature is much less than the effective Fermi temperature $T_F\sim 1/a^2$. For example, with an eight-dot array, the expression of the hopping strength $t=24e^{-1.0\sqrt{V}}\hbar^2/(ma^2)$ continues down to $V=ma^2V_0/\hbar^2 = 25$, thus we estimate the quantum-classical crossover temperature as
	\begin{equation}
        k_BT_F = \frac{0.32}{a^2},
	\end{equation}
	with $a$ expressed in Bohr radius and $k_BT_F$ in Rydberg. When $T\gg T_F$, the system is simply described by the classical Boltzmann distribution. The spatial density profile in this thermodynamic limit is obtained by
	\begin{equation}\label{eq12}
	\rho(y) \propto \int dx^N   \delta(x_1-y) \exp\left(-\frac{U(\{x_i\})}{k_BT}\right),
	\end{equation}
	where $U$ is the sum of the background potential and the interaction energy. For high $T$, the system is a classical liquid with uniform spatial density distribution. The spatial density profile in the classical regime for different sets of $a$ and $V_0$ are shown in Fig.~\ref{fig10}.
	
	In the classical limit, the hopping strength as a function of $a$ and $V_0$ or the non-interacting kinetic energy as a function of $a$ is replaced by the thermal energy which only depends on the temperature. As a result, changing $V_0$ has no effect on the charge distribution. On the other hand, decreasing $a$ increases the interaction energy $E_c\sim 1/a$ while the thermal energy is unchanged, thus effectively inducing a liquid-to-solid crossover. In the limit of very low $a$ only one configuration with the lowest interaction energy (for 4 electrons in an eight-dot array, the occupied 1-3-6-8 sites) survives. We emphasize that in the quantum system, such crossover happens for increasing $V_0$ at fixed $a$, or increasing $a$ at fixed $V_0$. Therefore, in experiments when the temperature is higher than $T_F$, the observed classical thermal states can have completely different behavior from the true quantum ground states. Thus, studying temperature dependence could also offer insight into the nature of collective quantum ground states of coupled quantum dot arrays.
	
	\section{Connection to the Delf Experiment}
	
	In Ref.~\cite{Delft2017}, the Delft group experimentally studied 3 gate-defined quantum dots to observe the inter-dot-tunneling-induced  `transition' from the individual Coulomb blockade (CB) behavior to the collective Coulomb blockade (CCB) behavior.  In the current section we consider spinful electrons in 3 coupled quantum dots in order to make connection between our work and this Delft experiment, which was interpreted as the solid state quantum simulation of the Fermi-Hubbard model.  We use the full continuum Coulomb model (our Eq.~\ref{eq1}) in contrast to the original predictions \cite{Cblockage,tunneling1998} and  of CCB (and the numerics in the Delft experiment) where the tight binding Hubbard model was used to predict the CB to CCB crossover with increasing electron hopping.
	
\begin{figure*}
	\begin{minipage}{0.04\textwidth}
		\rotatebox{90}{\hspace*{0.5cm} $\rho(x)a$ }
	\end{minipage}
	\begin{minipage}{0.3\textwidth}
		\centering
		\includegraphics[width=\textwidth]{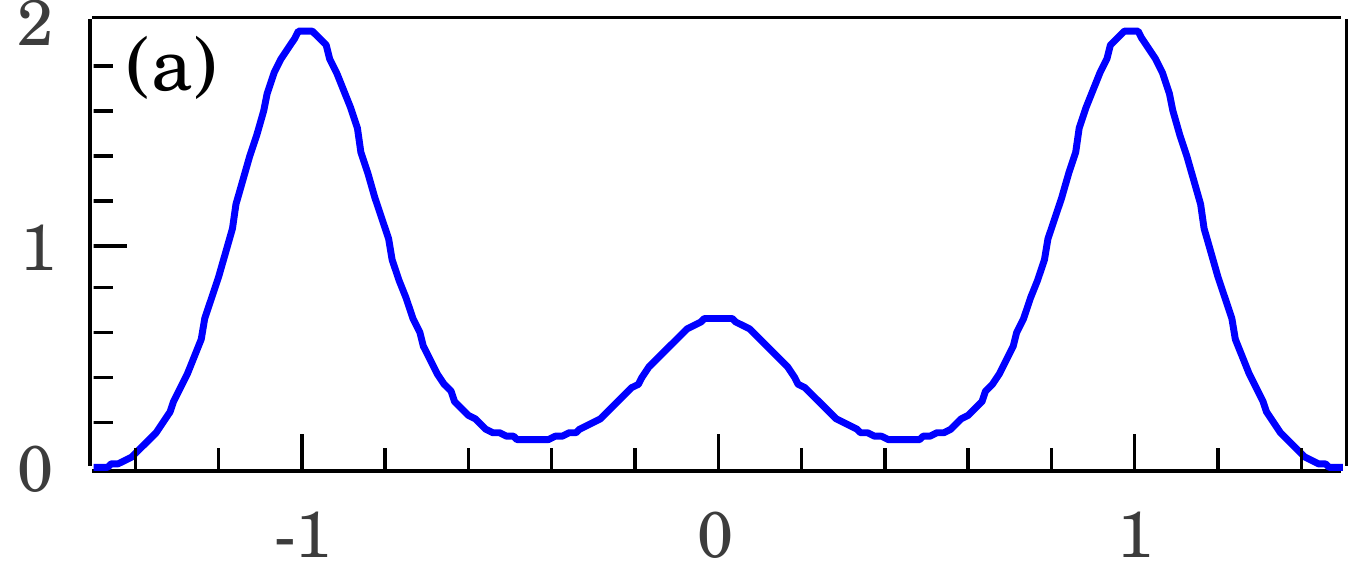}
		\includegraphics[width=\textwidth]{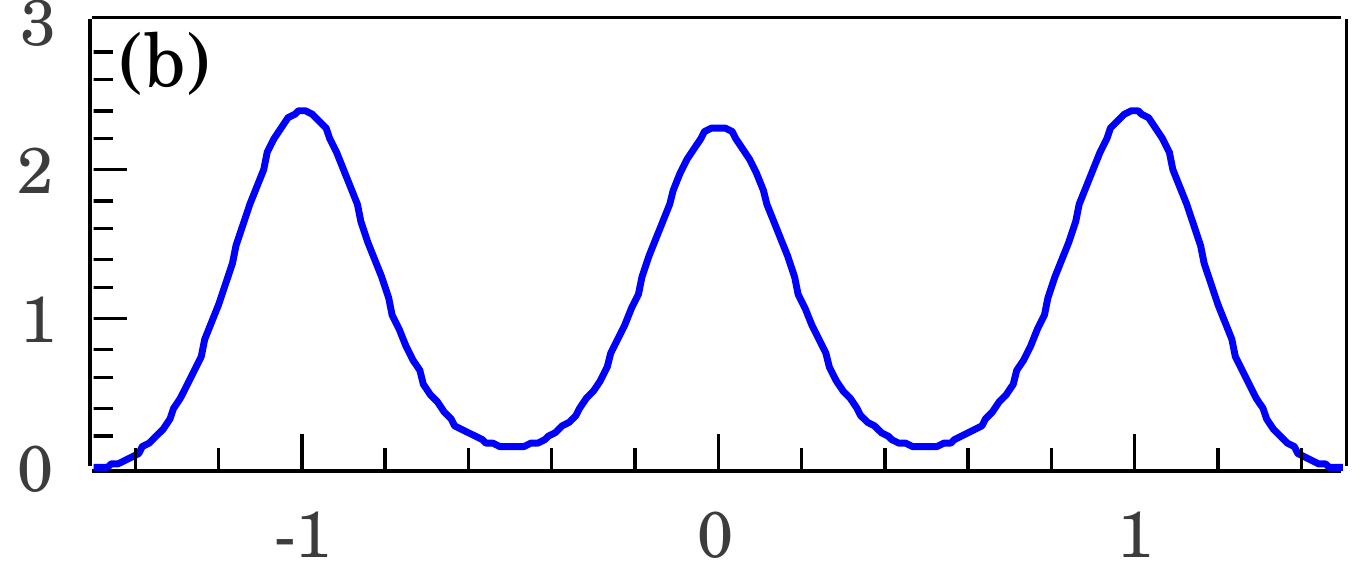}
		\includegraphics[width=\textwidth]{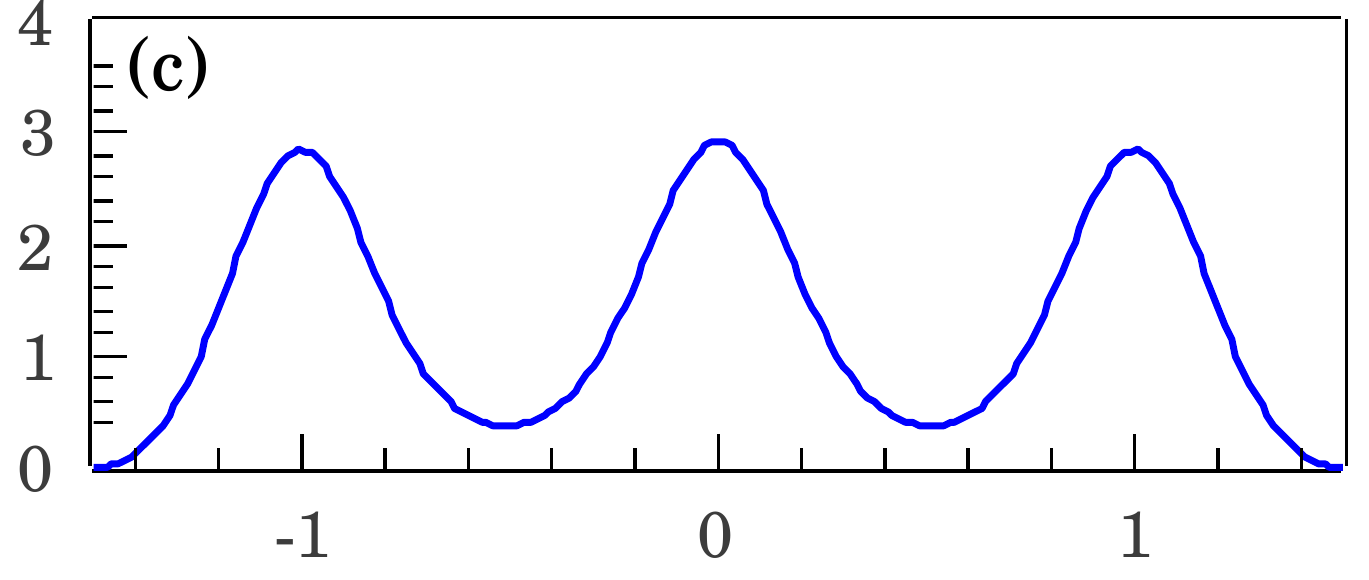}
	\end{minipage} 
	\begin{minipage}{0.3\textwidth}
		\centering
		\includegraphics[width=\textwidth]{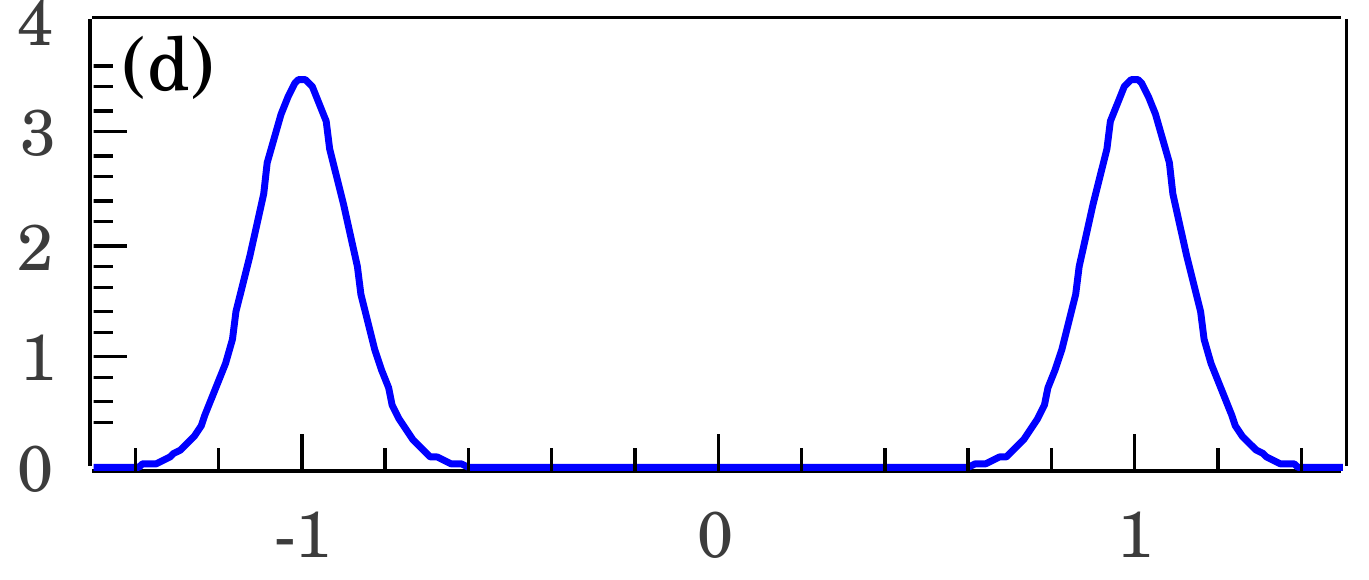}
		\includegraphics[width=\textwidth]{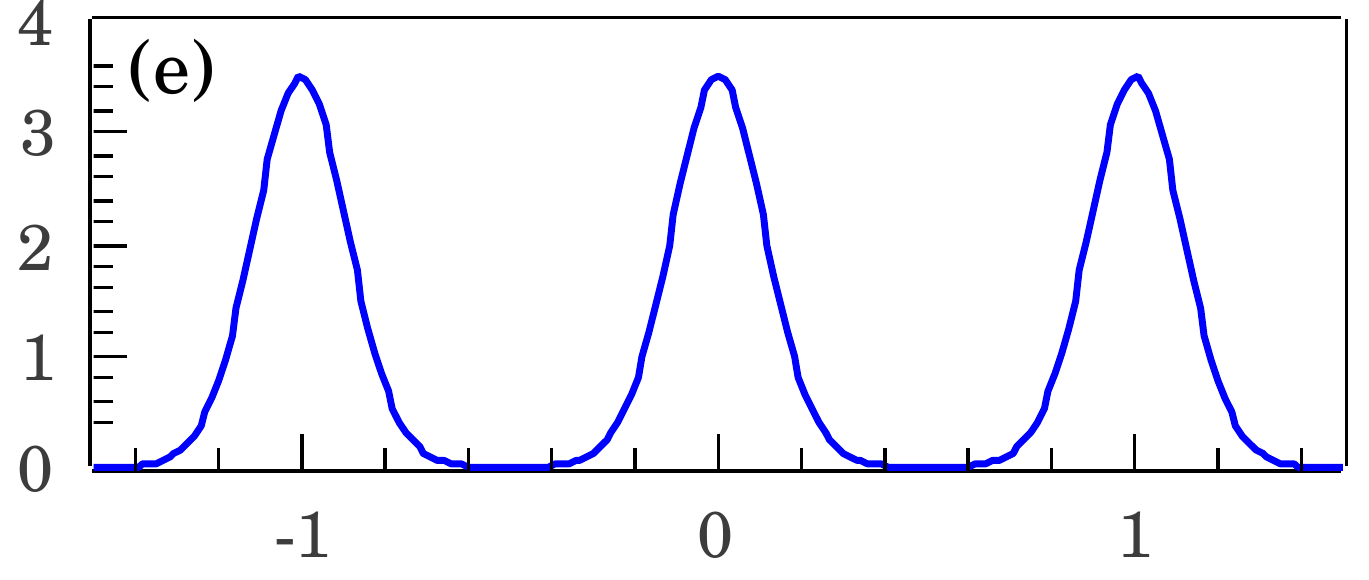}
		\includegraphics[width=\textwidth]{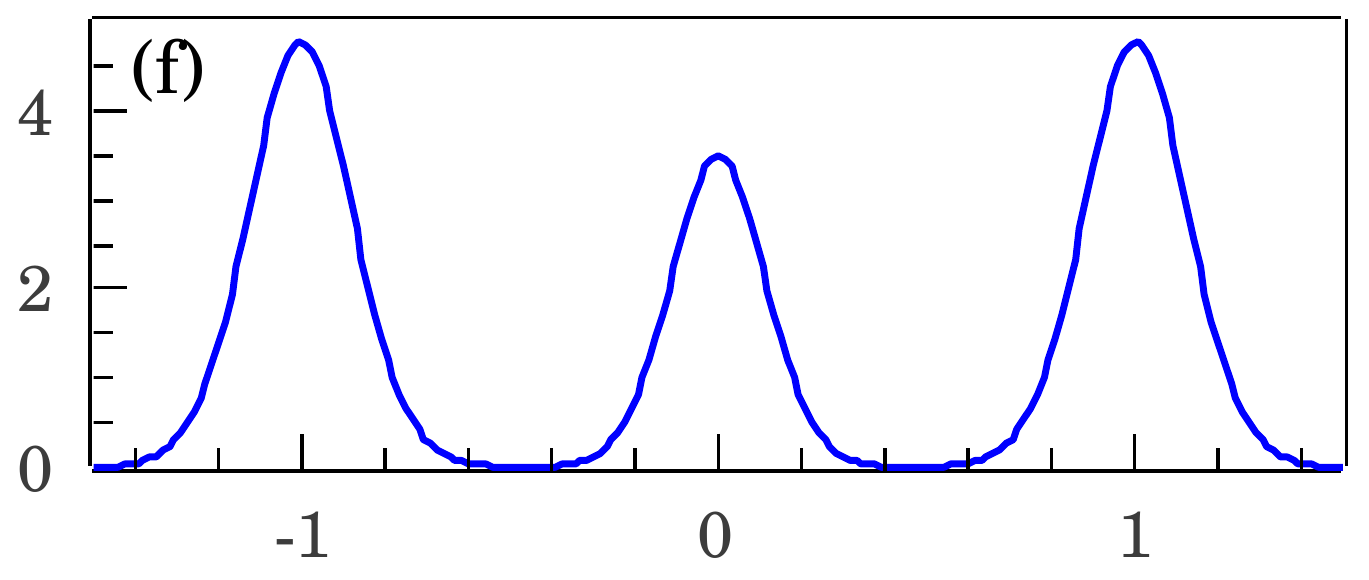}
	\end{minipage} 
	\begin{minipage}{0.3\textwidth}
		\centering
		\includegraphics[width=\textwidth]{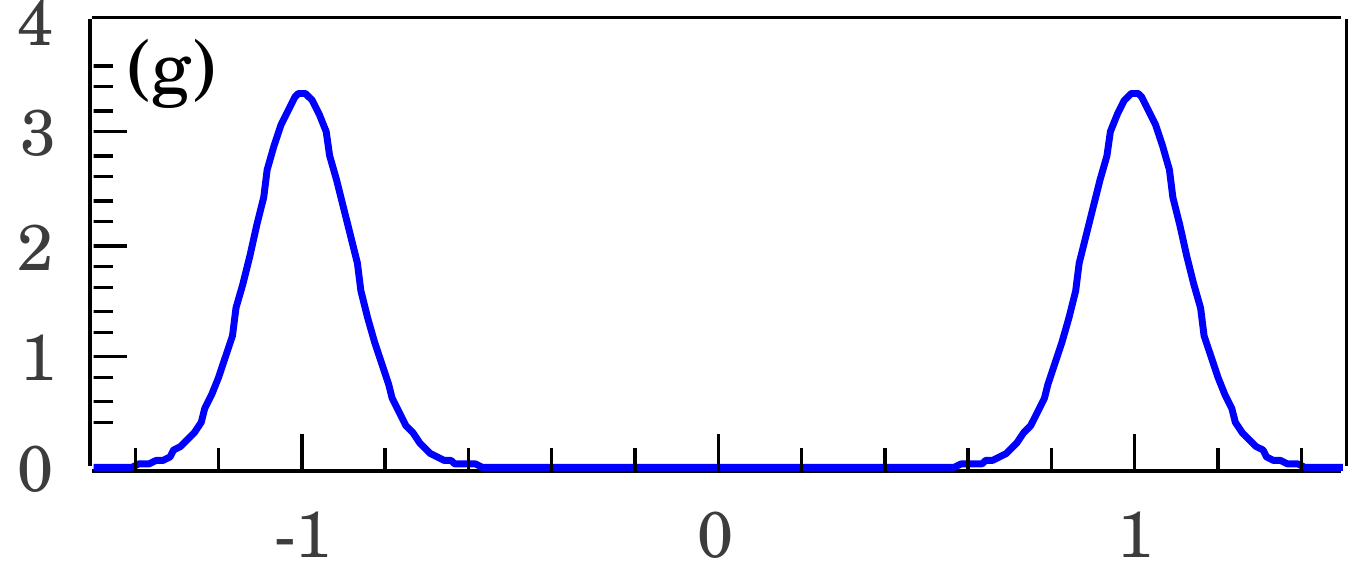}
		\includegraphics[width=\textwidth]{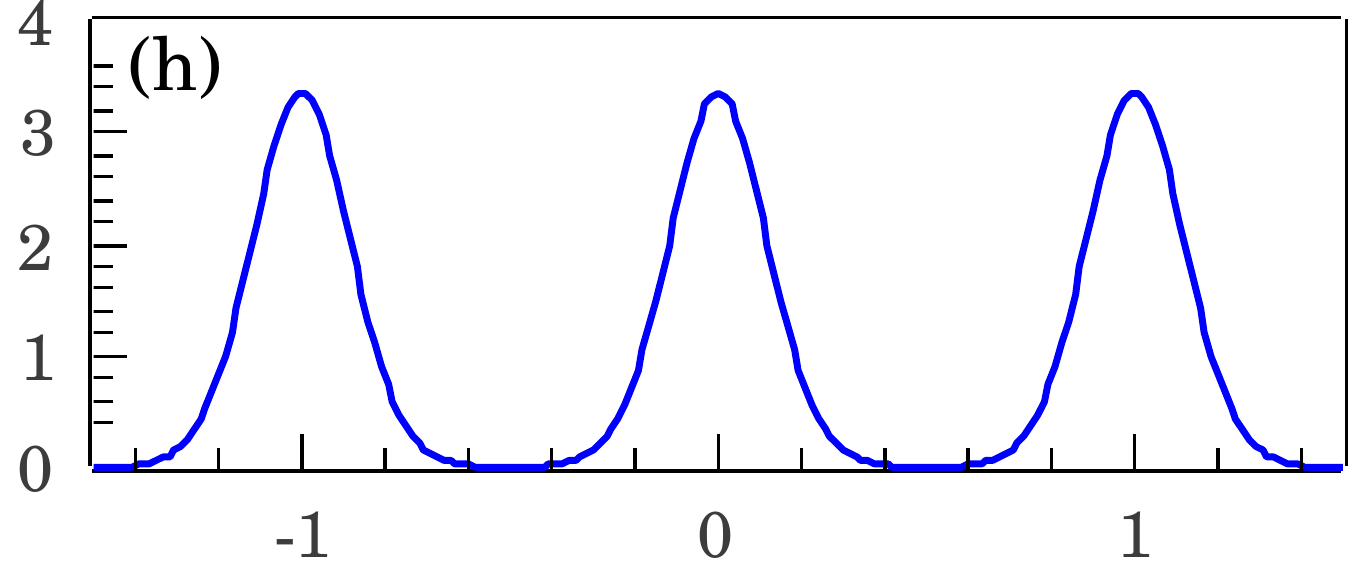}
		\includegraphics[width=\textwidth]{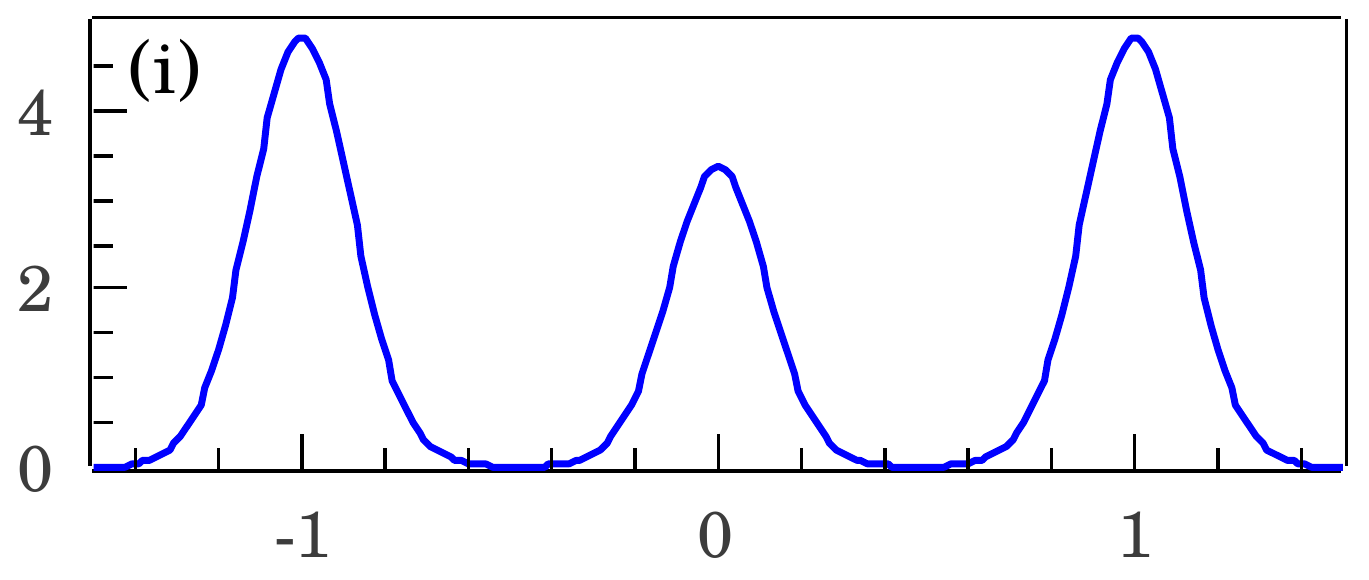}
	\end{minipage} 

   $x/a$
	\caption{Simulated spinful systems with $N=$2 (upper row), 3 (middle row) and 4 (lower row). We fix the Coulomb cut-off at $d=0.05a$ in all simulations and provide the ratio $t/U\approx 8\exp(-\sqrt{V})/(r_sV^{1/4})$ where $t$ is the hopping strength and $U$ is the on-site interaction energy ($r_s$ and $V$ are explained in the main text). Figures (a)-(c): $V_0=10~$Ry, $a=1.5~$a\textsubscript{B} and $t/U\approx0.1$, the system is a liquid, corresponding to the collective Coulomb blockade phase. Figures (d)-(f): $V_0=10$~Ry, $a=3$~a\textsubscript{B} and $t/U\approx0.001$. Figures (g)-(i): $V_0=35$~Ry, $a=1.5$~a\textsubscript{B} and $t/U\approx0.004$. At near zero $t/U$, the system is a Mott insulator, corresponding to the individual Coulomb blockade phase.}\label{fig3dot}
\end{figure*}

	In our work, the CB state is the Mott phase at very weak tunneling and the CCB state is the strong tunneling induced liquid phase.  In Fig.~\ref{fig3dot} we show our three-dot exact diagonalization results for spinful electrons based on Eq.\eqref{eq1}. For completeness, we show in Fig.~\ref{fig3dot} our exact three-dot results for $N=2,3,4$ spinful electrons as a function of increasing tunneling, choosing parameters which correspond approximately to the Delft experiment.  Our Fig.~\ref{fig3dot} clearly shows the smooth crossover nature of the hopping-induced transition from the CB (weak hopping) Mott phase to the CCB (strong hopping) liquid phase.  The individual electrons are strongly localized in the Mott phase leading to individual CB whereas the electrons are extended throughout all 3 dots in the CCB liquid phase.   It is instructive that the Mott CB to the liquid CCB phase crossover as a function of tunneling shows up for $N=2,3,4$ electrons in the same qualitative manner.  
	
\begin{figure*}
	\centering
	\begin{minipage}{0.04\textwidth}
		\rotatebox{90}{$\rho(x)a$}
	\end{minipage}	
	\begin{minipage}{0.95\textwidth}
		\centering
		\includegraphics[width=0.32\textwidth]{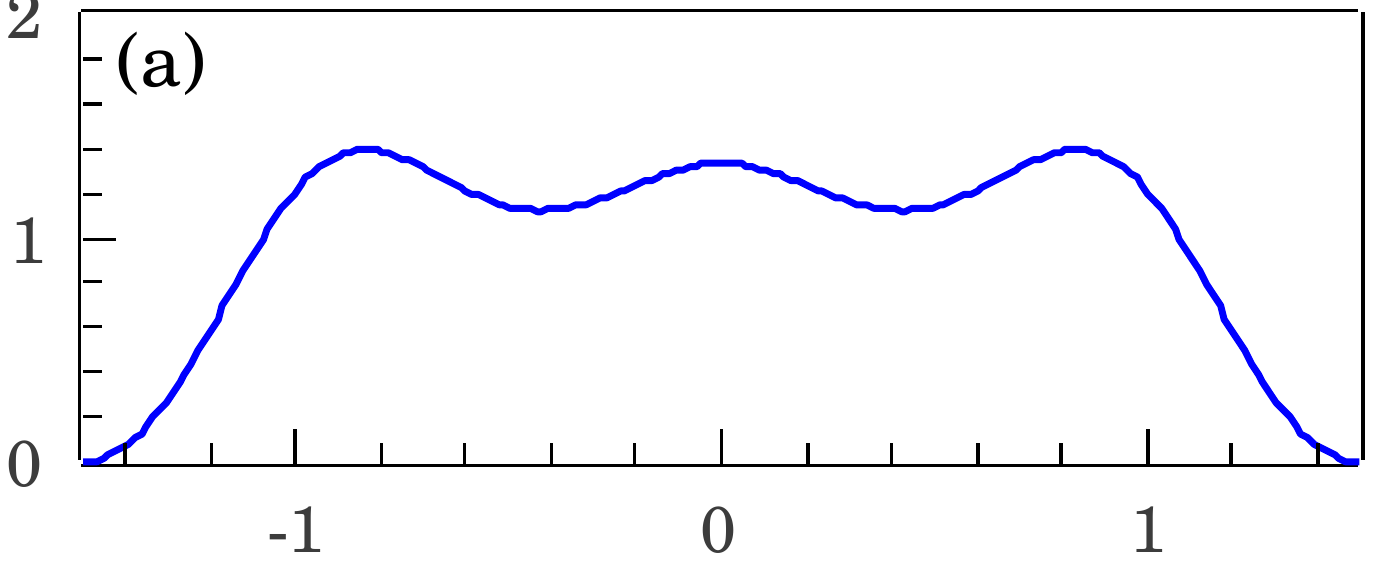}  	
		\includegraphics[width=0.32\textwidth]{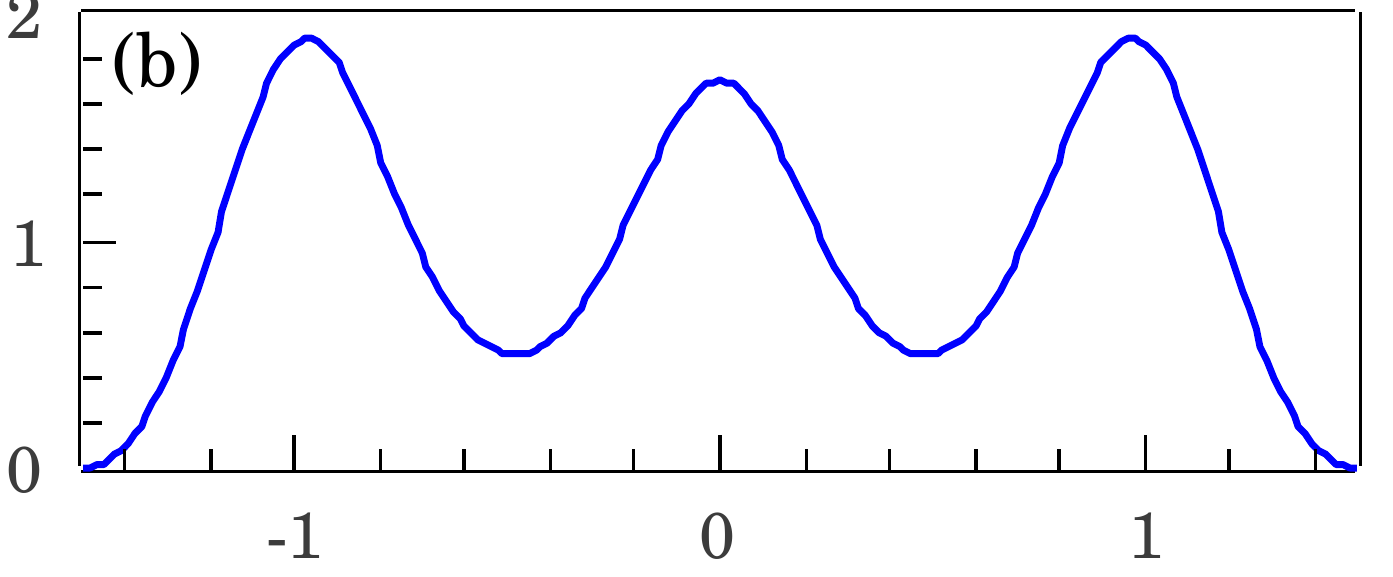}
		\includegraphics[width=0.32\textwidth]{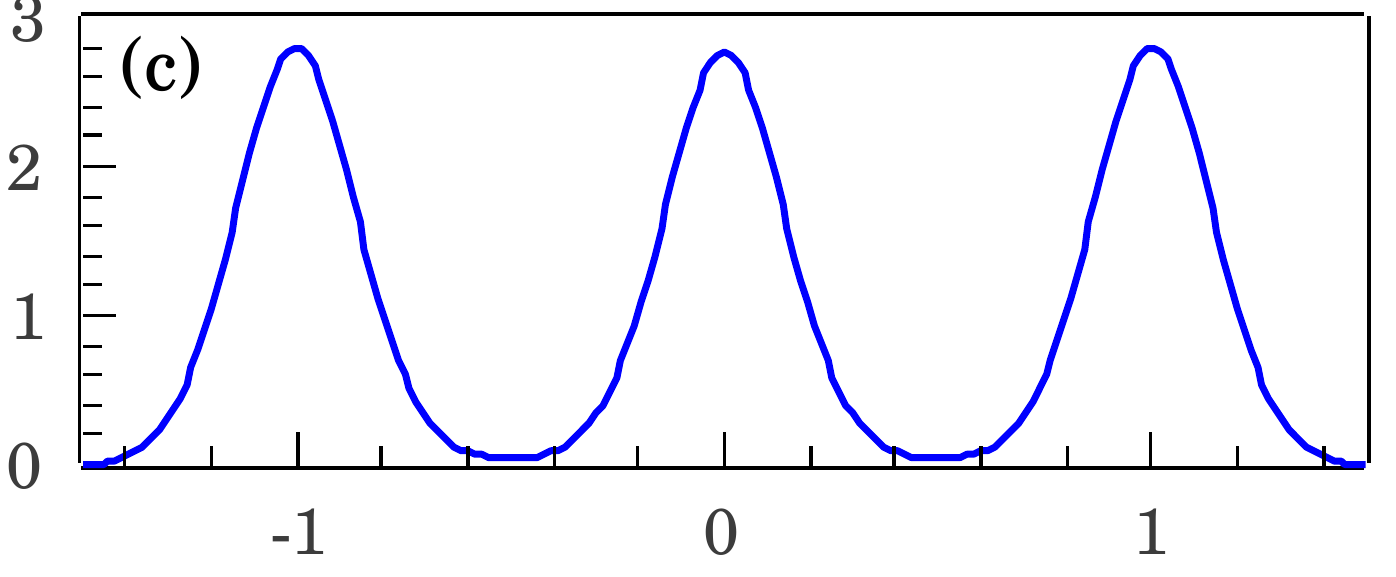} 
		
		\includegraphics[width=0.32\textwidth]{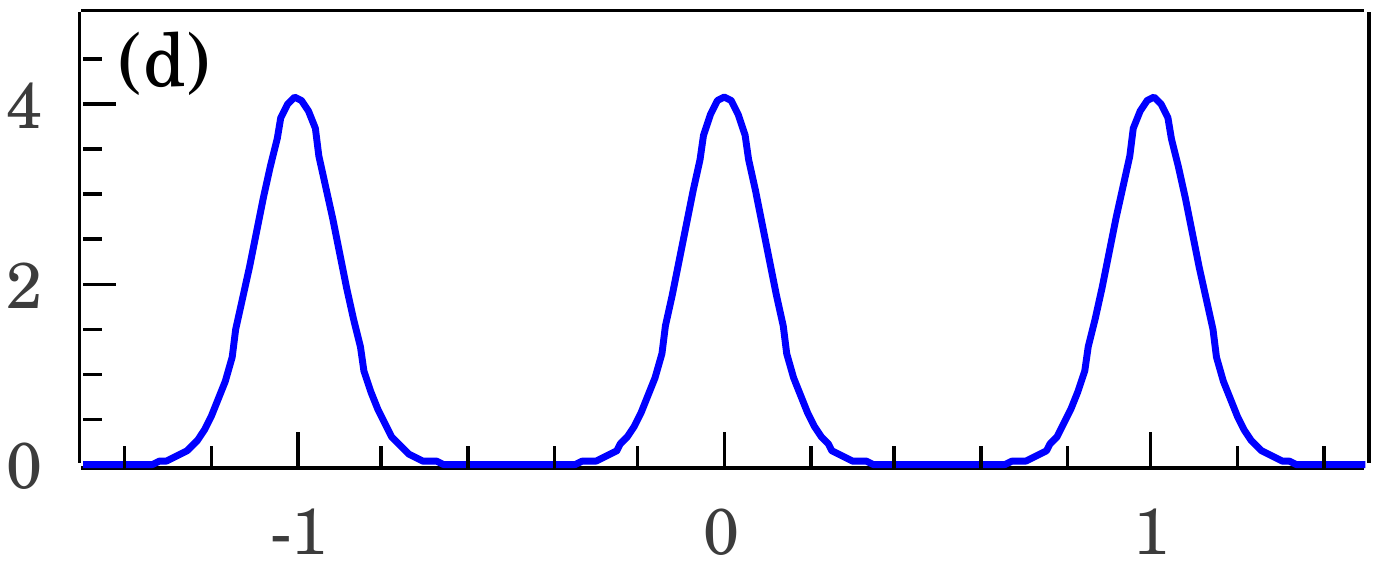}  	
		\includegraphics[width=0.32\textwidth]{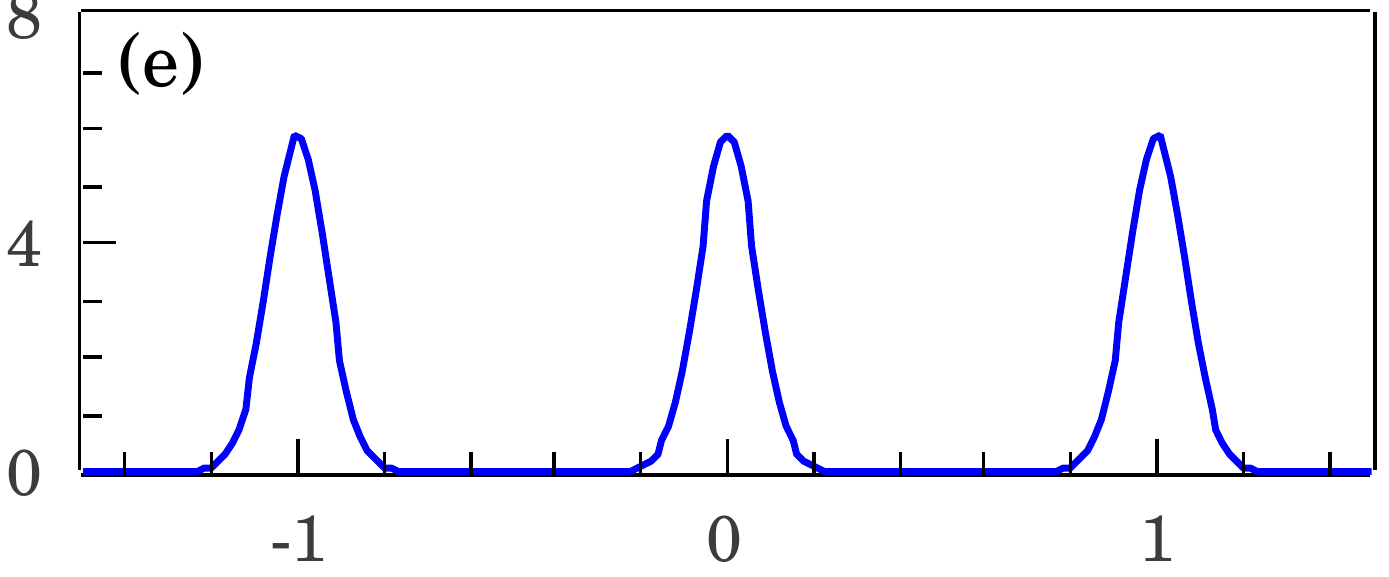}
		\includegraphics[width=0.32\textwidth]{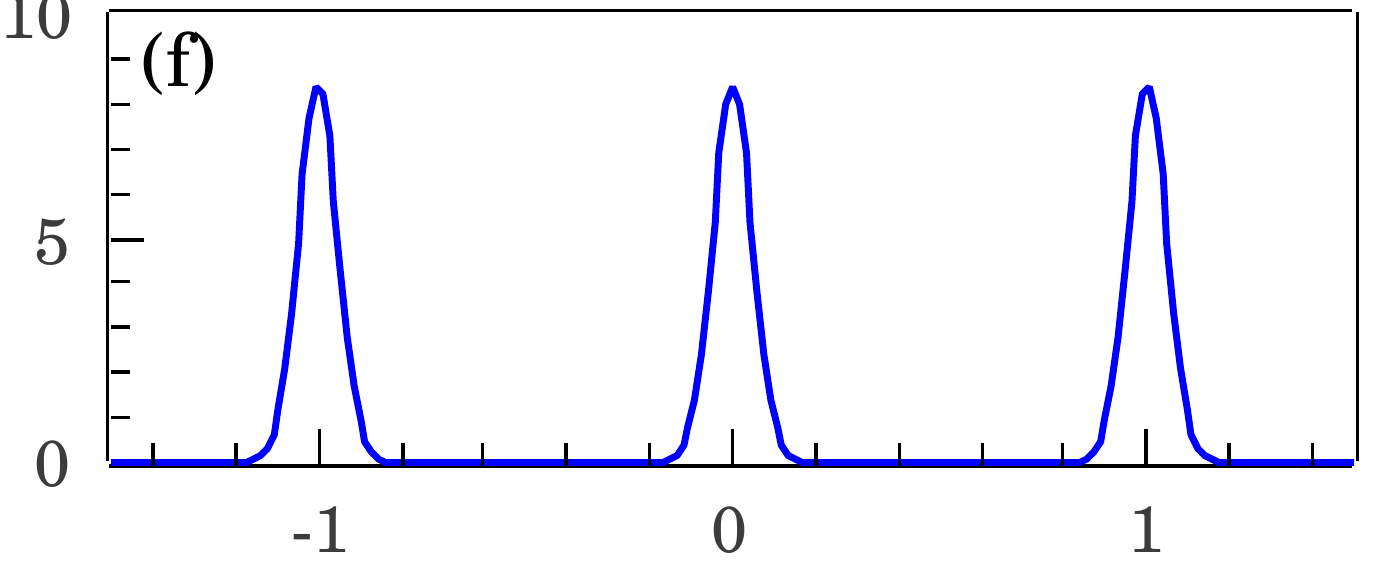}  	
		
		\includegraphics[width=0.32\textwidth]{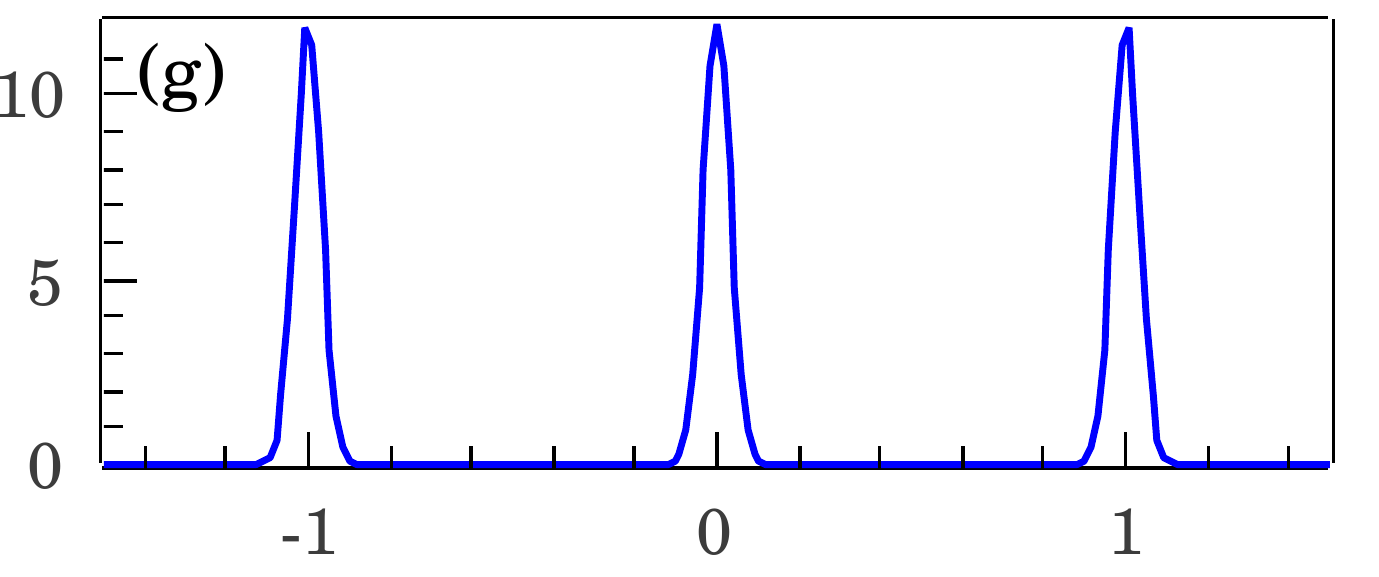}  	
        \includegraphics[width=0.32\textwidth]{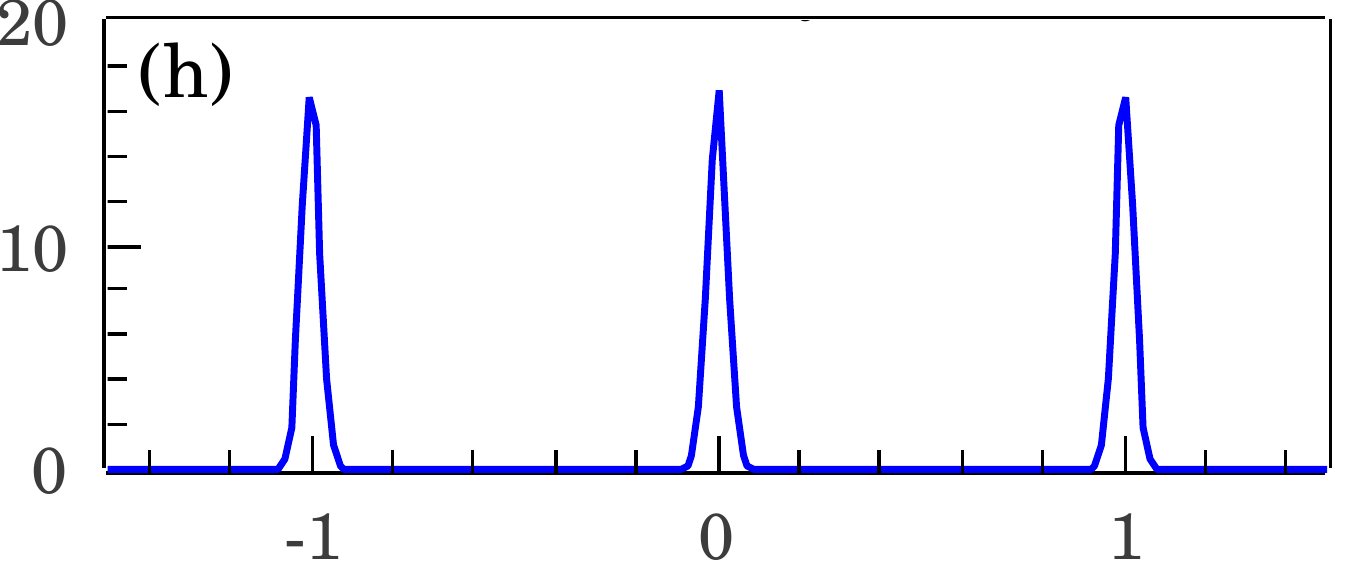}
        \includegraphics[width=0.32\textwidth]{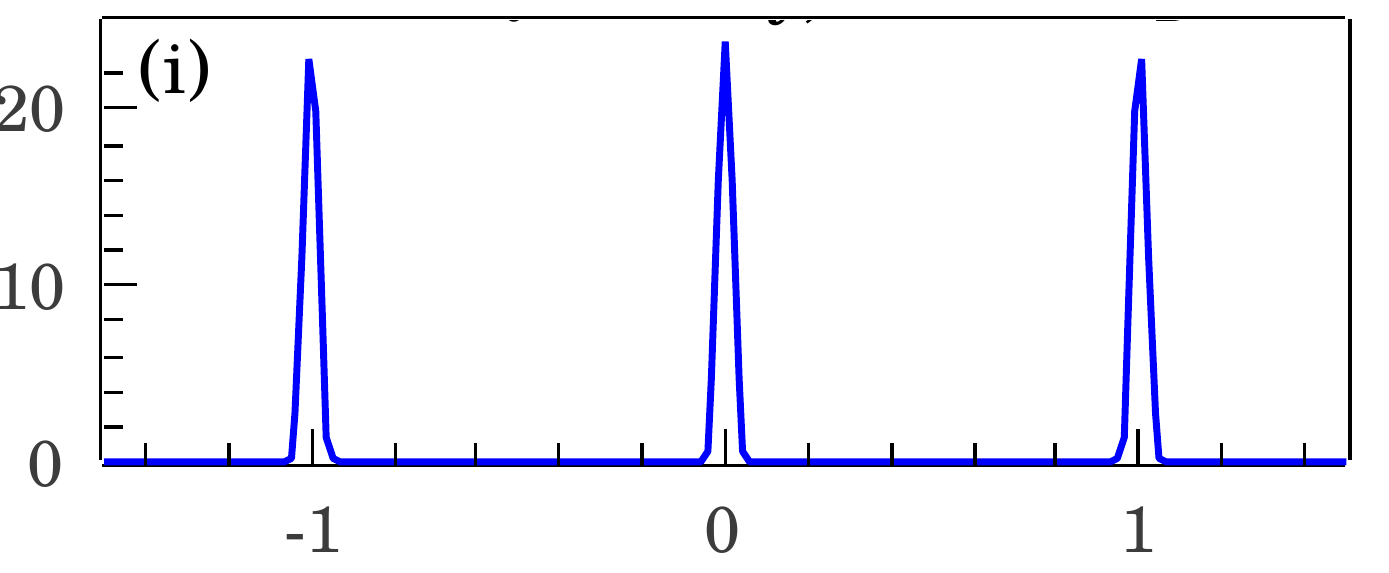}  				
	\end{minipage}
	
	$x/a$
	\caption{Simulated spatial density profile of 3 spinful electrons at fixed $V_0=10$~Ry, $d=0.05a$ and varying inter-dot spacing (a) $a=0.5$~a\textsubscript{B}, (b) $a=1$~a\textsubscript{B}, (c) $a=2$~a\textsubscript{B}, (d) $a=4$~a\textsubscript{B}, (e) $a=8$~a\textsubscript{B}, (f) $a=16$~a\textsubscript{B}, (g) $a=32$~a\textsubscript{B}, (h) $a=64$~a\textsubscript{B} and (i) $a=132$~a\textsubscript{B}. The system crossovers from the weakly interacting regime (CCB) at small $a$ to strongly interacting regime (CB) at large $a$. Given $V_0=10$~Ry, despite the strong localization at large $a$, the system is still an uncorrelated Mott insulator.}\label{figlargea}
\end{figure*}

	In Fig.~\eqref{figlargea} we focus on the half-filled three-dot and three-spinful electron system, showing results for a fixed $V_0$ (corresponding approximately to the Delft experiment), varying the inter-dot separation $a$, from very small $a$ corresponding to a weakly interacting system to a very large $a$ corresponding to a strongly interacting system.  The system crosses over from a CCB liquid state for strong tunneling (i.e. small $a$) to the CB Mott for larger $a$. According to our phase diagram, the Mott CB to the liquid CCB crossover takes place for $a\sim1$~a\textsubscript{B} in the Delft experiment where $V_0=10$.  This is consistent with the results shown in Fig.~\eqref{figlargea}. Also according to our calculated phase diagram, only Mott CB and liquid CCB phases are accessible for $V_0=10$~Ry, no matter how large $a$ is. Thus in the Delft experiment and indeed in all experimental systems fabricated so far, the correlated Mott phase and the Wigner phase remain inaccessible. To observe the correlated Mott or the Wigner phase the strength of $V_0$ must be two or more orders of magnitude smaller than in the Delft sample.
	
	 The clear observation of the Wigner phase in quantum dot arrays would thus necessitate developing quantum dot arrays with lower effective electron density, which should certainly be possible in the future.  Since the Wigner phase is relatively immune to disorder, the lower-density quantum dot arrays should manifest the density-modulated commensurate Wigner phase rather directly in the experiment.  We believe that such an observation will be easier if the electron density profile itself can be directly measured experimentally using well-known microscopic methods such as scanning tunneling microscope (STM) and/or atomic force microscope (AFM). We do emphasize, however, that the Wigner and Mott phases are not separated by any phase transition - it is simply a smooth crossover as the background periodic potential is decreased - the low-density periodic electron crystal in the absence of any background periodic potential is the Wigner phase whereas the corresponding sharply localized (at lattice sites) insulating electron phase in the presence of the lattice potential is the Mott phase.

	\section{Conclusion}
	  \begin{figure*}
	\centering
	\begin{minipage}{0.04\textwidth}
		\rotatebox{90}{$\rho(x)a$}
	\end{minipage}	
	\begin{minipage}{0.95\textwidth}
		\centering
		\includegraphics[width=0.32\textwidth]{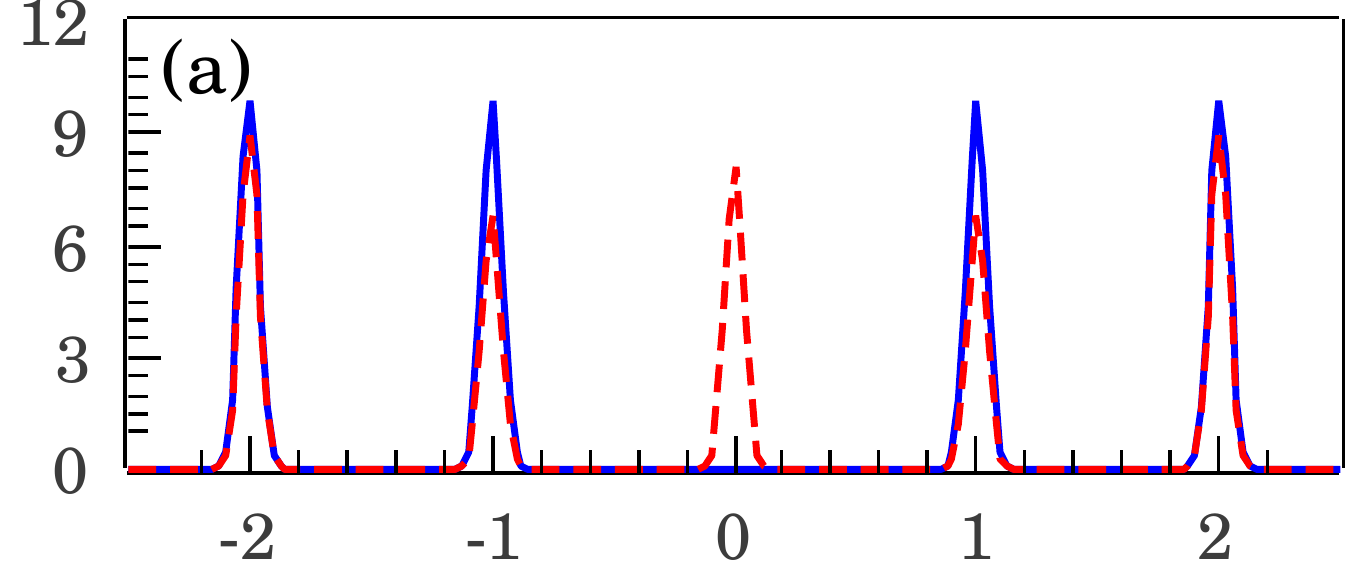}  	
		\includegraphics[width=0.32\textwidth]{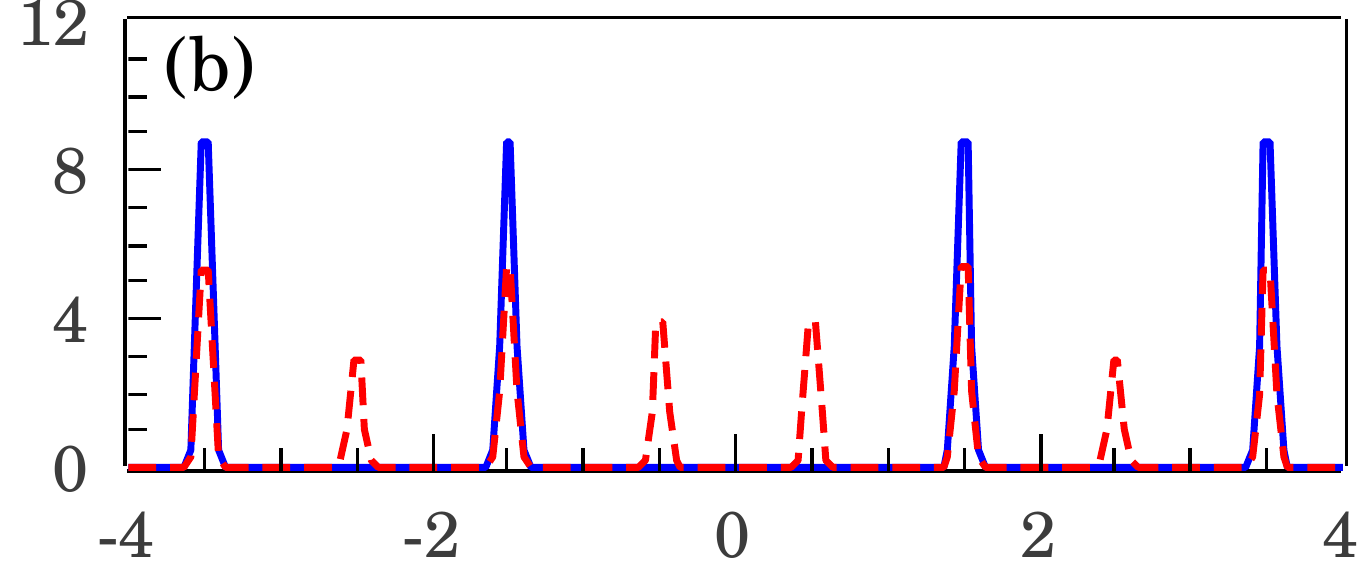}
		\includegraphics[width=0.32\textwidth]{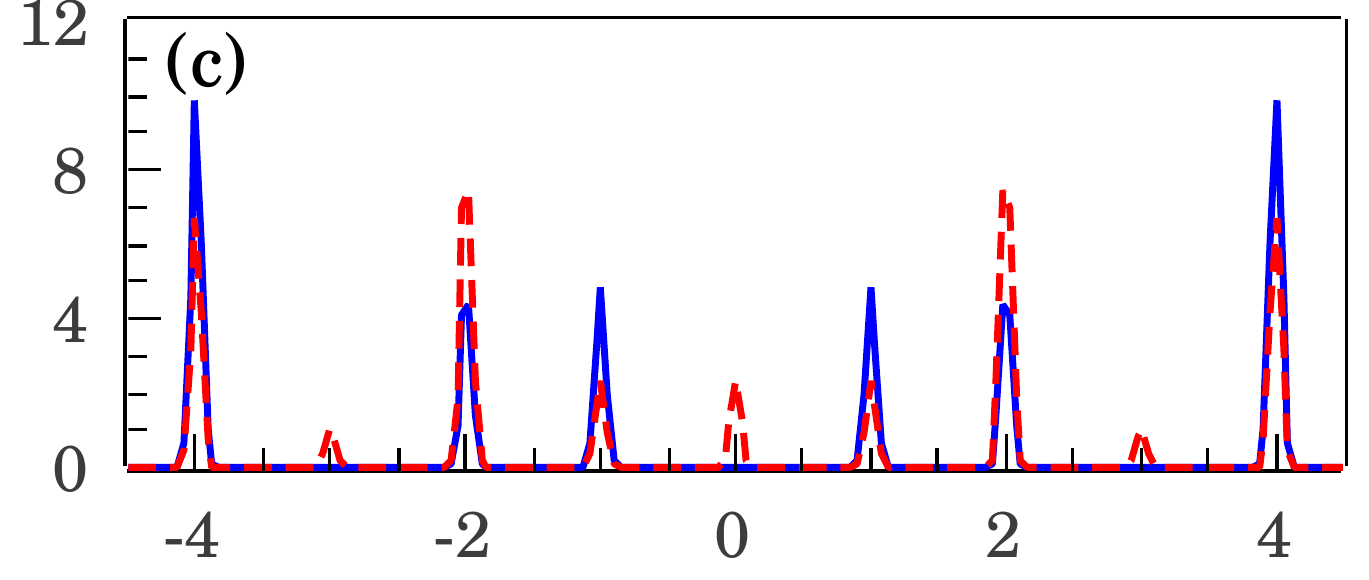} 
		
		\includegraphics[width=0.32\textwidth]{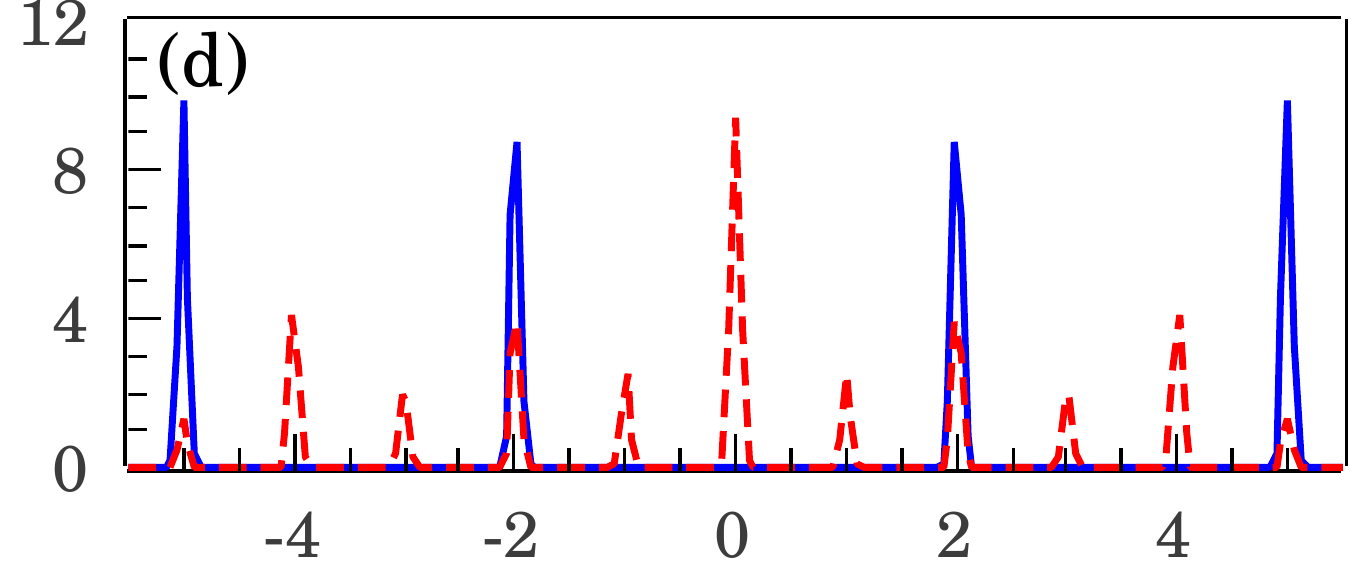}  	
		\includegraphics[width=0.32\textwidth]{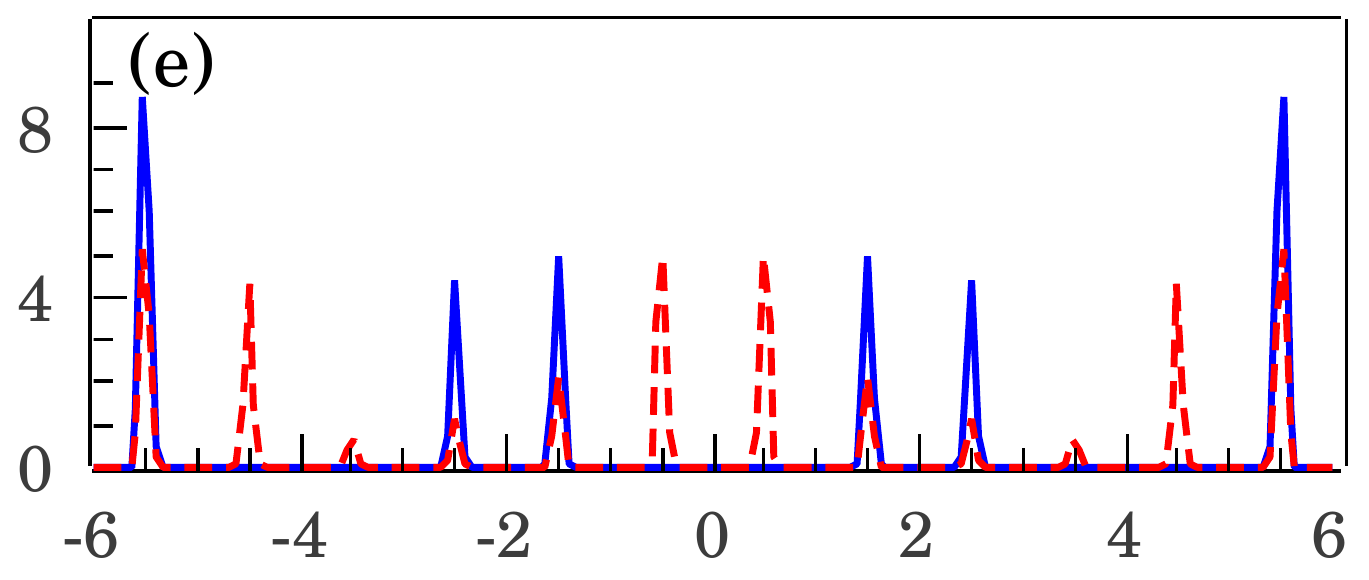}
		\includegraphics[width=0.32\textwidth]{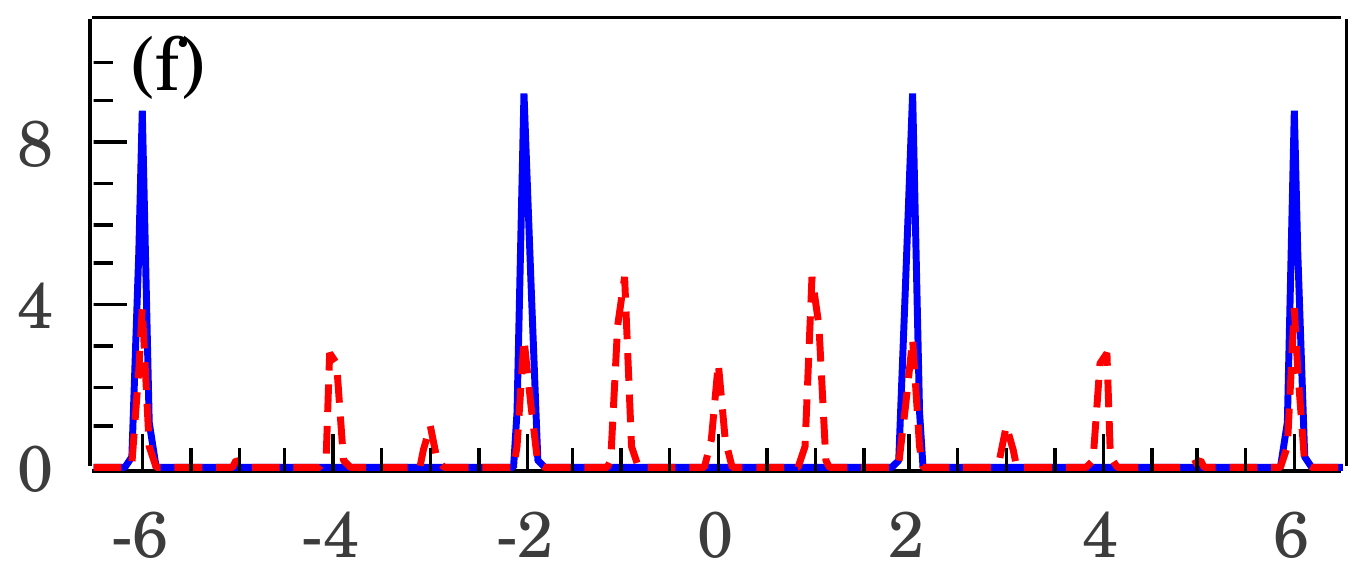}  		
	\end{minipage}
	
	$x/a$
	\caption{Mott phases of 4 electrons at $V_0=1$~Ry and $a=70$~a\textsubscript{B} for various number of dots (a) $N_d=5$, (b) $N_d=8$, (c) $N_d=9$, (d) $N_d=11$, (e) $N_d=12$ and (f) $N_d=13$. The blue (dark) solid lines are the density profiles of Coulomb interacting systems, while the red (gray) dashed lines represent the corresponding non-interacting systems. The number of peaks is larger than the number of electrons when $N_d$ is a multiple of 3 such as $N_d=9$ and 12. This is the Mott incommensurability discussed in the text}\label{fig11}
\end{figure*}	

	  \begin{figure*}
	\centering
	\begin{minipage}{0.04\textwidth}
		\rotatebox{90}{$\rho(x)a$}
	\end{minipage}	
	\begin{minipage}{0.95\textwidth}
		\centering
		\includegraphics[width=0.32\textwidth]{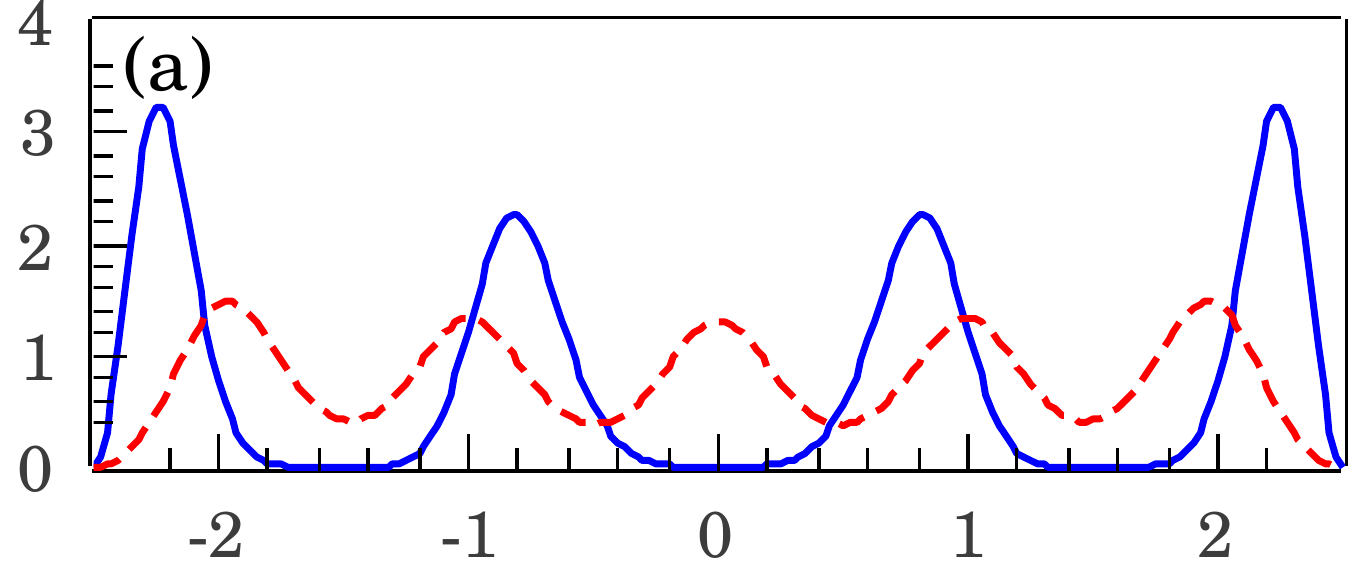}  	
		\includegraphics[width=0.32\textwidth]{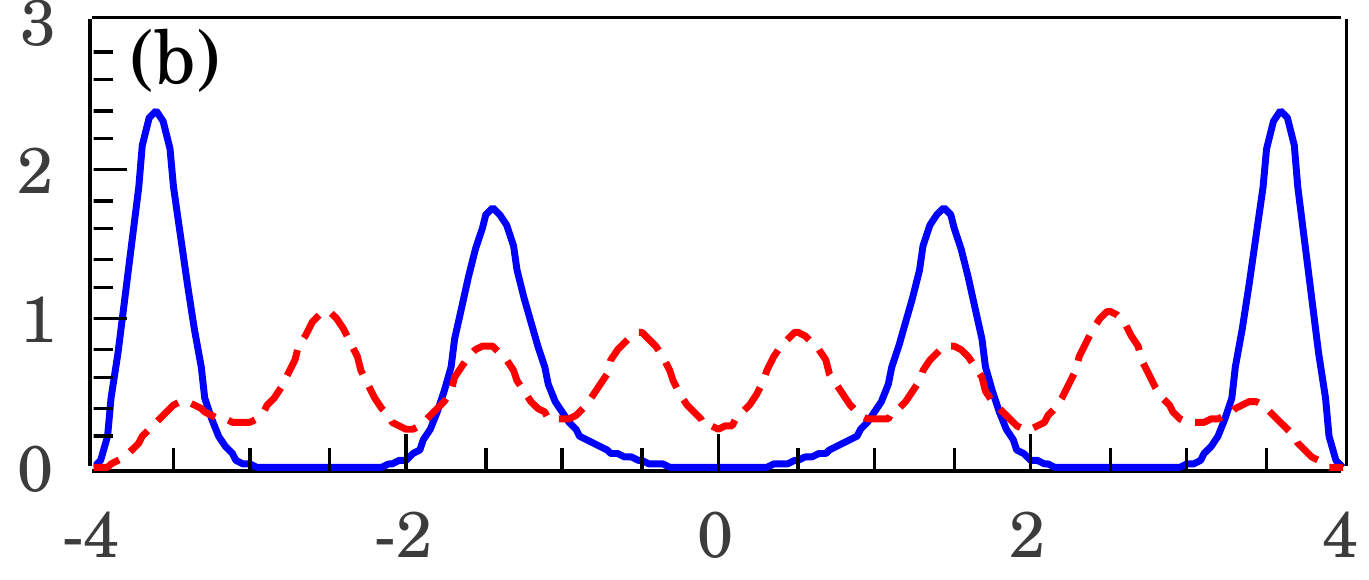}
		\includegraphics[width=0.32\textwidth]{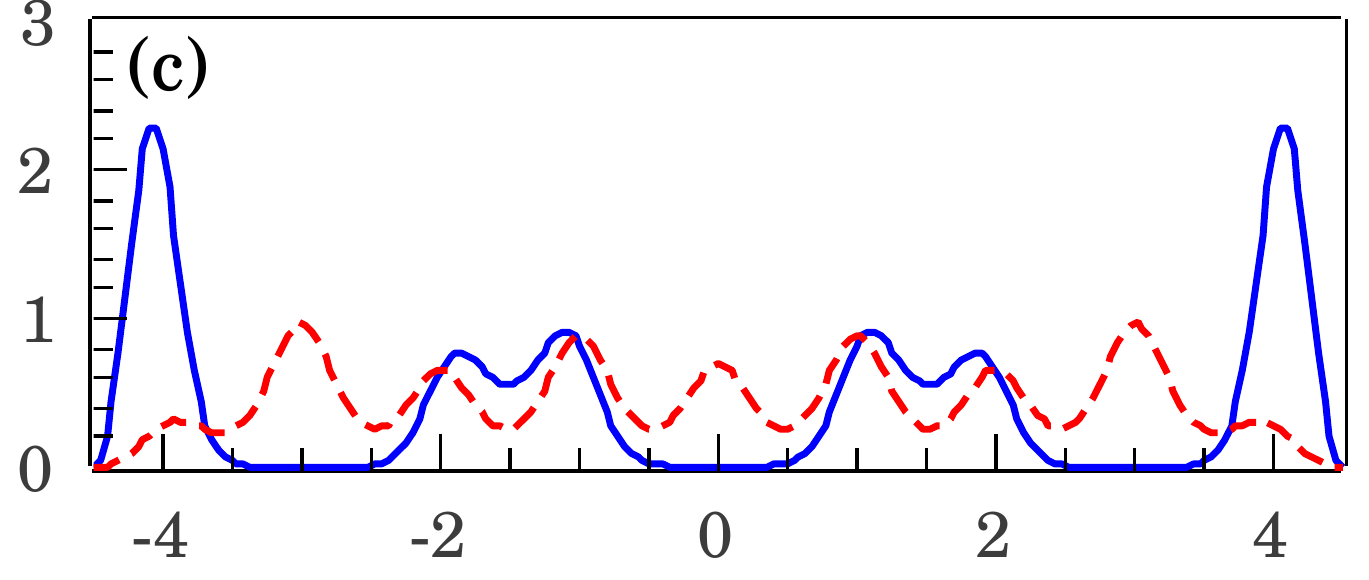} 
		
		\includegraphics[width=0.32\textwidth]{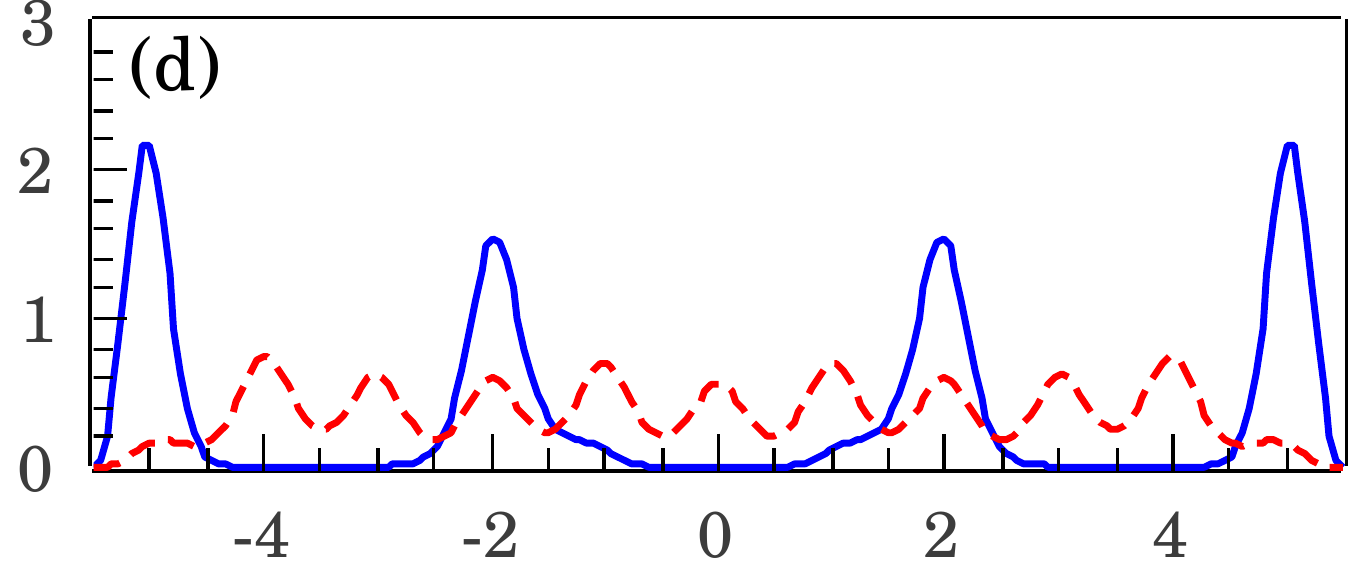}  	
		\includegraphics[width=0.32\textwidth]{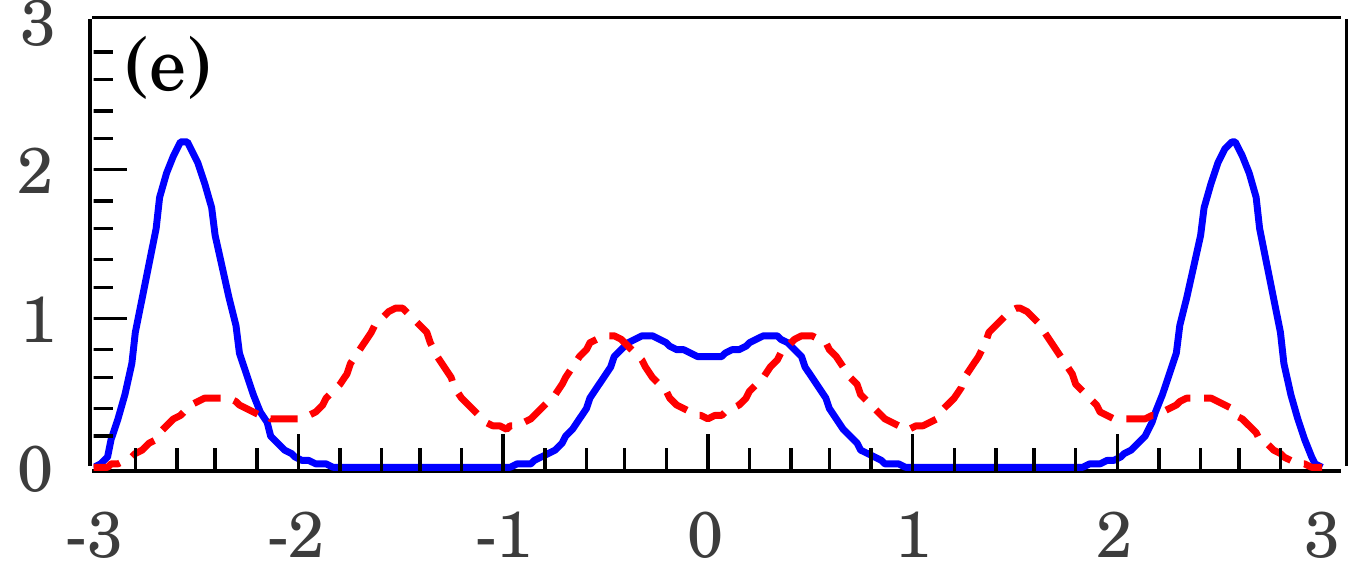}
		\includegraphics[width=0.32\textwidth]{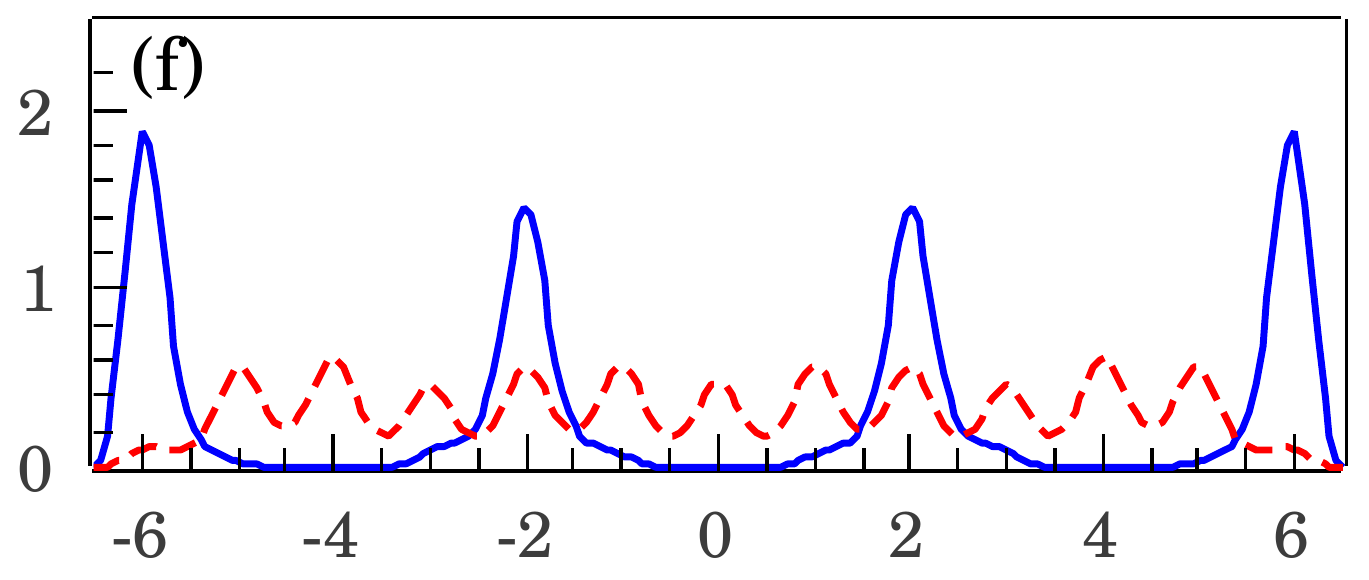}  		
	\end{minipage}
	
	$x/a$
	\caption{Similar to Fig.~\ref{fig11} but for the Wigner phases of 4 electrons at $V_0=0.001$~Ry and $a=100$~a\textsubscript{B}. The number of peaks always equals the number of electrons since the Wigner phase is determined entirely by the electron density.}\label{fig12}
\end{figure*}	

   \begin{figure*}
	\centering
	\begin{minipage}{0.04\textwidth}
		\rotatebox{90}{$\rho(x)a$}
	\end{minipage}	
	\begin{minipage}{0.95\textwidth}
		\centering	
		\includegraphics[width=0.32\textwidth]{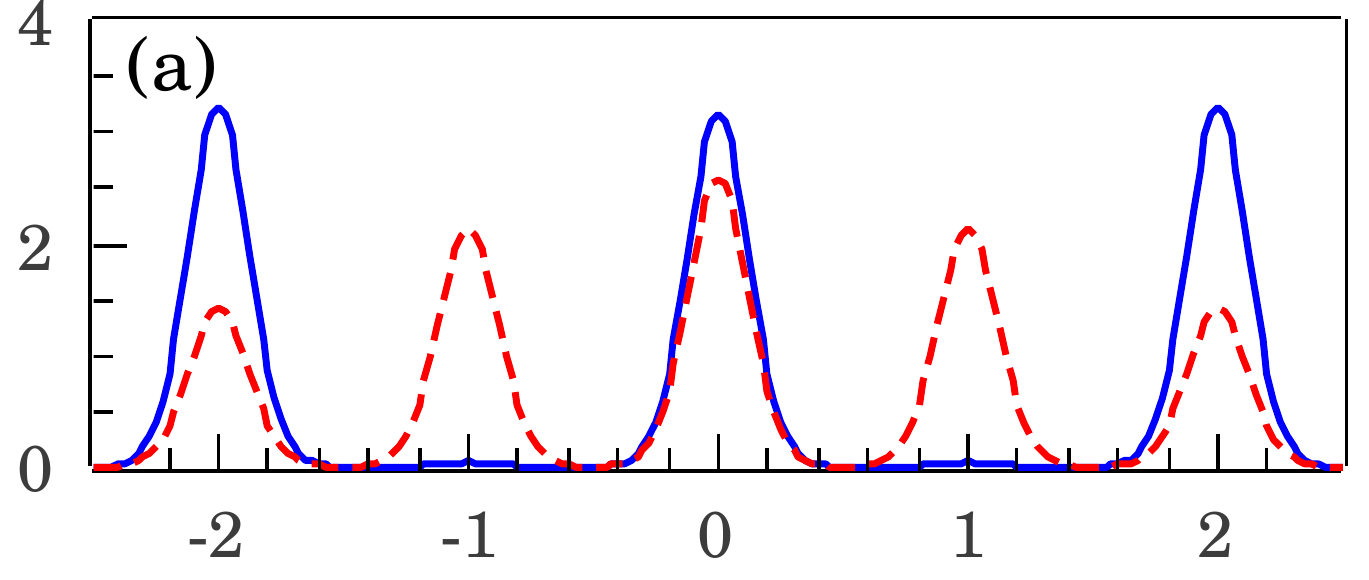}  	
		\includegraphics[width=0.32\textwidth]{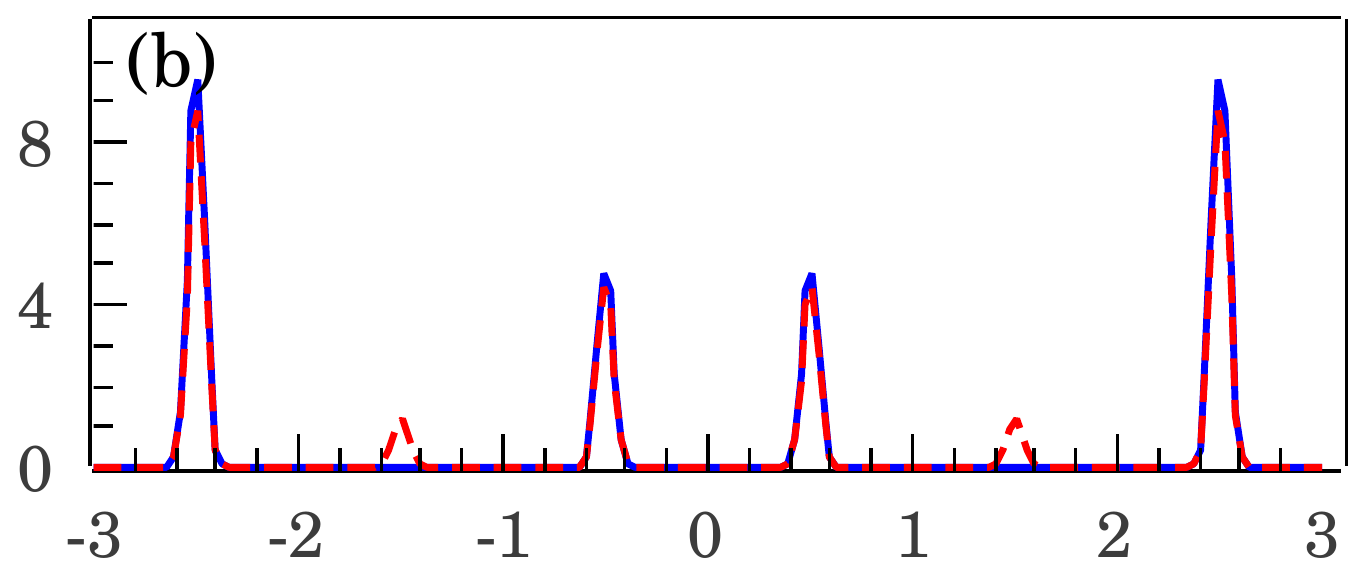}
		\includegraphics[width=0.32\textwidth]{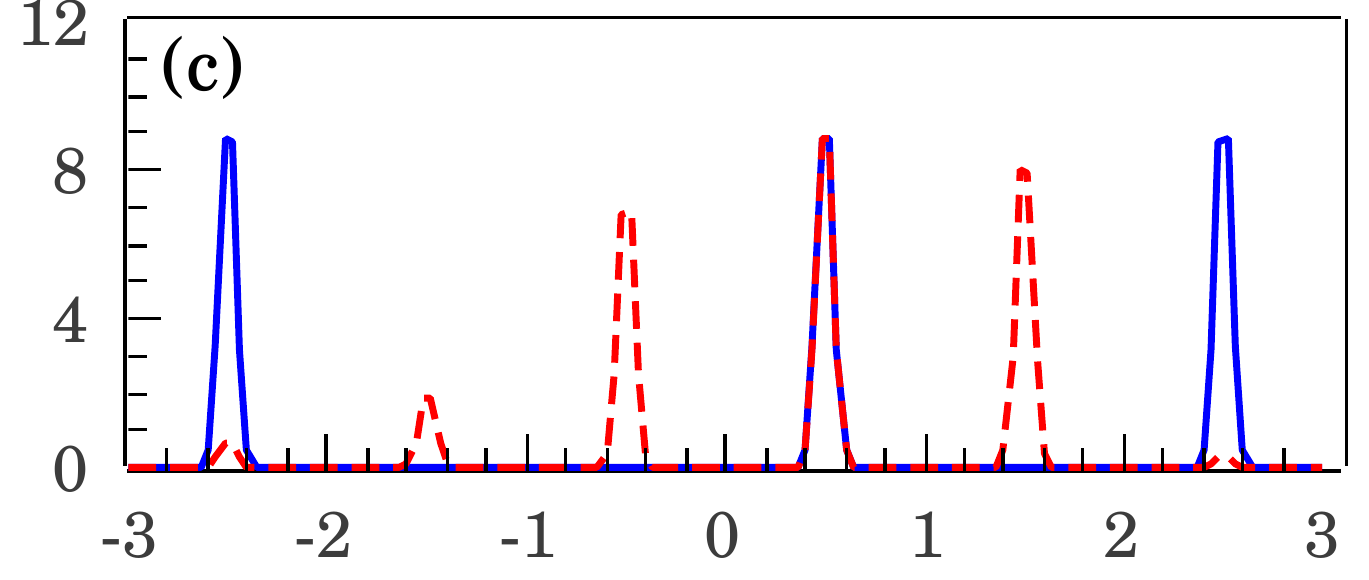} 
		
		\includegraphics[width=0.32\textwidth]{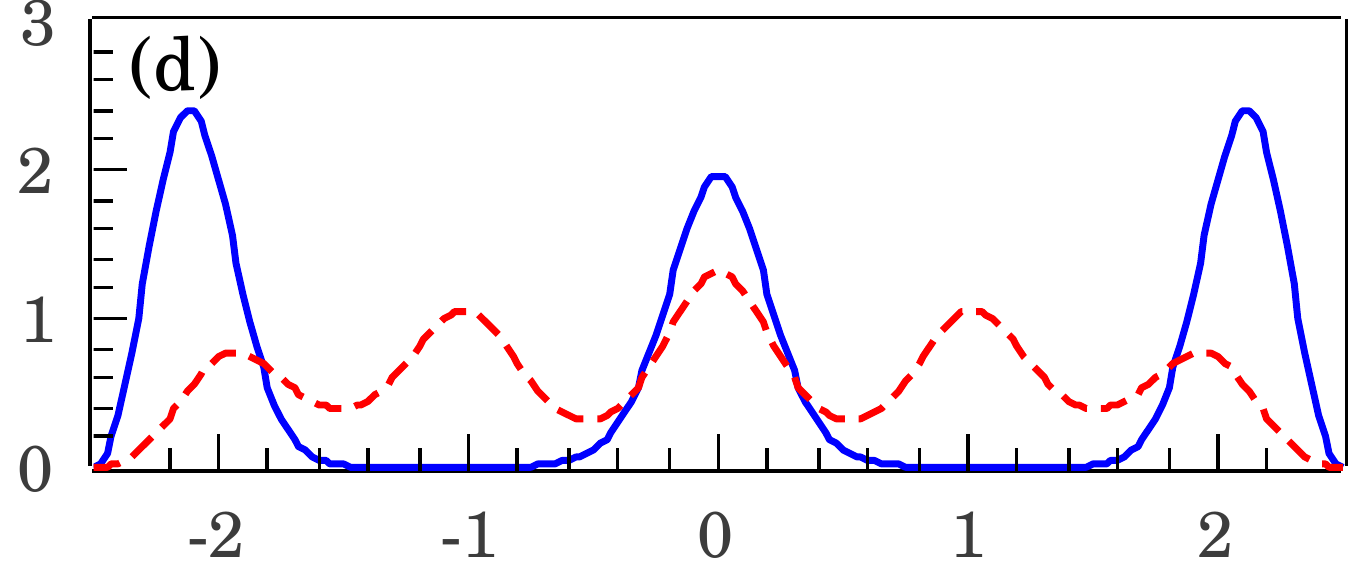} 
		\includegraphics[width=0.32\textwidth]{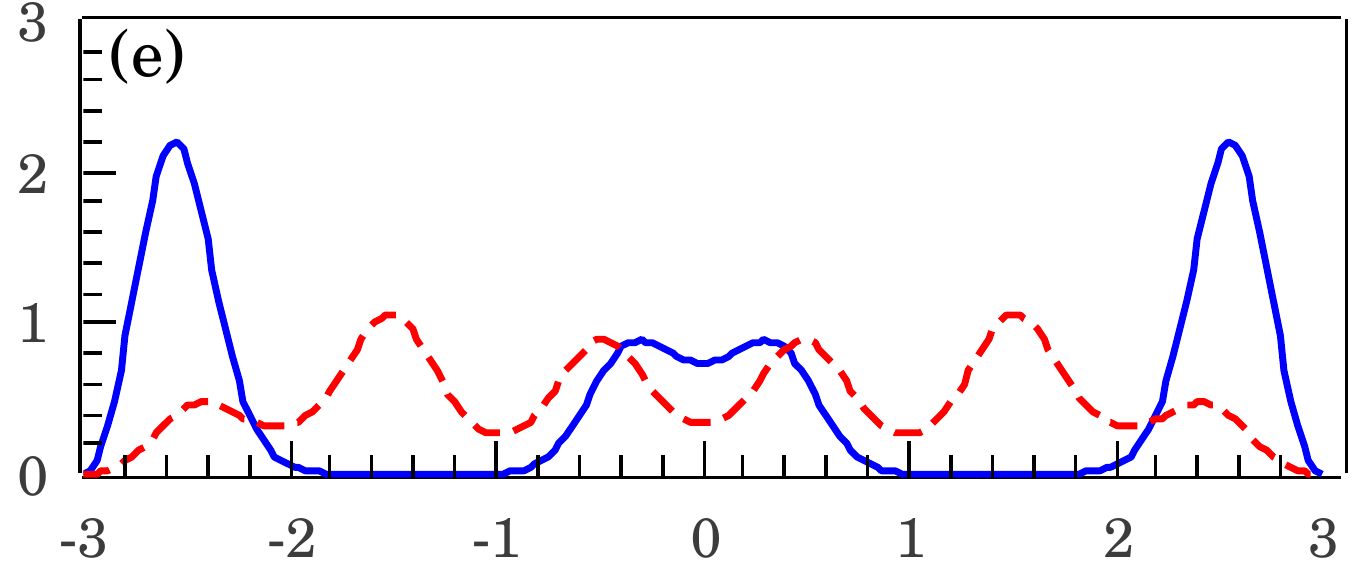} 
		\includegraphics[width=0.32\textwidth]{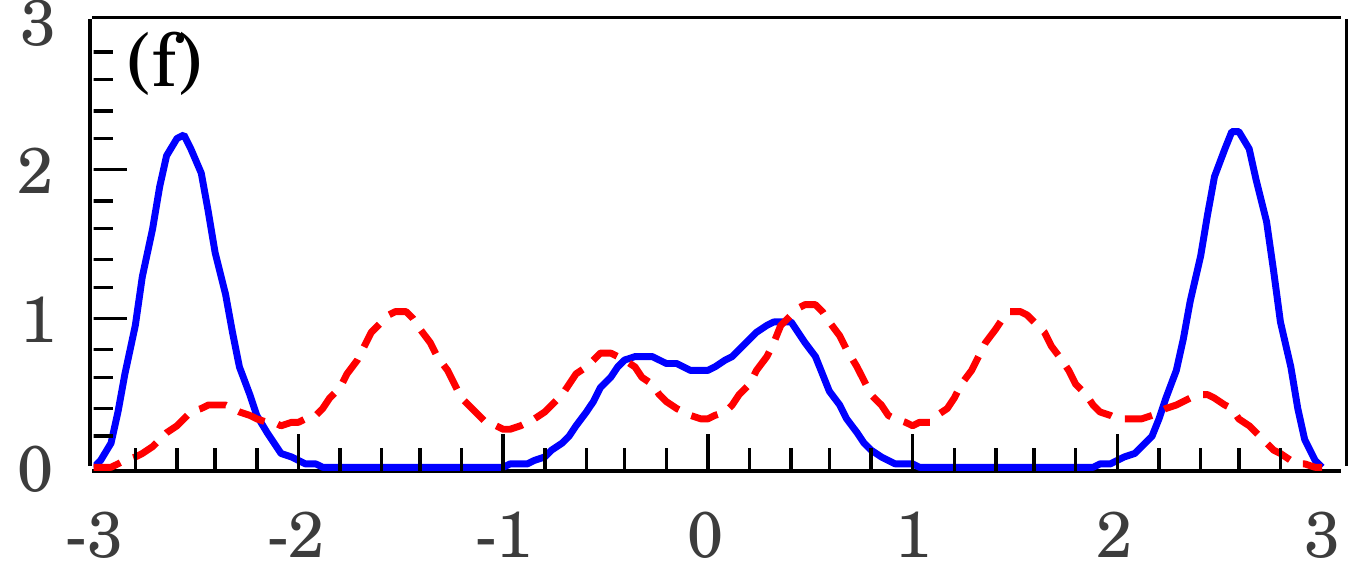} 
		
	\end{minipage}
	\vspace{0.1in}
	$x/a$
	\caption{Ground state spatial density profile of 3 spinless electrons in $N_d$-dot arrays. The  blue (dark) solid and red (gray) dashed lines represent interacting and non-interacting system's density profiles. Upper row: Mott insulator phase at $V_0=70$~Ry and $a=1$~a\textsubscript{B}. Lower row: Wigner crystal phase at $V_0=0.01$~Ry and $a=100$~a\textsubscript{B}. For Figs. (a) and (d), $N_d=5$; for Figs. (b) and (e), $N_d=6$; for Figs. (d) and (f), $N_d=6$ and the potential on the fourth dot enhanced by 10\% representing disorder effect.}\label{fig13}
\end{figure*}	 	 	

\begin{figure}
	\begin{minipage}{0.01\textwidth}
		\rotatebox{90}{$\Delta E~(\text{Ry})$}
	\end{minipage}
	\begin{minipage}{0.4\textwidth}	
		\centering
		\includegraphics[width=\textwidth]{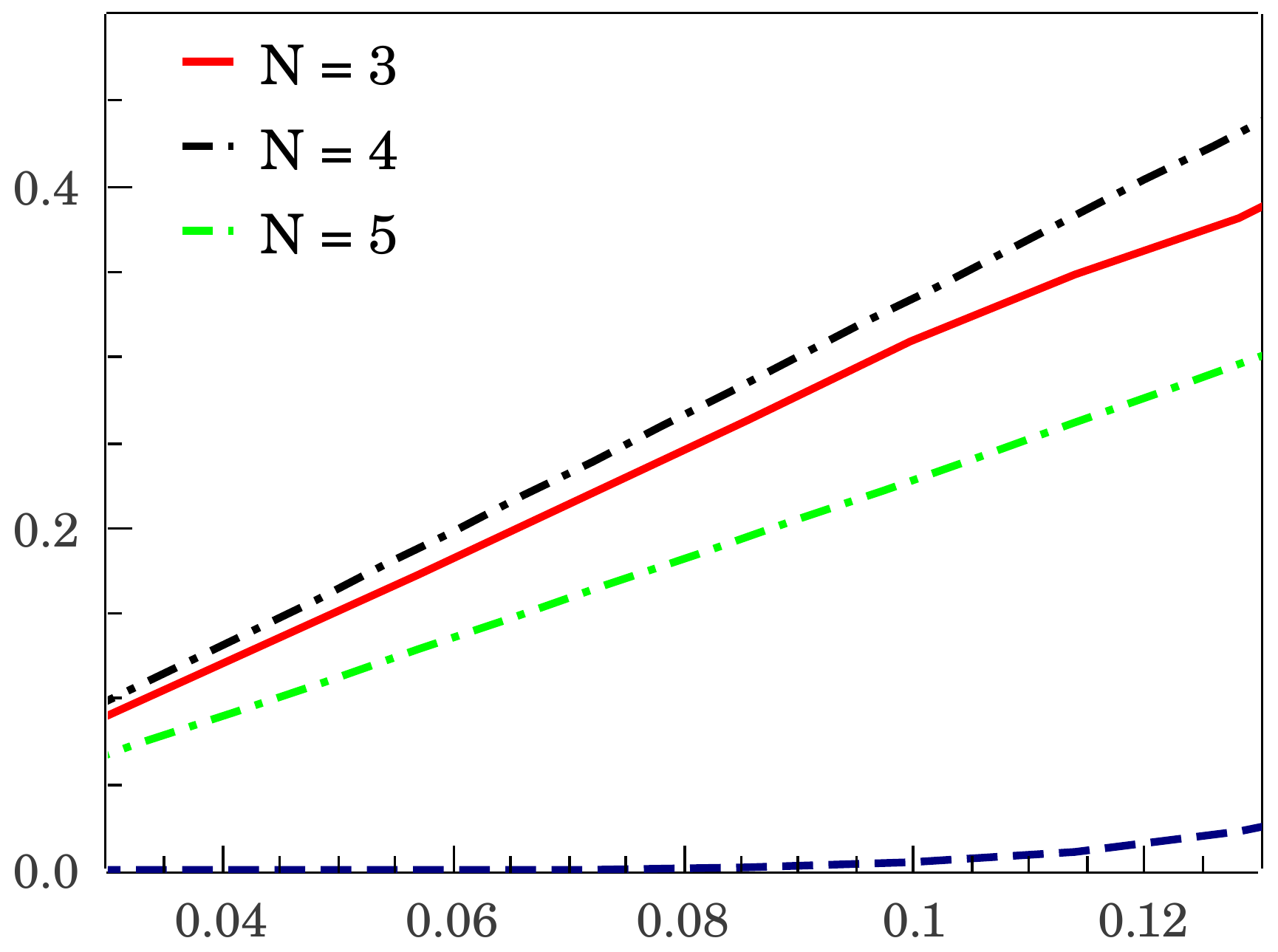}
		
		$n$~(a\textsubscript{B}\textsuperscript{-1})
	\end{minipage}
	\caption{Charge gap of an eight-dot array at fixed $V_0= 5$~Ry with $N=3$, 4 and 5 electrons. The dashed line at the bottom shows that charge gap of the non-interacting counterpart having 3 electrons. For Coulomb interacting systems, the charge gap is always non-zero for any fillings and finite densities and much larger than the non-interacting one.}\label{fig14}
\end{figure}	
    In this work, we consider a minimal continuum model of a finite coupled linear quantum dot array with a few interacting electrons to ascertain the suitability of such a semiconductor quantum dot system being used as a quantum emulator to study collective quantum ground states of systems interacting via the Coulomb interaction. The model has just two parameters, $V_0$ defining the effective background lattice or trapping confinement potential creating the dots which sets the scale for the kinetic energy through inter-dot electron hopping and the inter-electron separation $a$ defined by the electron number which sets the scale for Coulomb interaction.  We find that by varying $V_0$ and $a$, it is indeed possible to tune the system through three effective phases:  an electron liquid phase where the occupancy of all dots in the system are approximately equal indicating an extended effective metallic phase with a small charge gap and two insulating effective solid phases where the electrons are preferentially sharply localized at some of the dots leaving the other dots unoccupied.  The two solid phases, Mott and Wigner, occur at large $V_0$ and small $V_0$ respectively provided that the Coulomb interaction is strong.  The solid phases are strongly insulating with large charge gaps since electron hopping through the finite lattice is strongly inhibited by virtue of certain dots being permanently unoccupied in order to minimize the Coulomb energy. The main difference between Wigner and Mott phases, other than one (Mott) being in the strong lattice potential regime and the other (Wigner) being in the weak lattice potential regime, is that the Mott phase can in principle be incommensurate with the electron number with the number of sharp density peaks being different from the number of electrons whereas the Wigner phase is always commensurate with electron density with the number of density peaks being exactly equal to the number of electrons.  We find this interesting incommensuration to be rather fragile against background disorder. However, by averaging over many disorder configurations provided that the disorder strength is smaller than the typical Coulomb interaction energy, the incommensuration can be recovered, thus qualitatively distinguishing the Mott phase from the Wigner phase. Alternatively, of course, the experimentalists could focus on systems with little disorder, which are increasingly becoming available in semiconductor systems because of rapid advances in growth, lithography, and control techniques.

  The Mott (liquid) phase corresponds to the Coulomb blockade (collective Coulomb blockade) states with a large (small) charge gap in the finite array.  The crossover between these effective phases can be controlled by tuning the effective electron hopping.  Our work shows that small 1D arrays of coupled quantum dots can indeed be used as good solid state quantum emulators provided excellent electrostatic control is attained.
 
   We also construct an effective quantum phase diagram for the system in the $V_0-a$ space, emphasizing, however, that the physics here is purely crossover physics with no true phase transitions.  The Mott phase for large $V_0$ and $a$ goes over smoothly to the Wigner phase for vanishing $V_0$.  We show that density correlations can be used to distinguish between the Mott and Wigner phases, but this distinction is more a quantitative distinction rather than a qualitative difference since there is no phase transition separating the two phases - the Mott phase in the presence of the background lattice potential smoothly crosses over to the Wigner phase in the absence of the lattice potential. We also calculate the charge gap in the system, commenting on its different behavior in the kinetic energy versus the potential energy dominated regimes.

    Since the model parameters used for our simulations loosely correspond to Si and GaAs based quantum dot systems, we believe that some of our predictions can be experimentally verified in currently existing quantum dot arrays where the electrostatic control has now reached a very impressive level, allowing the degree of control necessary to observe the delicate interaction physics predicted in our work.   
    
     \section*{acknowledgments}
     This work is supported by the Laboratory for Physical Sciences.
	
	\appendix*
	\section{Additional simulation data}

In this Appendix we provide results for additional simulations of the quantum dot arrays to complement the results shown in the main text.

In particular, Fig.~\ref{fig11} presents results for 4 spinless electrons in an array of $N_d$ dots, staying in the effective Mott phase ($V_0=1$~Ry and $a=70$~a\textsubscript{B}) where $N_d$ varies from 5 to 13 - note that $N_d=4$ is the band insulator limit for 4 spinless electrons.  In each case, we compare the results with and without Coulomb interactions to show the clear effect of electron-electron interaction in producing the collective Mott-like ground state.  The incommensuration physics of the Mott phase can be seen in Fig.~\ref{fig11}(c) and (e) where the number of localized density peaks is more than the number of electrons (as it would be for the Wigner solid) and less than $N_d$ (as it is for the noninteracting case or the electron liquid phase).

In Fig.~\ref{fig12}, we show details in the Wigner phase, to be compared with Fig.~\ref{fig11} in the Mott phase, choosing a small $V_0=0.001$~Ry and a large $a=100$ ~a\textsubscript{B}, and comparing interacting and noninteracting situations.  The interacting system in this effective Wigner limit always has 4 peaks as determined by the electron number independent of $N_d$ in contrast to the Mott phase in Fig.~\ref{fig11}.  Thus, the number of peaks in the interacting system is capable of clearly distinguishing between Mott an Wigner phases although the sharply site-localized density profiles in these two phases are qualitatively similar.

To further emphasize the incommensuration physics in the Mott phase, we show results for 3 electrons in $N_d=5$ and 6 sites in Fig.~\ref{fig13} and comparing Wigner (small $V_0$) and Mott (large $V_0$) phases-- the number of localized density peaks in the Mott phase for $N_d=6$ is 4 (and not 3), indicating incommensuration with electron number.  In this figure, we also show in panels~\ref{fig13}(c) and (f), the effect of a weak disorder in the background potential by enhancing the potential depth on the 4\textsuperscript{th} site by 10\%.  The incommensuration in the Mott phase is suppressed by disorder in Fig.~\ref{fig13}(c) with only 3 peaks appearing here, but the Wigner result in Fig.~\ref{fig13}(f) is hardly affected by disorder.

In Fig.~\ref{fig14} we provide some additional results for the calculated charge gap.
		
   \bibliographystyle{apsrev4-1}
   \bibliography{reference}		
\end{document}